\documentstyle[psfig]{mn}                       

\newcommand\msun{\ifmmode{\hbox{M$_\odot$}}\else{M$_\odot$}\fi}
\newcommand\cat{Ca\,{\sc ii} triplet }

\title[The nIR Ca\,{\normalsize \it II} triplet: stellar population
synthesis models]
{Empirical Calibration of the near-IR Ca\,{\large \bf II} triplet -- IV. 
The stellar population synthesis models}

\author[Vazdekis et al.]  
 {A.~Vazdekis, $^{1}$\thanks{E-mail: vazdekis@ll.iac.es} 
  A.J.~Cenarro,$^2$ J.~Gorgas,$^2$  N.~Cardiel$^{2,3}$ and R.F.~Peletier$^4$\\
  $^1$Instituto de Astrof\'{\i}sica de Canarias, V\'{\i}a L\'actea s/n,
  La Laguna 38200, Tenerife, Spain\\
  $^2$Dept. de Astrof\'{\i}sica, Fac. de Ciencias
 F\'{\i}sicas, Universidad Complutense de Madrid, 28040 Madrid, Spain\\
  $^3$Calar Alto Observatory, CAHA, Apdo. 511, 04004, Almer\'{\i}a, Spain\\
  $^4$School of Physics and Astronomy, University of
 Nottingham, University Park, Nottingham NG7 2RD, UK}
\date{Accepted 2000 December 11. Received 2000 March 17}
\pagerange{\pageref{firstpage}--\pageref{lastpage}}
\pubyear{2000}

\def\LaTeX{L\kern-.36em\raise.3ex\hbox{a}\kern-.15em
    T\kern-.1667em\lower.7ex\hbox{E}\kern-.125emX}

\begin{document}

\label{firstpage}

\maketitle

\begin{abstract}

We present a new evolutionary stellar population synthesis model, which
predicts SEDs for single-age single-metallicity stellar populations, SSPs, at
resolution 1.5\AA~(FWHM) in the spectral region of the near-IR \cat feature.
The main ingredient of the model is a new extensive empirical stellar spectral
library that has been recently presented in Cenarro et al. (2001a,b; 2002),
which is composed of more than 600 stars with an unprecedented coverage of the
stellar atmospheric parameters.

Two main products of interest for stellar population analysis are presented.
The first is a spectral library for SSPs with metallicities
$-$1.7$<$[Fe/H]$<$+0.2, a large range of ages (0.1-18~Gyr) and IMF types. They
are well suited to model galaxy data, since the SSP spectra, with
flux-calibrated response curves, can be smoothed to the resolution of the
observational data, taking into account the internal velocity dispersion of the
galaxy, allowing the user to analyze the observed spectrum in its own system.
We also produce integrated absorption line indices (namely CaT$^*$, CaT and
PaT) for the same SSPs in the form of equivalent widths. 

We find the following behaviour for the \cat feature in old-aged SSPs:  {\it
i)} the strength of the CaT$^*$ index does not change much with time for all
metallicities for ages larger than $\sim$3~Gyr, {\it ii)} this index shows a
strong dependence with metallicity for values below [M/H] $\sim-0.5$ and {\it
iii)} for larger metallicities this feature does not show a significant
dependence either on age or on the metallicity, being more sensitive to changes
in the slope of power-like IMF shapes.

The SSP spectra have been calibrated with measurements for globular clusters of
Armandroff \& Zinn (1988), which are well reproduced, probing the validity of
using the integrated \cat feature for determining the metallicities of
these systems. Fitting the models to two early-type galaxies of different
luminosities (NGC~4478 and NGC~4365), we find that the \cat measurements
cannot be fitted unless a very dwarf-dominated IMF is imposed, or if the Ca
abundance is even lower than the Fe abundance. More details can be found in
Cenarro et al. (2003).

\end{abstract}

\begin{keywords}
stars: fundamental parameters  ---
globular clusters: general  ---
galaxies: abundances  ---
galaxies: elliptical and lenticular, cD  ---
galaxies: evolution  ---
galaxies: stellar content
\end{keywords}

\section{Introduction}

This is the fourth and last paper in a series devoted to understand the
behavior of the near-infrared \cat feature in stars and in stellar
populations.  The ultimate aim of this work is to use the strength of the \cat
lines in this spectral range to investigate the stellar content of early-type
galaxies, but the results are sufficiently general to be used in other areas
(starburst and active galaxies, globular clusters, or stellar astrophysics) as
well. In Cenarro et al. (2001a) (hereafter Paper~I) we presented a new
empirical stellar spectral library, mostly observed at the 1~m Jacobus Kapteyn
Telescope (JKT) at the Observatorio del Roque de Los Muchachos,
La Palma. The library is composed of 706 stars covering the \cat feature in
the near-infrared. In that paper we provided a new index definition for this
feature (CaT$^{*}$) that, among other advantages, minimizes the effects of the
Paschen series. For the stars in this library we derived a set of homogeneous
atmospheric parameters in Cenarro et al. (2001b) (hereafter Paper~II). In
Cenarro et al. (2002) (hereafter Paper~III) we described the behaviour of the
\cat through empirical fitting functions, which relate the strength of this
feature to the stellar atmospheric parameters. In the current paper, we make
use of these functions and the spectra of the stellar library to predict both
the strength of the \cat feature and spectral energy distributions, SEDs, in
the range $\lambda\lambda$8348.85-8950.65\AA~ for single-age,
single-metallicity stellar populations (SSPs) by means of evolutionary stellar
population synthesis modeling.

Traditionally, elliptical galaxies have been thought to be a uniform class of
objects, with global properties changing smoothly with mass and hosting old
and coeval stellar population. However, over the last decade, a growing body
of evidence is indicating that the formation processes and star formation
histories of, at least, an important fraction of early-type galaxies are more
complex and heterogeneous. The apparent age spread among elliptical galaxies
(Gonz\'{a}lez 1993; Faber et al. 1995; J{\o}rgensen 1999), the distribution of
element abundances (Worthey 1998; Peletier 1999; Trager et al. 2000a) and the
interpretation of the scaling relations (like the colour--magnitude or
Mg$_2$--$\sigma$ relations (Bower, Lucey \& Ellis 1992; Bender, Burstein \&
Faber 1993; Colless et al. 1999; Terlevich et al. 1999; Kuntschner 2000,
hereafter K00; Trager et al. 2000b, hereafter T00; Proctor \& Sansom 2002,
hereafter PS02), are some of the main issues in the present debate about the
evolutionary status of early-type galaxies. These studies have been possible
thanks to the comparison of observed data to the predictions of the, so
called, stellar population synthesis models. These models make use of a
theoretical isochrone, or H--R diagram, convert isochrone parameters to
observed ones, assuming empirical or theoretical prescriptions, and finally
integrate along the isochrone assuming an initial mass function, IMF
(e.g. Tinsley 1980; Bruzual 1983; Arimoto \& Yoshii 1986).

A large part of the discussion presented above is based on observations in the
Lick/IDS system (Worthey et al. 1994, hereafter W94, and references therein),
a set of 21 absorption line indices from 4100 to 6300~\AA. Based on high
quality data in that wavelength region, and stable stellar population models
we are left with the puzzling result that there is large scatter in the
luminosity-weighted mean age of the elliptical galaxies (e.g. T00). On the
other hand the colour-magnitude relation at $z=0.8$ continues to exist (Ellis
et al. 1997; Stanford et al. 1997), indicating that at least a large fraction
of elliptical galaxies in the local universe are coeval. However an
unambiguous analysis of the stellar populations has been hampered by the
fundamental age-metallicity degeneracy, i.e. the two effects cannot be fully
separated in the integrated spectrum of a composite stellar population. Some
newer, blue indices (Jones \& Worthey 1995; Worthey \& Ottaviani 1997,
hereafter WO97; Vazdekis \& Arimoto 1999, hereafter VA99; Gorgas et al. 1999)
could contribute to alleviate this situation.

In this set of papers we have increased the wavelength region that can be
studied, adding the tools to analyze a few important absorption lines in the
near-infrared. In fact, by extending the spectral coverage, we increase our
stellar population analysis constraining power, since the contribution of
different types of stars to the total luminosity varies as a function of the
spectral range. Furthermore, contrary to the ultraviolet region, where a large
fraction of the light can originate from just a few stars (e.g. Ponder et al.
1998), most of the stars contribute to the near-infrared \cat lines. Thanks to
the large stellar library and state-of-the-art evolutionary models provided in
this series of papers, observers will now be able to analyze more accurately
their galaxy \cat measurements, and compare them with the models that fit in
the blue. 

The \cat is one of the most prominent features in the near-IR spectrum of cool
stars and its potential to study the properties of stellar populations has
been extensively acknowledged in the literature. For instance, the \cat
strength has been found to correlate with globular cluster metallicity and,
therefore, it has been proposed as a metallicity indicator for old and coeval
stellar populations for the metallicity regime typical of galactic globular
clusters (Armandroff \& Zinn 1988, hereafter AZ88). This relation has been
recalibrated by Rutledge, Hesser \& Stetson (1997), which review the different
methods to measure cluster metallicities using this feature. Terlevich,
D\'{\i}az \& Terlevich (1990a) found that active galaxies exhibit \cat
strengths equal or larger than those found in normal ellipticals, which they
interpreted as due to the presence of red supergiant stars in the central
regions of these galaxies. The same approach has been followed by a number of
authors (Forbes, Boisson \& Ward 1992; Garc\'{\i}a--Vargas et al. 1993;
Gonz\'{a}lez Delgado \& P\'{e}rez 1996ab; Heckman et al. 1997; P\'{e}rez et
al. 2000). The suggested gravity sensitivity of the \cat has also been
proposed to constrain the dwarf/giant ratio in early-type galaxies (e.g. Cohen
1978; Faber \& French 1980; Carter, Visvanathan \& Pickles 1986; Alloin \&
Bica 1989). This feature has been measured in extensive galaxy samples
(e.g. Cohen 1979; Bica \& Alloin 1987; Terlevich et al. 1990b; Houdashelt
1995). Among the most interesting results derived from these studies is the
fact that the \cat strength does not seem to vary much among early-type
galaxies of different types, colours and luminosities. This result is at odds
with the strong metallicity correlation found for the globular clusters. We
refer the reader to \S~2 of Paper~I for a review of previous works on the
subject.

A reliable analysis of the \cat measurements in integrated spectra rests on
the comparison of the data with the predictions of stellar population
models. The accuracy of such predictions is highly dependent on the input
calibration of the \cat line-strengths in terms of the main atmospheric
stellar parameters, namely effective temperature, surface gravity and
metallicity. Such calibrations have been either theoretical, based on model
atmospheres, with their associated uncertainties (e.g. Smith \& Drake 1987,
1990; Erdelyi-Mendes \& Barbuy 1991; J{\o}rgensen, Carlsson \& Johnson 1992;
Chmielewski 2000), or based on empirical stellar libraries with a poor
coverage of the atmospheric parameter space (e.g. Jones, Alloin \& Jones 1984;
D\'{\i}az, Terlevich \& Terlevich 1989; Zhou 1991; Mallik 1994, 1997; Idiart,
Th\'{e}venin \& de Freitas Pacheco 1997, hereafter ITD). For a more detailed
discussion of these calibrations see Paper~I and III. The quality of this
calibration has been the major drawback of previous stellar population models
which have included predictions for the strength of the \cat feature (Vazdekis
et al. 1996, hereafter V96; ITD; Mayya 1997; Garc\'{\i}a--Vargas, Moll\'{a} \&
Bressan 1998; Moll\'{a} \& Garc\'{\i}a--Vargas 2000, hereafter MGV). In this
work we solve for these deficiencies by updating our model predictions on the
basis of the ample near-IR empirical stellar spectral library and calibrations
presented in Paper~I, II and III.  Therefore this library constitutes the main
ingredient of these new models.

Although this paper is not the first to present models for the \cat in
integrated stellar systems, it is the first to present integrated spectra at
intermediate resolution on the basis of an extensive empirical stellar
library.  Bruzual \& Charlot (2003) present spectra in this wavelength region
based on the theoretical library of Lejeune, Cuisinier \& Buser (1997) or the
observed library of Pickles (1998), both at a resolution of 10-20~\AA, i.e. an
order of magnitude larger than our models. On the other hand, Schiavon, Barbuy
\& Bruzual (2000, hereafter SBB) predicted SSP spectra at high resolution
employing a theoretical stellar library.

We must keep in mind that the empirical library used here, as well as any
empirical library, implicitly includes the chemical enrichment history of the
solar neighbourhood. For instance current models use scaled-solar abundance
ratios. However there are indications that Ca abundance is not enhanced
compared to Fe in elliptical galaxies (O'Connell 1976; Vazdekis et al. 1997,
hereafter V97;
MGV; Peletier et al. 1999; Vazdekis et al.
2001a, hereafter V01A; PS02). This discrepancy did mostly show up when
investigating the Ca~4227 line, which seems to tracks Fe lines in galaxies.
Nucleosynthesis theory (Woosley \& Weaver 1995) predicts that Ca is an
$\alpha$-element, i.e. is mainly produced in SN Type II and therefore should
follow Mg. 

The stellar population model that we describe in this paper is an improved
version, as well as an extension to the near-IR spectral range, of the model
presented in V96 and Vazdekis (1999, hereafter V99). Among the major
changes introduced here, apart of the stellar library, is the implementation
of two new IMF-shapes by Kroupa (2001, hereafter K01) (see \S~\ref{sec:imfs}),
the inclusion of the new scaled-solar isochrones of the Padova group (Girardi
et al. 2000, hereafter G00) (see
\S~\ref{sec:isochrones}), and the transformation of their theoretical
parameters to the observational plane (i.e. fluxes and colors) on the basis of
almost fully empirical photometric stellar libraries, such as the one of
Alonso, Arribas \& Mart\'{\i}nez-Roger (1999) (see
\S~\ref{sec:transformations}). We also have significantly improved the way in
which we grid our stellar library for computing a stellar spectrum of a given
set of atmospheric parameters (see \S~\ref{sec:modelsIR}).  A full description
of the behaviour of the \cat in SSPs of ages larger than 1~Gyr is provided in
\S~\ref{sec:behavior}, whilst in \S~\ref{sec:intermediate} we extend our model
predictions to some younger stellar populations, where the Asymptotic Giant
Branch (AGB) contribution to the total luminosity in this spectral range is
very important. In \S~\ref{sec:veldisp} we discuss the behaviour of the \cat
feature as a function of the spectral resolution and galaxy velocity
dispersion. Detailed comparisons with previous papers are made in
\S~\ref{sec:comparison}. In \S~\ref{sec:GC-G} we establish the validity of
these new models by making a direct comparison with galactic globular clusters
and with two galaxies of very different luminosities. We use NGC~4478 and
NGC~4365, for which we have data both in the \cat spectral region and in the
blue (V01A). A detailed analysis of galaxy spectra, however, is given in 
Cenarro et al. (2003, hereafter C03), in which we present new \cat data for a
large sample of elliptical galaxies, and show to what extent our view of the
stellar populations in elliptical galaxies may change as a result of the
indices in this wavelength region. In \S~\ref{sec:conclusions} we write our
conclusions. Finally, we provide three appendices with further details on the
model construction as well as an analysis of the main uncertainties affecting
our model predictions.

\section{The Models} 
\label{sec:models}

The predictions that we present in this paper for the near-IR spectral range
are calculated on the basis of the evolutionary stellar population synthesis
model presented in V96 and V99. In this section
we describe the main ingredients of these models stressing those updates that
have been performed since the original version. In \S~\ref{sec:imfs} we
summarize the IMF types used by our models and introduce two new IMF shapes
considered in this paper. In \S~\ref{sec:isochrones} we describe the properties
of the new isochrones of G00 that are employed here, showing
the relevant differences with respect to the ones used in V96
(mostly Bertelli et al. 1994, hereafter B94). In \S~\ref{sec:transformations} we
describe how we transform the theoretical parameters of these isochrones to the
observational plane, which is performed on the basis of extensive empirical
stellar libraries rather than using model stellar atmospheres. Finally, in
\S~\ref{sec:summaryprevious} we provide a brief summary of the main predictions
of our models.

\subsection{The initial mass function}
\label{sec:imfs}

In this paper we adopt the IMF shapes described in V96 (i.e
unimodal and bimodal) and the two IMFs recently introduced by K01. To
calculate the number of stars in the mass interval $m$ to $m+dm$ we use the
following analytical approaches for the IMF:

\begin{itemize}

\item {\it Unimodal:} a power-law function characterized by its slope $\mu$ as
a free parameter

\begin{equation}
\Phi(m)\propto m^{-(\mu+1)}.
\end{equation}

\noindent
It is worth noting that the Salpeter (1955) IMF corresponds to $\mu=1.3$.

\item {\it Bimodal:} similar to the unimodal IMF for stars with masses above
$0.6~{\rm M}_{\odot}$, but decreasing the number of the stars with lower masses by
means of a transition to a shallower slope (see V96). This IMF
is also characterized by the slope $\mu$.

\item {\it Kroupa (2001) universal:} a multi-part power-law IMF, which is
similar to the Salpeter (1955) IMF for stars masses above $0.5~{\rm M}_{\odot}$, but
with a decreasing contribution of lower masses by means of two flatter
segments.

\item {\it Kroupa (2001) revised:} a multi-part power-law IMF, in which the
systematic effects due to unresolved binaries on the single-star IMF have been
taken into account. This IMF is steeper than the universal IMF by
$\Delta\mu\sim0.5$ in the mass range $0.08~{\rm M}_{\odot} \le m < 1~{\rm M}_{\odot}$.

\end{itemize}

\begin{figure}
\centerline{\psfig{file=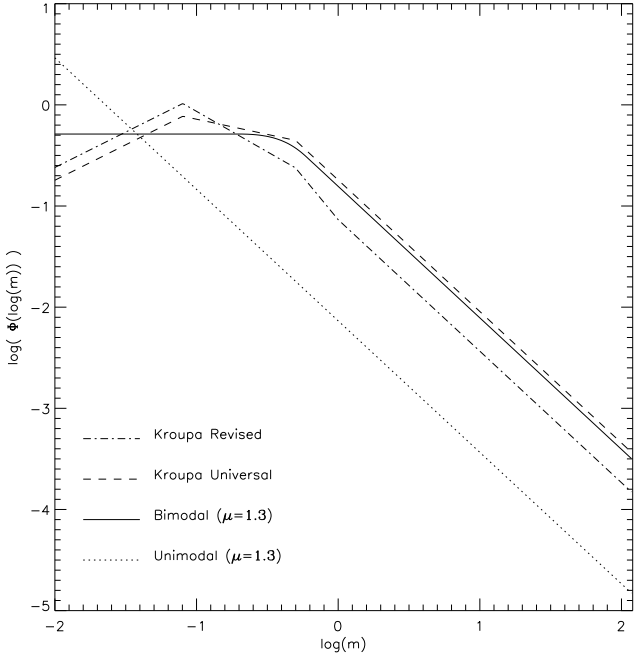,width=3.6in}}
\caption{
IMF types used by our models. Although not plotted here the slope of the
power-law ($\mu$) is allowed to vary for the unimodal and bimodal cases
} 
\label{fig:IMF} 
\end{figure}

These IMF types are plotted in Figure~\ref{fig:IMF}. Further details of the IMF
definitions are given in Appendix~\ref{ap:IMF}. We also refer the reader to
V96 and K01 for a full description of these IMFs.  

\subsection{Isochrones}
\label{sec:isochrones} 

The new model makes use of the updated version of the theoretical isochrones of
the Padova group (G00), whereas in V96 and
V99 we used the B94 isochrones and the stellar
tracks of Pols et al. (1995) for very low-mass stars (${\rm M} < 0.6 {\rm
M}_{\odot}$). These isochrones cover a wide range of ages and  metallicities
and include the latest stages of the stellar evolution through the thermally
pulsing AGB regime to the point of complete envelope ejection (employing a
synthetic prescription). The G00 set has a larger age resolution for ages lower
than 10~Gyr, however the largest metallicity covered is $Z=0.03$ (instead of
$Z=0.05$ as in B94). Hereafter we will refer to the metallicity
following the relation 

\begin{equation} 
{\rm [M/H]}=\log(Z/Z_{\odot}),
\end{equation}

\noindent  
where $Z_{\odot}=0.019$ according to the reference value adopted by G00. 
The lower mass cutoff of the new isochrones has been extended down
to $0.15 {\rm M}_{\odot}$. The input physics of the isochrones have been updated with
an improved version of the equation of state, the opacities of Alexander \&
Ferguson (1994) (which make the Red Giant Branch, RGB, slightly hotter than in
B94 isochrones, see Figure~\ref{fig:isochrones}) and a milder
convective overshoot scheme. A helium fraction was adopted according to the 
relation: $Y\approx0.23+2.25Z$. 

It is worth noting that, although giant elliptical galaxies show non-solar
abundance trends (e.g. Peletier 1989; Worthey, Faber \& Gonz\'{a}lez 1992; K00;
V01A), only scaled-solar abundance ratios are adopted for this set of
isochrones. In particular, these galaxies seem to show an enhancement of the
$\alpha$-elements (e.g. O, Ne, Mg, Si, S, etc.) with respect to Fe. For a
given total metallicity, the $\alpha$-enhanced mixtures yield lower opacities,
which translates into an increase of the temperature of the stars in the Main
Sequence (MS) and the RGB phases (e.g. Salaris \& Weiss 1998, hereafter SW98;
VandenBerg et al. 2000; Salasnich et al. 2000, hereafter S00; Kim et
al. 2002).  Atomic diffusion is not included in the present set of isochrones,
although this phenomenon decreases the temperatures of the stars of the
turnoff (e.g.  Salaris, Groenewegen \& Weiss 2000), an effect which has been
found to be useful to be able to fit the low line-strength values of the
Balmer lines in the spectra of metal-rich globular clusters (Vazdekis et al.
2001b, hereafter V01B).

\subsection{Transformation to the observational plane} 
\label{sec:transformations}

To transform the theoretical parameters of the isochrones to the observational
plane, i.e. colours and magnitudes, we make use of relations inferred on the
basis of extensive empirical photometric stellar libraries (rather than by
implementing theoretical stellar atmospheric spectra) to obtain each colour as
a function of the temperature, metallicity and gravity. We use the
metallicity-dependent empirical relations of Alonso, Arribas \&
Mart\'{\i}nez-Roger (1996, 1999), for dwarfs and giants respectively (each
stellar sample is composed of $\sim$~500 stars). The derived temperature scale
is based on the IR-Flux method and, therefore, is marginally dependent on the
model atmospheres. This treatment for the giants is the most important
difference with respect to the models of V96, where we used
the empirical calibration of Ridgway et al. (1980) to obtain the $V-K$ colour
from $T_{\rm eff}$, after which we applied the colour-colour conversions of
Johnson (1966). The empirical (not the theoretical) compilation of Lejeune,
Cuisinier \& Buser (1997, 1998) (and references therein) are used for the
coolest dwarfs ($T_{\rm eff}\le4000$~K) and giants ($T_{\rm eff}\le3500$~K),
respectively, for solar metallicity; a semi-empirical approach was applied to
other metallicities on the basis of these relations and the  model atmospheres
of Bessell et al. (1989, 1991) and the library of Fluks et al. (1994). The
empirical compilation of Lejeune et al. was also used for stars
with temperatures above $\sim$~8000~K.

Finally, we use the metal-dependent bolometric correction given by Alonso,
Arribas \& Mart\'{\i}nez-Roger (1995, 1999) for dwarfs and giants,
respectively. For the Sun we adopt the bolometric correction 
$BC_{\odot}=-0.12$, with a bolometric magnitude of 4.70 according to these
authors.

\subsection{Summary of the previous model predictions} 
\label{sec:summaryprevious}

Using data obtained at different spectral ranges, or combining spectroscopic
and photometric information, increases our constraining power for interpreting
the stellar populations (e.g. V97). Furthermore, the use of a
set of model predictions for a given spectral range, such as the ones presented
in this paper and those of other authors for a different spectral range, might
drive to systematic effects due to differences in the adopted model
prescriptions (e.g. Charlot, Worthey \& Bressan 1996). Furthermore, it is well
known that the ages or metallicities inferred by means of stellar population
synthesis modeling should be taken on a relative basis (e.g. V01B).
Therefore we briefly summarize below our model predictions for those
interested in applying the models presented in this paper.

Broadband colours, mass-to-light ratios and line-strengths for the main
features in the Lick/IDS system were predicted in V96. A new
version of these models include the H$\delta$ and H$\gamma$ indices of WO97,
as well as the break at $\sim$4000~\AA~ of Gorgas et al.
(1999). SSP spectral energy distributions in the blue and visual at resolution
1.8\AA~(FWHM) were presented in V99. New H$\gamma$ index
predictions can be found in VA99 and V01B.
Finally, optical and near-IR surface brightness fluctuations magnitudes and
colours (including the WFPC2-HST
filter system) were presented in Blakeslee, Vazdekis \& Ajhar (2001), where we
include a description of the updated models. All these predictions are
available at the web page: {\tt http://www.iac.es/galeria/vazdekis/}.

\section{New Model Predictions for the near-infrared}
\label{sec:modelsIR}

Using the model ingredients described in \S~\ref{sec:models} and the new
stellar spectral library presented in Paper~I and II, we follow two different
approaches two build up two set of model predictions. First we compute
spectral energy distributions at resolution (FWHM=1.5\AA) for different
metallicities, ages and IMFs. In \S~\ref{sec:stars} we further the preparation
of our stellar library for stellar population synthesis
modeling, whereas in \S~\ref{sec:ss} and Appendix~\ref{ap:boxes} we describe
our model and computing details. In Appendix~\ref{ap:uncertain} we describe
some tests that we have performed to show the main uncertainties affecting the
obtained model predictions, as well as to probe the robustness of our approach.
For the second set of model predictions, which is described in
\S~\ref{sec:ff}, we make use of the empirical fitting functions presented in
Paper~III (based on the same stellar spectral library) to calculate
the strengths of the \cat feature by means of the index definitions given in
Paper~I. Finally, in \S~\ref{sec:summaryIR} we summarize these two sets of model
predictions and describe the spectral properties and parameter coverage of the
synthesized SSP spectral energy distributions.

\subsection{The near-IR empirical stellar spectral library}
\label{sec:stars}

The main ingredient of the models presented in this paper is a new stellar
library covering the spectral range around the near-IR \cat feature at
resolution 1.5~\AA~(FWHM). A full description of this library was given in
Paper~I, where we also propose a new index definition for this feature (i.e.
CaT$^*$). In order to make this stellar library useful for stellar population
synthesis modeling, we require an homogeneous set of accurate atmospheric
parameters ($T_{\rm eff}$, $\log g$, [Fe/H]) for the stellar sample (see
Paper~II). As pointed out in V99, this step is particularly
important to avoid systematic trends in the parameters among different authors.
In this section we give further details on the preparation of the stellar
library for its implementation in the model. This requires us to identify those
stars whose spectra might not be properly representing a given set of
atmospheric parameters (\S~\ref{sec:library}). Another important step is the
characterization of the parameter coverage of the stellar library. This step,
which is described in \S~\ref{sec:coverage}, allows us to to understand the
limitations of the models presented in \S~\ref{sec:modelsIR}. 

\subsubsection{Preparation of the stellar library}
\label{sec:library}

To optimize the stellar library we have performed a second selection on the
stars to use for the stellar population modeling. For this purpose we
identified all the spectroscopic binaries, making use of the SIMBAD database,
as well as those stars with high signal of variability ($\Delta
V > 0.10$~mag). We used the Combined General Catalogue of Variable Stars of
Kholopov et al. (1998) (the electronically-readable version provided at CDS). 
Moreover, we removed from the sample a number of these stars that were found
irrelevant since the stellar library contains a reasonable large number of
stars with similar atmospheric parameters with no such high variability
feature. The removed stars were: HD~10476, HD~36079, HD~113139, HD~113226,
HD~125454, HD~138481, HD~153210 and HD~205435. We refer the reader to
\S~\ref{sec:AdoptedSpectrum} and Appendix~\ref{ap:boxes} for a full
description of how we deal with these  stars when calculating a representative
stellar spectrum for a given set of atmospheric parameters.

A number of stars were removed due to the high residuals obtained when
calculating the \cat fitting functions (see Paper~III).  These stars were:
HD~17491, HD~35601, HD~42475, HD~115604, HD~120933, HD~121447, HD~138279,
HD~181615, HD~217476 and HD~222107.

We did remove from the original stellar sample most of the stars for which at
least one of the three atmospheric parameters was lacking (see Table~5 of
Paper~II). However we kept 47 of these stars, with unknown metallicities, but
whose temperatures were either larger than 9000~K or smaller than 4000~K.

We are interested in achieving the largest possible spectral resolution for
our model predictions, i.e. FWHM~=~1.5~\AA. We therefore removed from the
library the few stars observed with instrumental configurations providing
$\sim$~2.1~\AA, as described in Table~2 of Paper~I  (i.e. those corresponding to
the observing runs 5 and 6). These stars were those corresponding to NGC~188
as well as NGC~7789 676, NGC~7789 859, NGC~7789 875, NGC~7789 897 and NGC~7789
971. The remaining stars whose spectra were obtained at resolutions slightly
better than 1.5~\AA~ were broadened to match this resolution (see Paper~I).

For different technical reasons we removed from the sample an additional number
of stars. In particular, a subset of stars were discarded because their spectra
were of low signal-to-noise (i.e. M~5~II-53, M~92~I-10 and M~92~XII-24).
M~71~1-63 and HD~232979 were removed for being affected by cosmic rays and
M~67~F119 for the poor flux calibration quality. 

As a result of this cleaning process we obtained a subsample composed of 616
stars. A final step was to revise the atmospheric parameters presented in
Paper~II for a number of stars. This is motivated by the evidence that their
line-strength values significantly deviate from the expected general trends
predicted by the fitting functions. In particular, we modified the $T_{\rm
eff}$ of HD~167006 from 3470~K to 3650~K in order to bring it in better
agreement with the value predicted by a new set of fitting functions, which
calculates the slope of the spectrum around the \cat feature due to the TiO
molecular bands (Cenarro 2002). Finally, we also changed
the metallicity adopted for the stars of M~71 from $-$0.70 (Carretta \& Gratton
1997) to $-$0.84 in order to achieve a better agreement with the predictions of
the fitting functions presented in Paper~III for the \cat feature. As described
in \S~3.3 of Paper~III, the stars of this globular cluster showed systematic
residuals in comparison to the strengths predicted by the fitting functions
(i.e. $\Delta {\rm CaT}^*=-0.35$), which translates to the metallicity value
that we adopt here. We refer the reader to that paper for an extensive
discussion on the subject.

\begin{figure}
\centerline{\psfig{file=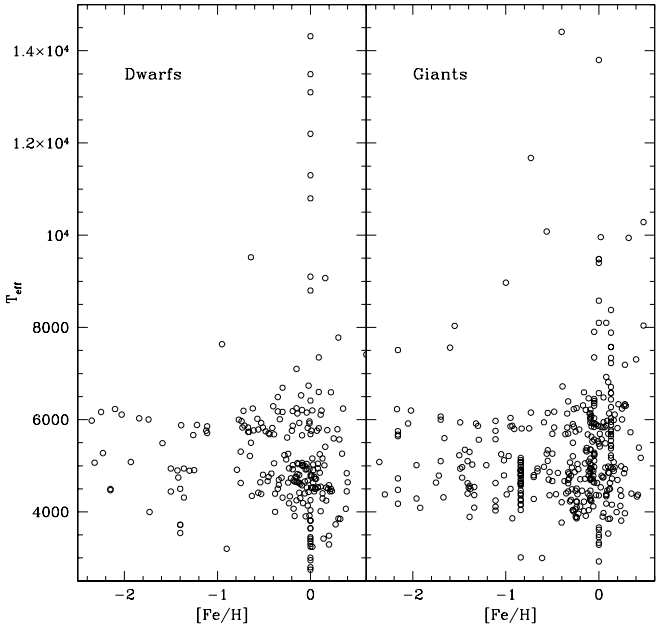,width=3.6in}}
\caption{The fundamental parameter coverage of the subsample of selected stars}
\label{fig:PARAM_CAT}
\end{figure}

\subsubsection{Stellar atmospheric parameter coverage}
\label{sec:coverage}

Figure~\ref{fig:PARAM_CAT} shows the parameter coverage of the selected
subsample of stars. The adopted values for the T$_{\rm eff}$, $\log g$ and
[Fe/H] of the stars of our sample were extensively discussed in Paper~II. This
figure includes the corrections described in \S~\ref{sec:library}. The figure
shows that all types of stars are well represented for solar metallicity. We
also see a good coverage of typical dwarfs and giants of $5000<T_{\rm
eff}<6500$~K and $4000<T_{\rm eff}<5500$~K respectively, for metallicities in
the range $-0.7\leq{\rm [Fe/H]}\leq+0.2$.

Unfortunately, very metal rich stars, particularly dwarfs, are scarce in the
sample preventing us to provide predictions for SSPs of metallicities larger
than +0.2. We also note that very cool dwarfs and giants of non solar
metallicity are scarce. These dwarfs are particularly useful for predicting
stellar populations with very steep IMFs, i.e. dwarf-dominated, whereas the
effect of these low metallicity giants will be discussed below. An important
gap is the absence of metal-poor dwarfs of $T_{\rm eff}>6500$~K. The lack of
these stars do not allow us to predict SSPs of ages younger than $\sim$6~Gyr
for [Fe/H] $\sim-0.7$ and younger than $\sim2.5$~Gyr for [Fe/H] $\sim-0.4$,
since they will be representing the predicted Turnoff for these stellar
populations. Overall we conclude that the metallicity range where we are
mostly safe is $-0.7\leq{\rm [Fe/H]}\leq+0.2$. With more caveats we also can predict
stellar populations for lower metallicities where the contribution of dwarf
stars with temperatures above $T_{\rm eff}>6500$~K is not significant, and the
Horizontal Branch is not too blue (note in Fig.~\ref{fig:PARAM_CAT} the lack
of metal-poor giants with temperatures larger than $\sim6000$~K). Therefore in
this metallicity range we mostly need to focus on SSPs with ages larger than
$\sim10$~Gyr, but smaller than $\sim13$~Gyr. 

\subsection{Spectral synthesis of SSPs}
\label{sec:ss}

Predicting spectral energy distributions for stellar populations at
intermediate or moderately high resolution, rather than the strengths of a
given number of features (see \S~\ref{sec:ff}), requires a complete stellar
spectral library where all type of stars are well represented, and whose
spectra are of high quality and flux calibrated. Theoretical libraries are
difficult to achieve due to the limitations of the input physics and the
computing power. On the other hand, the empirical libraries actually suffer
from the difficulties in observing a good number of representative stars of all
types whose atmospheric parameters are accurately known. Spectral energy
distributions at very low resolutions for stellar populations covering a large
range of ages and metallicities were predicted, either based on the Kurucz
(1992) theoretical stellar library (e.g. Bressan, Chiosi \& Fagotto 1994;
Kodama \& Arimoto 1997; Kurth, Fritze-V.Alvensleben \& Fricke 1999; Poggianti
et al. 2001) or empirical stellar libraries (Bruzual \& Charlot 1993).  However
the low dispersion of the predicted spectra did not allow to measure reliable
line-strengths. 

More recently, the availability of such libraries at higher dispersion has made
it possible to follow the spectral synthesis approach for the blue spectral
range. In V99, we used the extensive empirical stellar library of
Jones (1999), composed of more than 600 stars, to predict spectra of SSPs at
resolution 1.8\AA~(FWHM). This approach allowed us to analyse the entire
spectrum of a galaxy at one time and to measure line-strengths at the
resolution imposed by the instrumental configuration of the observational data,
or to galaxy internal velocity dispersion. This represents a great advantage
over previous models, such as those of Worthey (1994) or V96,
which only predicted the strengths of a given number of features under
a determined instrumental configuration and resolution, i.e. the Lick/IDS
system (Gorgas et al. 1993; W94). Note that the old models
required the observers to transform their data to the characteristics of the
model predictions. A different approach was followed by SBB,
who synthesized SEDs of SSPs for the near-IR spectral range on
the basis of a fully synthetic stellar spectral library by SBB.
In the following, we describe in detail how we make use of our stellar
library for predicting near-IR spectra at resolution 1.5\AA~(FWHM) for SSPs of
different ages, metallicities and IMF slopes. Although we have followed a
procedure similar to that of V99 for the optical spectral range we
have updated the method as described below.

To predict any integrated observable (e.g. colours, line-strengths, SEDs)
of an SSP of age $t$ and metallicity [M/H] the synthesis code yields a
stellar distribution on the basis of the adopted isochrones and IMF (see
V96 for more details). Therefore, the integrated spectrum of
the SSP, $S_{\lambda}(t,{\rm [M/H]})$, is calculated in the following way:

\begin{eqnarray}
S_{\lambda}(t,{\rm [M/H]})&=&\int_{m_{\rm l}}^{m_{\rm t}}
S_{\lambda}(m,t,{\rm [M/H]})N(m,t)\times \nonumber \\
&& F_{\Delta\lambda_{\rm ref}}(m,t,{\rm [M/H]})dm,
\label{eq:SSP}
\end{eqnarray}

\noindent 
where $S_{\lambda}(m,t,{\rm [M/H]})$ is the empirical spectrum corresponding to a
star of mass $m$ and metallicity [M/H] which is alive at the age assumed for
the stellar population $t$. $F_{\Delta\lambda_{\rm ref}}(m,t,{\rm [M/H]})$ is its
corresponding absolute flux at a certain wavelength reference interval,
$\Delta\lambda_{\rm ref}$, and $N(m,t)$ is the number of this type of stars, which
depends on the adopted IMF. $m_{\rm l}$ and $m_{\rm t}$ are the stars with the
smallest and largest stellar masses, respectively, which are alive in the SSP.
The upper mass limit depends on the age of the stellar population. 

The spectrum to be assigned to each of the stars required by Eq.~\ref{eq:SSP}
is selected from the empirical stellar database, whereas the corresponding
absolute flux is assigned following the prescriptions of our code, which are
based on the extensive empirical photometric stellar libraries described in
\S~\ref{sec:transformations}. In \S~\ref{sec:flux} and
\S~\ref{sec:AdoptedSpectrum} we describe how we calculate
$F_{\Delta\lambda_{\rm ref}}(m,t,{\rm [M/H]})$ and $S_{\lambda}(m,t,{\rm [M/H]})$,
respectively.

\subsubsection{Assigned stellar flux}
\label{sec:flux}

We need to find a common reference wavelength interval, $\Delta\lambda_{\rm ref}$,
from which we can scale S$_{\lambda}(m,t,{\rm [M/H]})$. We selected  for this
purpose the spectral range $\lambda\lambda$8475-8807\AA~around the peak of
the I filter of Johnson (1966). We have chosen the Johnson (1966) I filter
because its effective wavelength is closer to the spectral range covered by
our stellar spectra than the I filter of the Cousins system (Bessell 1979). We
divide our stellar spectra by the average flux in the selected wavelength
reference interval. To calculate the absolute flux in this spectral region for
each of the stars requested by the code, $F_{\Delta\lambda_{\rm ref}}(m,t,{\rm [M/H]})$,
we first used equation (6) of Code et al. (1976) which provides a relation
between the absolute $V$ flux and the calibrated bolometric correction in order
to infer equation (2) of V99 from which we derive the following
relation:

\begin{equation}
\frac{F_{I}}{F_{{\rm bol}_{\odot}}} = \frac{10^{+0.4(V-I)}}{10^{+0.4(M_{V}-3.762)}},
\end{equation}

\noindent
where $F_{I}$ is the absolute flux in the Johnson I-band. The next step is to
calculate the absolute flux per angstrom in the selected reference interval:

\begin{equation}
F_{\Delta\lambda_{\rm ref}}=\frac{ f_{n} F_{I} }{\Delta\lambda_{\rm ref}},
\end{equation}

\noindent  
where $f_{n}$ represents the fraction of $F_{I}$ corresponding to the
normalization wavelength interval, i.e. $\Delta\lambda_{\rm ref}$=8475--8807~\AA.
This factor is calculated on the basis of the empirical stellar spectral
library of Pickles (1998), for which we have compared the flux in the I filter
and in the normalization region, obtaining the following relations

\begin{equation}
 \begin{array}{r@{}c@{}l@{}@{}l@{}l@{}}
f_{n} & = & -1.620675+0.496916\log{T_{\rm eff}} , & \log{T_{\rm eff}} < 3.55 ,
& {\rm dwarfs}\\
f_{n} & = & -0.163463+0.086852\log{T_{\rm eff}} , & \log{T_{\rm eff}} < 3.55 ,
& {\rm giants} \\
f_{n} & = & +0.156952-0.003324\log{T_{\rm eff}} , & \log{T_{\rm eff}} \ge 3.55 &. \\
 \end{array}
 \label{eq:fitfn}
\end{equation}

\begin{figure}
\centerline{\psfig{file=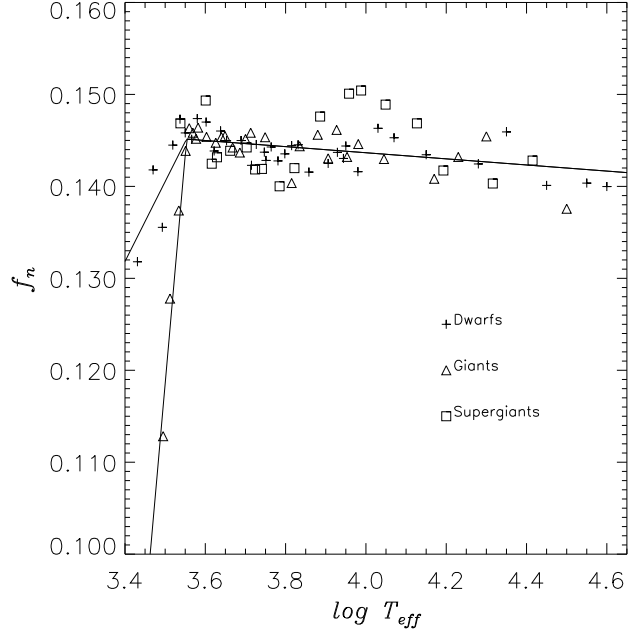,width=3.6in}}
\caption{
Normalization wavelength interval flux correcting factor versus $\log T_{\rm eff}$. 
The fit is given in Eq.~\ref{eq:fitfn}.
} 
\label{fig:FACTOR} 
\end{figure}

\noindent
Note that $f_{n}$ is nearly constant, except for M stars with temperatures
lower than $\sim3500$~K. See Figure~\ref{fig:FACTOR}.

\begin{figure*}
\centerline{\psfig{file=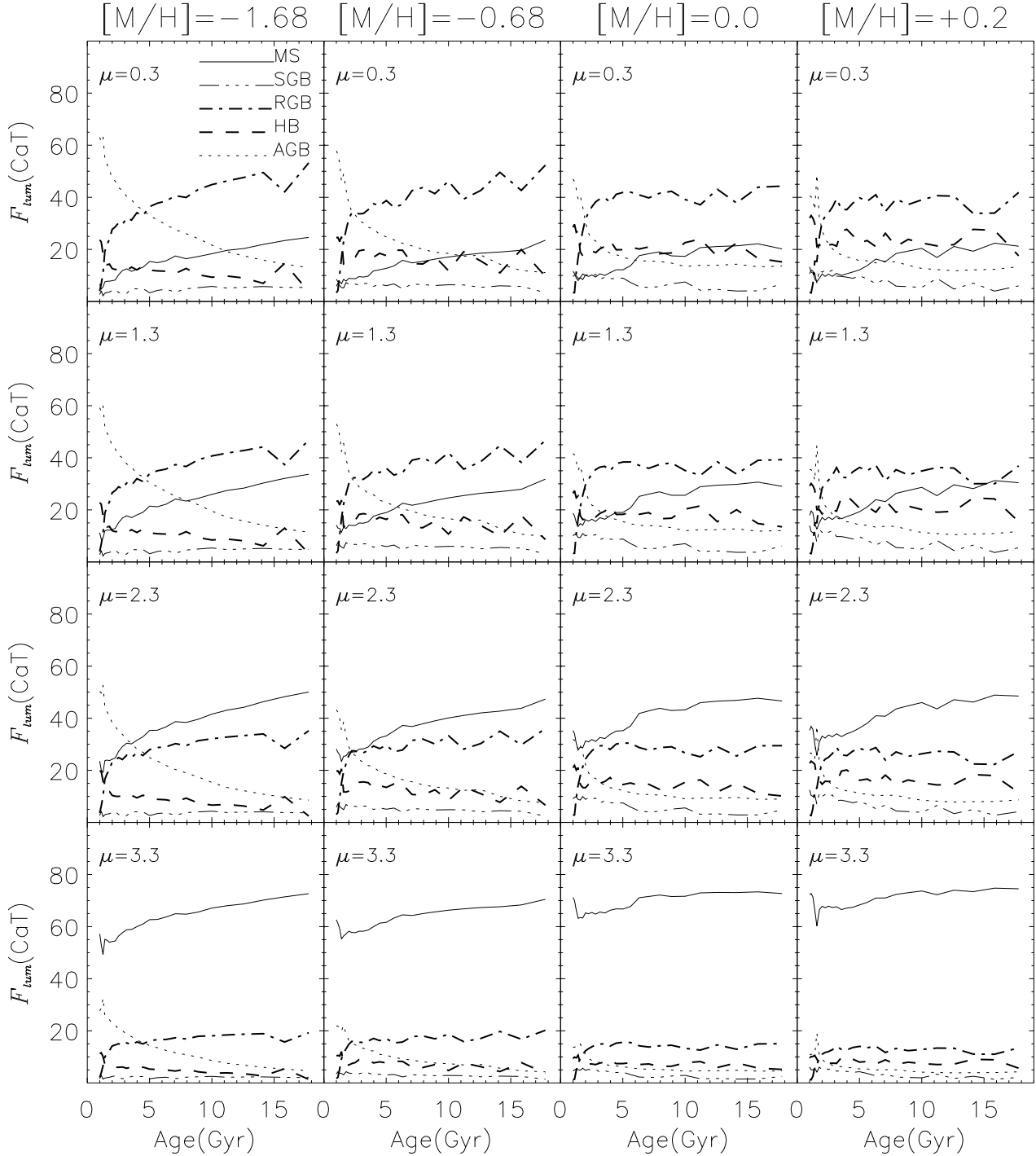,width=7.6in}}
\caption{
Fraction of luminosity (in percentage) of the different stellar evolutionary
phases in the \cat spectral region for SSPs of unimodal IMF of different
slopes (from top to bottom panels), metallicities (from left to right panels)
and ages (in each panel).
}
\label{fig:cat_contr}
\end{figure*}

Figure~\ref{fig:cat_contr} illustrates the contributions of the different
evolutionary phases to $S_{\lambda}(t,{\rm [M/H]})$ in the selected normalization
wavelength interval, according to the values calculated for
$F_{\Delta\lambda_{ref}}$ for each star. The figure shows the fraction of
luminosity (in percentage) of these evolutionary stages for SSPs of several IMF
slopes (unimodal), metallicities and ages. The RGB, MS and HB are
the main contributors to this spectral range. The weight of the AGB decays very
rapidly as the stellar population evolves, whereas the RGB is completely
build-up when the age reaches $\sim2$~Gyr. The contribution of the Sub Giant
Branch (SGB) is the smallest one in all the diagrams. The effect of the IMF is
mostly reflected in the relative contribution of the MS and RGB.

Finally, it is worth recalling that the fact that the effective wavelength of
the Johnson I filter experiments a significant shift as a function of the
spectral type, makes our scaling approach more secure than if we had chosen, as
a reference interval, a significantly narrower spectral range. In fact we have
tested to scale the stellar spectra with F$_{I}$ (i.e. non corrected by
$f_{n}$), selecting as a normalization region either the pseudocontinuum
$\lambda\lambda$8474-8484\AA~(i.e. $c1$, see Table~\ref{eq:defCaTPaT}) or
8776-8792\AA~(i.e. $c5$). For the synthesized SSP spectrum, e.g. for 12.6~Gyr
and solar metallicity, the slope of the continuum in the wavelength range
$\lambda\lambda$8500-8850\AA~ (measured as ${\rm Flux}_{c5}/{\rm Flux}_{c1}$) varies by
$\sim4$\%, and the value of the CaT$^{*}$ index by $\sim0.14$\AA~ (which
represents $\sim$2\% of the index  strength). This effect is due to the fact
that the selection of a redder continuum emphasizes the contribution of redder
stars in Eq.~\ref{eq:SSP}.

\subsubsection{Adopted stellar spectrum}
\label{sec:AdoptedSpectrum}

In Appendix~\ref{ap:boxes} we describe the method that we have followed for
obtaining from the whole subsample of selected stars, a representative
spectrum, $S_{\lambda}(m,t,{\rm [M/H]})$, for a given set of atmospheric parameters
$\theta$, $\log g$ and [M/H]. It is worth noting that rather than T$_{\rm eff}$ we
have chosen to work with $\theta$ ($\equiv5040/T_{\rm eff}$) in order to follow
the scale adopted in Paper~III for calculating the empirical fitting functions
for the \cat feature. We have tested however that using this parameter does not
affect the resulting integrated SSP spectrum.

Basically our approach consists in finding all the stars whose parameters are
enclosed within a given box in the 3-parametrical space. The method ensures the
presence of stars in several directions to alleviate the effect of asymmetries
in the distribution of stars around the point. For example, for metallicity
larger than solar, more stars are usually found with lower, rather than higher,
metallicity. The size of the selected finding box is taken to be a function of
the density of stars around the requested point, i.e. the larger the density
the smaller the box. Moreover, in the more populated parametrical regions, the
typical uncertainties in the determination of the atmospheric parameters are
usually smaller than in regions with lower densities (see Paper~II), in
agreement with the adopted criterion. When needed, the starting box is enlarged
until finding suitable stars. 

The obtained stars are combined according to their parameters $\theta$, $\log
g$ and [M/H] and to the signal-to-noise of their spectra. However, the method
ensures that spectroscopic binaries and stars with anomalous signatures of
variability do not significantly contribute to the average spectrum. Finally,
we apply a minimal correction to these weights in order to ensure the obtention
of a spectrum whose atmospheric parameters matches the requested ones.

\subsection{Line-strength index synthesis}
\label{sec:ff}

The most widely employed approach for modeling and studying stellar populations
using absorption lines is based on the, so-called, empirical fitting
functions. These functions describe the strength of, previously defined,
spectral features in terms of the main atmospheric parameters. These
calibrations are directly implemented into the stellar population models to
derive the index values for stellar populations of different ages,
metallicities and IMFs. The main advantage of this approach over the spectral
synthesis, which provides full spectra, is that the later method relies on the
availability of a complete stellar library at the appropriate spectral
resolution. The difficulties in achieving such libraries in the most recent
past made the fitting functions method the only possible approach to provide
robust diagnostics for the stellar populations (e.g. Buzzoni 1993; Worthey
1994; V96; Tantalo, Chiosi \& Bressan 1998; Kurth et al.
1999; Maraston \& Thomas 2000; Poggianti et al. 2001; Bruzual \& Charlot 2003).
The predictions based on these models have been widely used
along the last decade (see Trager et al. 1998 for an extensive review).

In the blue and visible part of the spectrum, we predicted in V96
the most important absorption line-strengths of the Lick/IDS system, at
intermediate resolution FWHM $\sim9$\AA. We used the empirical fitting
functions of Gorgas et al. (1993), W94 and WO97.
Moreover we also predicted in that paper the strengths
of the \cat on the basis of the stellar spectral library and the index
definition of D\'{\i}az et al. (1989). \cat strengths were
also calculated by Garc\'{\i}a--Vargas et al. (1998), MGV
and ITD
(see \S~\ref{sec:comparison} for a full comparison with these models). It is
worth noting that the use of these predictions requires one to adapt the data
to the characteristics of the instrumental-dependent stellar spectral library
employed by these models (see WO97 for an extensive review
of the method).   

Since we now have a better stellar library and models, our previous predictions
for the \cat presented in V96 are superseeded by the ones
calculated in this section. Here we adopt for the \cat feature the
index definition given in Paper~I (CaT$^{*}$). This new index removes the H
Paschen lines contamination and it is, therefore, a reliable indicator of pure
\cat strength. The index is given by CaT$^{*}={\rm CaT}-0.93~{\rm PaT}$, where
the CaT and PaT indices measure the strengths of the raw calcium triplet and
of three pure H Paschen lines, respectively (see Table~\ref{eq:defCaTPaT}).
All these indices are computed for a nominal resolution FWHM $=1.5$\AA. Further
details, as well as relations to transform this index to other popular
definitions for the \cat feature can be found in Paper~I.

\begin{table}
\centering{
\caption{Bandpass limits for the generic indices CaT and PaT.}
\label{eq:defCaTPaT}
\begin{tabular}{@{}ccc@{}}
\hline                    
CaT central        & PaT central        & Continuum        \\                     
bandpasses (\AA)   & bandpasses (\AA)   & bandpasses (\AA) \\ 
\hline                     
Ca1 8484.0--8513.0 & Pa1 8461.0--8474.0 & $c1$ 8474.0--8484.0   \\       
Ca2 8522.0--8562.0 & Pa2 8577.0--8619.0 & $c2$ 8563.0--8577.0   \\       
Ca3 8642.0--8682.0 & Pa3 8730.0--8772.0 & $c3$ 8619.0--8642.0   \\
                   &                    & $c4$ 8700.0--8725.0   \\
                   &                    & $c5$ 8776.0--8792.0   \\ 
\hline
\end{tabular}
}
\end{table}

Empirical fitting functions for the CaT and PaT indices were calculated in
Paper~III. These functions are polynomia that relate the strength of these
indices to the stellar atmospheric parameters ($\theta,\log g,{\rm [M/H]}$) derived
in Paper~II. The new fitting functions reveal a complex behaviour of the \cat
features as a function of the atmospheric parameters. In particular, for hot
and cold stars, the temperature and luminosity class are the main driving
parameters, whereas, in the mid-temperature regime, the three atmospheric
parameters play an important role. Among the advantages of the new fitting
functions over previous predictions is that they include the whole range of
effective temperatures and, in particular, cold stars. We refer the interested
reader to Paper~III for a full description of these functions.

The implementation of these fitting functions in our stellar population code is
readily done as follows. A flux-weighted index $i$ for a SSP of age ($t$) and
metallicity ([M/H]) is given by:

\begin{equation}
i_{\rm SSP}=\frac{\int_{m_{\rm l}}^{m_{\rm t}}i(m,t,{\rm
[M/H]})N(m,t)F_{\Delta\lambda_{\rm ref}}(m,t,{\rm [M/H]})dm}
{\int_{m_{\rm l}}^{m_{\rm t}}N(m,t)F_{\Delta\lambda_{\rm ref}}(m,t,{\rm [M/H]})dm},
\end{equation}

\noindent 
where the flux $F_{\Delta\lambda_{\rm ref}}(m,t,{\rm [M/H]})$ is used to scale the
contribution of a star of mass $m$ and age $t$ in a given evolutionary stage.
We adopt the flux corresponding to the scaling wavelength interval used for
the spectral synthesis, i.e. $\Delta\lambda_{\rm ref}$, since it is centered on
these features. 

\subsection{Summary of the new model predictions for the near-IR}
\label{sec:summaryIR}

We summarize in this section the new model predictions presented here for the
near-IR spectral range. There are two main products interesting for
the stellar population analyses: {\it i)} the spectral library for SSPs, for
which we list in Table~\ref{tab:SSPspectra} the main properties of the
synthesized SSP spectra, as well as the model parameters coverage according to
the limitations described in \S~\ref{sec:coverage}, and {\it ii)} the CaT$^*$,
CaT and PaT index strengths predicted on the basis of the empirical fitting
functions, which are listed in Table~\ref{tab:index-values}. It is worth
noting that the index values given in this table for SSPs of different ages,
metallicities and IMF types and slopes, are calculated at resolution
1.5 \AA~(FWHM). Therefore, users interested in applying these predictions should
correct their index measurements to the ones they would have obtained at the
resolution of the models presented here (see the relations given in
\S~\ref{sec:veldisp}). 

These two set of model predictions can either be obtained from the authors
or from the following web pages: 
{\tt http://www.ucm.es/info/Astrof/ellipt/CATRIPLET.html},
{\tt http://www.nottingham.ac.uk/$\sim$ppzrfp/CATRIPLET.html},
and {\tt http://www.iac.es/galeria/vazdekis/}.

\begin{table*}
\centering{
\caption{Spectral properties and parameter coverage of the synthesized model SEDs}
\label{tab:SSPspectra}
\begin{tabular}{lc}
\hline 
\multicolumn{2}{c}{Spectral properties}\\
\hline                   
Spectral range     & $\lambda\lambda$8348.85--8950.65 \AA            \\	
Spectral resolution& FWHM $=1.5$\AA, $\sigma=22.2$ km~s$^{-1}$\\	
Linear dispersion  & 0.85 \AA/pix (29.358 km~s$^{-1}$)\\
Continuum shape    & Flux-scaled                    \\
\hline
\multicolumn{2}{c}{SSPs parameter coverage}\\
\hline
IMF type           & Unimodal, Bimodal, Kroupa universal, Kroupa revised\\
IMF slope (for unimodal and bimodal) & 0.3 -- 3.3 \\
Metallicity        & $-1.68$, $-1.28$, $-0.68$, $-0.38$, $0.0$, $+0.20$\\
Age ([M/H] $<-0.68$)&   $10.0\le t\le17.78$~Gyr	   \\
Age ([M/H] $=-0.68$)  & $5.62\le t\le17.78$~Gyr	   \\
Age ([M/H] $=-0.38$)  & $2.51\le t\le17.78$~Gyr	   \\
Age ([M/H] $= 0.0$)   & $0.1\le t\le17.78$~Gyr       \\
Age ([M/H] $=+0.2$)   & $1.0\le t\le17.78$~Gyr       \\
\hline
\end{tabular}
}
\end{table*}

\begin{table*}
\begin{center}
\caption{
Predicted CaT$^*$ and PaT indices for SSPs of different ages and metallicities
(indicated in the first two columns) for different IMF shapes and slopes
($\mu$): unimodal, bimodal (according to the definitions given in V96)
and the universal and revised IMFs of K01 (see
\S~\ref{sec:imfs} and Appendix~\ref{ap:IMF}). The Salpeter (1955) solar
neighbourhood IMF is given by a unimodal IMF of slope 1.3. The tabulated index
values were calculated on the basis of the empirical fitting functions given in
Paper~III (see \S~\ref{sec:ff}). The spectral resolution is 1.5\AA~(FWHM). The
Paschen contaminated \cat index, CaT, can be easily calculated by applying the
relation: CaT = CaT$^*$ + 0.93~PaT (see Paper~I).
}
\label{tab:index-values}
\begin{tabular}{@{}c@{$\;$}cc@{$\;$}cc@{$\;$}cc@{$\;$}cc@{$\;$}cc@{$\;$}c@{$\;$}cc@{$\;$}cc@{$\;$}c@{$\;$}cc@{$\;$}c@{$\;$}c@{}}
\hline
\multicolumn{2}{c}{}&\multicolumn{8}{c}{Unimodal}&&\multicolumn{4}{c}{Bimodal}&&\multicolumn{2}{c}{K.
Universal}&&\multicolumn{2}{c}{K. Revised}\\
\multicolumn{1}{c}{[M/H]}&\multicolumn{1}{c}{Age}&\multicolumn{2}{c}{$\mu$ = 0.3}&\multicolumn{2}{c}{$\mu$ = 1.3}&\multicolumn{2}{c}{$\mu$ = 2.3}&\multicolumn{2}{c}{$\mu$ =
3.3}&&\multicolumn{2}{c}{$\mu$ = 1.3}&\multicolumn{2}{c}{$\mu$ = 2.3}&&\multicolumn{2}{c}{}&&\multicolumn{2}{c}{}\\
\cline{3-10}\cline{12-15}\cline{17-18}\cline{20-21}
       &(Gyr)& CaT$^*$&   PaT   &  CaT$^*$&   PaT   &  CaT$^*$&   PaT   &  CaT$^*$& PaT     &&    CaT$^*$&   PaT &  CaT$^*$&   PaT  &&  CaT$^*$&   PaT&&  CaT$^*$& PaT   \\
\hline
 $-1.68$ & 10.00  &  4.283 & 0.783   &   4.339 & 0.756   &   4.564 & 0.698   &   5.104 & 0.606     &&	4.290 & 0.765	&   4.331 & 0.735   && 4.289 &  0.766	&&4.320 & 0.752    \\
 $-1.68$ & 11.22  &  4.257 & 0.737   &   4.326 & 0.710   &   4.568 & 0.658   &   5.115 & 0.582     &&	4.272 & 0.718	&   4.321 & 0.690   && 4.271 &  0.719	&&4.307 & 0.705    \\
 $-1.68$ & 12.59  &  4.244 & 0.703   &   4.326 & 0.676   &   4.581 & 0.629   &   5.131 & 0.565     &&	4.266 & 0.684	&   4.321 & 0.657   && 4.265 &  0.685	&&4.306 & 0.672    \\
 $-1.68$ & 14.13  &  4.249 & 0.749   &   4.342 & 0.714   &   4.611 & 0.655   &   5.159 & 0.578     &&	4.277 & 0.724	&   4.337 & 0.691   && 4.275 &  0.725	&&4.320 & 0.709    \\
 $-1.68$ & 15.85  &  4.274 & 0.809   &   4.375 & 0.763   &   4.652 & 0.690   &   5.192 & 0.597     &&	4.304 & 0.778	&   4.367 & 0.737   && 4.303 &  0.779	&&4.351 & 0.758    \\
 $-1.68$ & 17.78  &  4.327 & 0.855   &   4.434 & 0.801   &   4.714 & 0.715   &   5.235 & 0.610     &&	4.358 & 0.820	&   4.420 & 0.774   && 4.356 &  0.821	&&4.406 & 0.797    \\
\\
 $-1.28$ & 10.00  &  5.447 & 0.760   &   5.444 & 0.740   &   5.529 & 0.696   &   5.783 & 0.619     &&	5.418 & 0.747	&   5.406 & 0.725   && 5.417 &  0.747	&&5.420 & 0.737     \\
 $-1.28$ & 11.22  &  5.451 & 0.723   &   5.455 & 0.703   &   5.549 & 0.664   &   5.801 & 0.599     &&	5.428 & 0.709	&   5.422 & 0.689   && 5.427 &  0.710	&&5.433 & 0.700     \\
 $-1.28$ & 12.59  &  5.448 & 0.695   &   5.460 & 0.676   &   5.563 & 0.641   &   5.816 & 0.586     &&	5.430 & 0.682	&   5.429 & 0.664   && 5.429 &  0.682	&&5.438 & 0.673     \\
 $-1.28$ & 14.13  &  5.443 & 0.678   &   5.463 & 0.659   &   5.574 & 0.626   &   5.829 & 0.577     &&	5.431 & 0.665	&   5.436 & 0.646   && 5.430 &  0.665	&&5.442 & 0.656     \\
 $-1.28$ & 15.85  &  5.383 & 0.682   &   5.420 & 0.660   &   5.555 & 0.624   &   5.830 & 0.574     &&	5.380 & 0.666	&   5.396 & 0.646   && 5.378 &  0.667	&&5.398 & 0.657     \\
 $-1.28$ & 17.78  &  5.395 & 0.821   &   5.437 & 0.781   &   5.578 & 0.716   &   5.850 & 0.626     &&	5.395 & 0.795	&   5.413 & 0.761   && 5.393 &  0.796	&&5.414 & 0.778     \\
\\
 $-0.68$ &  5.62  &  6.768 & 0.951   &   6.671 & 0.930   &   6.527 & 0.865   &   6.289 & 0.711     &&	6.680 & 0.939	&   6.579 & 0.912   && 6.680 &  0.939	&&6.656 & 0.926     \\
 $-0.68$ &  6.31  &  6.807 & 0.913   &   6.711 & 0.892   &   6.564 & 0.830   &   6.314 & 0.690     &&	6.722 & 0.900	&   6.624 & 0.874   && 6.722 &  0.900	&&6.694 & 0.888     \\
 $-0.68$ &  7.08  &  6.835 & 0.880   &   6.741 & 0.859   &   6.592 & 0.801   &   6.335 & 0.672     &&	6.754 & 0.867	&   6.659 & 0.843   && 6.754 &  0.867	&&6.720 & 0.856     \\
 $-0.68$ &  7.94  &  6.863 & 0.851   &   6.770 & 0.829   &   6.618 & 0.773   &   6.352 & 0.655     &&	6.785 & 0.837	&   6.693 & 0.814   && 6.785 &  0.837	&&6.744 & 0.826     \\
 $-0.68$ &  8.91  &  6.897 & 0.825   &   6.804 & 0.804   &   6.648 & 0.751   &   6.372 & 0.642     &&	6.821 & 0.812	&   6.730 & 0.791   && 6.821 &  0.812	&&6.780 & 0.802     \\
 $-0.68$ & 10.00  &  6.925 & 0.803   &   6.832 & 0.783   &   6.672 & 0.732   &   6.389 & 0.630     &&	6.851 & 0.791	&   6.762 & 0.771   && 6.851 &  0.791	&&6.810 & 0.781     \\
 $-0.68$ & 11.22  &  6.961 & 0.784   &   6.867 & 0.764   &   6.701 & 0.716   &   6.409 & 0.621     &&	6.889 & 0.772	&   6.801 & 0.753   && 6.889 &  0.772	&&6.848 & 0.762     \\
 $-0.68$ & 12.59  &  6.999 & 0.765   &   6.904 & 0.746   &   6.732 & 0.701   &   6.430 & 0.613     &&	6.929 & 0.754	&   6.842 & 0.737   && 6.929 &  0.754	&&6.887 & 0.745     \\
 $-0.68$ & 14.13  &  7.026 & 0.754   &   6.930 & 0.735   &   6.754 & 0.691   &   6.446 & 0.607     &&	6.958 & 0.743	&   6.872 & 0.727   && 6.958 &  0.743	&&6.916 & 0.734     \\
 $-0.68$ & 15.85  &  7.033 & 0.745   &   6.937 & 0.726   &   6.759 & 0.684   &   6.452 & 0.603     &&	6.967 & 0.735	&   6.885 & 0.720   && 6.968 &  0.735	&&6.925 & 0.726     \\
 $-0.68$ & 17.78  &  7.033 & 0.746   &   6.936 & 0.726   &   6.757 & 0.682   &   6.453 & 0.602     &&	6.970 & 0.735	&   6.890 & 0.720   && 6.971 &  0.735	&&6.928 & 0.726     \\
\\
 $-0.38$ &  2.51  &  7.096 & 1.188   &   6.963 & 1.181   &   6.720 & 1.105   &   6.162 & 0.830     &&	6.976 & 1.188	&   6.823 & 1.163   && 6.976 &  1.189	&&6.962 & 1.178     \\
 $-0.38$ &  2.82  &  7.111 & 1.135   &   6.984 & 1.123   &   6.743 & 1.046   &   6.180 & 0.790     &&	6.999 & 1.131	&   6.854 & 1.103   && 6.999 &  1.131	&&6.983 & 1.120     \\
 $-0.38$ &  3.16  &  7.140 & 1.074   &   7.014 & 1.058   &   6.768 & 0.982   &   6.193 & 0.746     &&	7.031 & 1.066	&   6.890 & 1.037   && 7.031 &  1.067	&&7.012 & 1.056     \\
 $-0.38$ &  3.55  &  7.163 & 1.042   &   7.042 & 1.024   &   6.796 & 0.947   &   6.214 & 0.723     &&	7.060 & 1.032	&   6.925 & 1.002   && 7.060 &  1.032	&&7.039 & 1.021     \\
 $-0.38$ &  3.98  &  7.192 & 1.024   &   7.073 & 1.004   &   6.826 & 0.927   &   6.236 & 0.710     &&	7.093 & 1.012	&   6.963 & 0.981   && 7.094 &  1.012	&&7.070 & 1.001     \\
 $-0.38$ &  4.47  &  7.223 & 1.000   &   7.107 & 0.978   &   6.855 & 0.901   &   6.256 & 0.693     &&	7.129 & 0.986	&   7.002 & 0.956   && 7.130 &  0.987	&&7.103 & 0.975     \\
 $-0.38$ &  5.01  &  7.232 & 0.973   &   7.114 & 0.948   &   6.855 & 0.870   &   6.249 & 0.670     &&	7.140 & 0.957	&   7.015 & 0.926   && 7.140 &  0.958	&&7.109 & 0.945     \\
 $-0.38$ &  5.62  &  7.263 & 0.949   &   7.146 & 0.923   &   6.882 & 0.845   &   6.266 & 0.654     &&	7.174 & 0.932	&   7.052 & 0.901   && 7.175 &  0.933	&&7.140 & 0.920     \\
 $-0.38$ &  6.31  &  7.294 & 0.926   &   7.174 & 0.898   &   6.897 & 0.821   &   6.268 & 0.637     &&	7.205 & 0.908	&   7.082 & 0.877   && 7.206 &  0.909	&&7.166 & 0.896     \\
 $-0.38$ &  7.08  &  7.276 & 0.911   &   7.154 & 0.882   &   6.871 & 0.802   &   6.244 & 0.622     &&	7.189 & 0.893	&   7.068 & 0.862   && 7.190 &  0.894	&&7.145 & 0.880     \\
 $-0.38$ &  7.94  &  7.301 & 0.897   &   7.176 & 0.867   &   6.884 & 0.788   &   6.250 & 0.613     &&	7.215 & 0.879	&   7.094 & 0.848   && 7.216 &  0.879	&&7.165 & 0.865     \\
 $-0.38$ &  8.91  &  7.319 & 0.886   &   7.192 & 0.856   &   6.894 & 0.776   &   6.256 & 0.606     &&	7.234 & 0.868	&   7.115 & 0.837   && 7.236 &  0.868	&&7.178 & 0.853     \\
 $-0.38$ & 10.00  &  7.347 & 0.877   &   7.218 & 0.846   &   6.912 & 0.767   &   6.269 & 0.601     &&	7.264 & 0.859	&   7.145 & 0.830   && 7.265 &  0.859	&&7.204 & 0.844     \\
 $-0.38$ & 11.22  &  7.350 & 0.875   &   7.221 & 0.844   &   6.913 & 0.764   &   6.274 & 0.600     &&	7.270 & 0.857	&   7.155 & 0.828   && 7.271 &  0.858	&&7.210 & 0.843     \\
 $-0.38$ & 12.59  &  7.347 & 0.875   &   7.217 & 0.842   &   6.907 & 0.761   &   6.271 & 0.598     &&	7.270 & 0.857	&   7.158 & 0.828   && 7.272 &  0.858	&&7.210 & 0.842     \\
 $-0.38$ & 14.13  &  7.357 & 0.872   &   7.224 & 0.838   &   6.906 & 0.756   &   6.271 & 0.594     &&	7.281 & 0.854	&   7.170 & 0.825   && 7.283 &  0.855	&&7.220 & 0.838     \\
 $-0.38$ & 15.85  &  7.382 & 0.863   &   7.239 & 0.828   &   6.907 & 0.744   &   6.266 & 0.586     &&	7.304 & 0.845	&   7.190 & 0.816   && 7.306 &  0.846	&&7.240 & 0.829     \\
 $-0.38$ & 17.78  &  7.423 & 0.863   &   7.276 & 0.827   &   6.935 & 0.743   &   6.290 & 0.587     &&	7.346 & 0.845	&   7.232 & 0.817   && 7.348 &  0.846	&&7.280 & 0.829     \\
\hline
\end{tabular}
\end{center}
\end{table*}

\begin{table*}
\begin{center}
\contcaption{}
\begin{tabular}{@{}c@{$\;$}cc@{$\;$}cc@{$\;$}cc@{$\;$}cc@{$\;$}cc@{$\;$}c@{$\;$}cc@{$\;$}cc@{$\;$}c@{$\;$}cc@{$\;$}c@{$\;$}c@{}}
\hline
\multicolumn{2}{c}{}&\multicolumn{8}{c}{Unimodal}&&\multicolumn{4}{c}{Bimodal}&&\multicolumn{2}{c}{K.
Universal}&&\multicolumn{2}{c}{K. Revised}\\
\multicolumn{1}{c}{[M/H]}&\multicolumn{1}{c}{Age}&\multicolumn{2}{c}{$\mu$ = 0.3}&\multicolumn{2}{c}{$\mu$ = 1.3}&\multicolumn{2}{c}{$\mu$ = 2.3}&\multicolumn{2}{c}{$\mu$ =
3.3}&&\multicolumn{2}{c}{$\mu$ = 1.3}&\multicolumn{2}{c}{$\mu$ = 2.3}&&\multicolumn{2}{c}{}&&\multicolumn{2}{c}{}\\
\cline{3-10}\cline{12-15}\cline{17-18}\cline{20-21}
       &(Gyr)& CaT$^*$&   PaT   &  CaT$^*$&   PaT   &  CaT$^*$&   PaT   &  CaT$^*$& PaT     &&    CaT$^*$&   PaT &  CaT$^*$&   PaT  &&  CaT$^*$&   PaT&&  CaT$^*$& PaT   \\
\hline
  0.00 &  1.00  &  6.169 & 1.998   &   6.024 & 2.068   &   5.819 & 1.951   &   5.302 & 1.234     && 	6.031 & 2.078	&   5.900 & 2.062&& 6.031 &  2.078  &&  6.029 & 2.064\\
  0.00 &  1.12  &  6.317 & 1.807   &   6.188 & 1.857   &   5.978 & 1.746   &   5.397 & 1.119     && 	6.196 & 1.866	&   6.072 & 1.846&& 6.196 &  1.867  &&  6.192 & 1.853\\
  0.00 &  1.26  &  6.646 & 1.584   &   6.531 & 1.614   &   6.301 & 1.522   &   5.599 & 1.013     && 	6.541 & 1.622	&   6.411 & 1.604&& 6.541 &  1.623  &&  6.534 & 1.612\\
  0.00 &  1.41  &  6.963 & 1.397   &   6.857 & 1.414   &   6.608 & 1.336   &   5.805 & 0.919     && 	6.869 & 1.420	&   6.733 & 1.403&& 6.869 &  1.421  &&  6.859 & 1.412\\
  0.00 &  1.58  &  6.965 & 1.354   &   6.867 & 1.355   &   6.622 & 1.263   &   5.814 & 0.864     && 	6.881 & 1.362	&   6.758 & 1.332&& 6.882 &  1.363  &&  6.870 & 1.353\\
  0.00 &  1.78  &  7.100 & 1.271   &   6.998 & 1.265   &   6.731 & 1.171   &   5.871 & 0.803     && 	7.014 & 1.272	&   6.884 & 1.237&& 7.015 &  1.273  &&  7.000 & 1.263\\
  0.00 &  2.00  &  7.235 & 1.192   &   7.124 & 1.180   &   6.828 & 1.086   &   5.912 & 0.745     && 	7.144 & 1.188	&   7.002 & 1.152&& 7.145 &  1.188  &&  7.126 & 1.179\\
  0.00 &  2.24  &  7.346 & 1.128   &   7.231 & 1.113   &   6.915 & 1.020   &   5.958 & 0.704     && 	7.253 & 1.121	&   7.105 & 1.084&& 7.254 &  1.121  &&  7.233 & 1.111\\
  0.00 &  2.51  &  7.397 & 1.079   &   7.281 & 1.060   &   6.953 & 0.967   &   5.975 & 0.670     && 	7.306 & 1.068	&   7.157 & 1.030&& 7.307 &  1.068  &&  7.282 & 1.058\\
  0.00 &  2.82  &  7.435 & 1.043   &   7.320 & 1.021   &   6.983 & 0.928   &   5.991 & 0.645     && 	7.347 & 1.029	&   7.200 & 0.990&& 7.348 &  1.029  &&  7.321 & 1.019\\
  0.00 &  3.16  &  7.494 & 1.012   &   7.375 & 0.988   &   7.024 & 0.895   &   6.011 & 0.624     && 	7.406 & 0.996	&   7.255 & 0.957&& 7.407 &  0.997  &&  7.376 & 0.986\\
  0.00 &  3.55  &  7.506 & 0.999   &   7.386 & 0.972   &   7.027 & 0.878   &   6.011 & 0.612     && 	7.420 & 0.981	&   7.271 & 0.942&& 7.421 &  0.982  &&  7.387 & 0.971\\
  0.00 &  3.98  &  7.532 & 0.984   &   7.408 & 0.955   &   7.034 & 0.858   &   6.004 & 0.598     && 	7.445 & 0.965	&   7.294 & 0.924&& 7.447 &  0.965  &&  7.408 & 0.953\\
  0.00 &  4.47  &  7.515 & 0.994   &   7.390 & 0.963   &   7.013 & 0.862   &   5.992 & 0.599     && 	7.430 & 0.973	&   7.282 & 0.931&& 7.432 &  0.974  &&  7.391 & 0.961\\
  0.00 &  5.01  &  7.525 & 0.983   &   7.399 & 0.951   &   7.014 & 0.849   &   5.994 & 0.591     && 	7.442 & 0.962	&   7.295 & 0.919&& 7.444 &  0.962  &&  7.399 & 0.949\\
  0.00 &  5.62  &  7.531 & 0.974   &   7.402 & 0.940   &   7.007 & 0.836   &   5.985 & 0.583     && 	7.449 & 0.951	&   7.302 & 0.909&& 7.451 &  0.952  &&  7.402 & 0.938\\
  0.00 &  6.31  &  7.493 & 0.965   &   7.355 & 0.926   &   6.937 & 0.816   &   5.914 & 0.563     && 	7.408 & 0.940	&   7.257 & 0.895&& 7.411 &  0.941  &&  7.355 & 0.925\\
  0.00 &  7.08  &  7.481 & 0.963   &   7.339 & 0.922   &   6.914 & 0.809   &   5.898 & 0.558     && 	7.397 & 0.937	&   7.247 & 0.891&& 7.400 &  0.938  &&  7.340 & 0.920\\
  0.00 &  7.94  &  7.467 & 0.954   &   7.321 & 0.912   &   6.886 & 0.797   &   5.876 & 0.550     && 	7.383 & 0.928	&   7.233 & 0.882&& 7.386 &  0.929  &&  7.322 & 0.910\\
  0.00 &  8.91  &  7.465 & 0.945   &   7.315 & 0.902   &   6.871 & 0.786   &   5.867 & 0.543     && 	7.381 & 0.919	&   7.232 & 0.873&& 7.385 &  0.920  &&  7.316 & 0.900\\
  0.00 & 10.00  &  7.459 & 0.938   &   7.303 & 0.893   &   6.851 & 0.776   &   5.852 & 0.537     && 	7.375 & 0.911	&   7.225 & 0.866&& 7.379 &  0.912  &&  7.304 & 0.891\\
  0.00 & 11.22  &  7.431 & 0.935   &   7.274 & 0.890   &   6.819 & 0.771   &   5.836 & 0.534     && 	7.349 & 0.909	&   7.203 & 0.863&& 7.353 &  0.910  &&  7.275 & 0.887\\
  0.00 & 12.59  &  7.413 & 0.934   &   7.251 & 0.887   &   6.792 & 0.766   &   5.820 & 0.532     && 	7.332 & 0.907	&   7.187 & 0.861&& 7.335 &  0.908  &&  7.254 & 0.884\\
  0.00 & 14.13  &  7.390 & 0.934   &   7.226 & 0.885   &   6.765 & 0.763   &   5.808 & 0.531     && 	7.311 & 0.907	&   7.170 & 0.862&& 7.315 &  0.908  &&  7.233 & 0.884\\
  0.00 & 15.85  &  7.364 & 0.934   &   7.196 & 0.884   &   6.732 & 0.760   &   5.789 & 0.529     && 	7.286 & 0.907	&   7.148 & 0.862&& 7.290 &  0.908  &&  7.207 & 0.884\\
  0.00 & 17.78  &  7.335 & 0.933   &   7.162 & 0.881   &   6.695 & 0.756   &   5.768 & 0.527     && 	7.258 & 0.906	&   7.124 & 0.862&& 7.263 &  0.907  &&  7.178 & 0.882\\
\\
  0.20 &  1.00  &  6.677 & 1.697   &   6.559 & 1.746   &   6.313 & 1.638   &   5.484 & 1.034     && 	6.571 & 1.755	&   6.447 & 1.731&& 6.571 &  1.755  &&  6.563 & 1.743\\
  0.20 &  1.12  &  6.812 & 1.523   &   6.705 & 1.556   &   6.449 & 1.455   &   5.564 & 0.930     && 	6.718 & 1.564	&   6.597 & 1.539&& 6.719 &  1.564  &&  6.708 & 1.553\\
  0.20 &  1.26  &  6.981 & 1.372   &   6.884 & 1.393   &   6.616 & 1.300   &   5.667 & 0.847     && 	6.899 & 1.400	&   6.776 & 1.374&& 6.899 &  1.400  &&  6.887 & 1.390\\
  0.20 &  1.41  &  7.350 & 1.208   &   7.254 & 1.220   &   6.960 & 1.148   &   5.904 & 0.780     && 	7.271 & 1.226	&   7.130 & 1.208&& 7.271 &  1.226  &&  7.256 & 1.218\\
  0.20 &  1.58  &  7.262 & 1.121   &   7.204 & 1.124   &   6.974 & 1.063   &   6.011 & 0.755     && 	7.219 & 1.128	&   7.124 & 1.111&& 7.219 &  1.129  &&  7.206 & 1.123\\
  0.20 &  1.78  &  7.531 & 1.094   &   7.426 & 1.089   &   7.092 & 1.007   &   5.963 & 0.684     && 	7.448 & 1.095	&   7.297 & 1.066&& 7.449 &  1.095  &&  7.427 & 1.087\\
  0.20 &  2.00  &  7.619 & 1.052   &   7.506 & 1.041   &   7.149 & 0.956   &   5.982 & 0.648     && 	7.532 & 1.048	&   7.372 & 1.015&& 7.533 &  1.048  &&  7.507 & 1.040\\
  0.20 &  2.24  &  7.685 & 1.021   &   7.570 & 1.004   &   7.196 & 0.917   &   6.004 & 0.622     && 	7.598 & 1.011	&   7.434 & 0.976&& 7.599 &  1.012  &&  7.570 & 1.003\\
  0.20 &  2.51  &  7.717 & 1.009   &   7.599 & 0.987   &   7.209 & 0.895   &   6.001 & 0.605     && 	7.630 & 0.995	&   7.464 & 0.956&& 7.631 &  0.995  &&  7.598 & 0.986\\
  0.20 &  2.82  &  7.744 & 1.000   &   7.622 & 0.974   &   7.221 & 0.879   &   6.001 & 0.593     && 	7.657 & 0.982	&   7.490 & 0.941&& 7.658 &  0.983  &&  7.622 & 0.973\\
  0.20 &  3.16  &  7.751 & 0.989   &   7.630 & 0.961   &   7.224 & 0.864   &   6.008 & 0.584     && 	7.667 & 0.969	&   7.503 & 0.927&& 7.669 &  0.970  &&  7.630 & 0.960\\
  0.20 &  3.55  &  7.717 & 0.977   &   7.600 & 0.948   &   7.200 & 0.851   &   6.003 & 0.578     && 	7.639 & 0.957	&   7.483 & 0.915&& 7.640 &  0.957  &&  7.600 & 0.947\\
  0.20 &  3.98  &  7.680 & 0.965   &   7.562 & 0.934   &   7.155 & 0.835   &   5.968 & 0.566     && 	7.604 & 0.943	&   7.451 & 0.901&& 7.606 &  0.944  &&  7.562 & 0.933\\
  0.20 &  4.47  &  7.664 & 0.962   &   7.545 & 0.930   &   7.132 & 0.829   &   5.954 & 0.563     && 	7.589 & 0.940	&   7.438 & 0.897&& 7.591 &  0.941  &&  7.545 & 0.929\\
  0.20 &  5.01  &  7.627 & 0.966   &   7.504 & 0.931   &   7.081 & 0.826   &   5.911 & 0.558     && 	7.552 & 0.943	&   7.401 & 0.899&& 7.554 &  0.943  &&  7.504 & 0.930\\
  0.20 &  5.62  &  7.612 & 0.957   &   7.485 & 0.921   &   7.052 & 0.814   &   5.886 & 0.549     && 	7.537 & 0.933	&   7.386 & 0.889&& 7.540 &  0.934  &&  7.486 & 0.920\\
  0.20 &  6.31  &  7.590 & 0.948   &   7.456 & 0.910   &   7.007 & 0.799   &   5.843 & 0.537     && 	7.514 & 0.923	&   7.361 & 0.878&& 7.517 &  0.924  &&  7.457 & 0.908\\
  0.20 &  7.08  &  7.570 & 0.930   &   7.430 & 0.891   &   6.967 & 0.780   &   5.809 & 0.524     && 	7.492 & 0.905	&   7.338 & 0.861&& 7.495 &  0.906  &&  7.431 & 0.890\\
  0.20 &  7.94  &  7.516 & 0.860   &   7.375 & 0.826   &   6.912 & 0.727   &   5.774 & 0.495     && 	7.441 & 0.840	&   7.291 & 0.804&& 7.444 &  0.841  &&  7.378 & 0.826\\
  0.20 &  8.91  &  7.524 & 0.866   &   7.374 & 0.830   &   6.893 & 0.726   &   5.755 & 0.492     && 	7.446 & 0.845	&   7.292 & 0.807&& 7.449 &  0.846  &&  7.377 & 0.829\\
  0.20 & 10.00  &  7.536 & 0.896   &   7.375 & 0.855   &   6.878 & 0.742   &   5.738 & 0.499     && 	7.454 & 0.872	&   7.295 & 0.830&& 7.457 &  0.873  &&  7.378 & 0.854\\
  0.20 & 11.22  &  7.519 & 0.883   &   7.353 & 0.842   &   6.849 & 0.729   &   5.721 & 0.492     && 	7.436 & 0.860	&   7.278 & 0.819&& 7.440 &  0.861  &&  7.357 & 0.841\\
  0.20 & 12.59  &  7.484 & 0.877   &   7.313 & 0.835   &   6.803 & 0.721   &   5.692 & 0.487     && 	7.402 & 0.854	&   7.246 & 0.814&& 7.406 &  0.856  &&  7.318 & 0.834\\
  0.20 & 14.13  &  7.450 & 0.864   &   7.274 & 0.822   &   6.757 & 0.708   &   5.662 & 0.479     && 	7.369 & 0.842	&   7.214 & 0.803&& 7.373 &  0.843  &&  7.283 & 0.822\\
  0.20 & 15.85  &  7.399 & 0.861   &   7.218 & 0.817   &   6.698 & 0.701   &   5.625 & 0.474     && 	7.319 & 0.839	&   7.167 & 0.800&& 7.324 &  0.840  &&  7.232 & 0.818\\
  0.20 & 17.78  &  7.346 & 0.855   &   7.160 & 0.809   &   6.636 & 0.691   &   5.584 & 0.468     && 	7.267 & 0.833	&   7.119 & 0.794&& 7.272 &  0.834  &&  7.179 & 0.811\\
\hline
\end{tabular}
\end{center}
\end{table*}

\section{The Behavior of the Ca{\sc II} Triplet Feature in SSPs}
\label{sec:behavior}

In this section we focus on describing the behaviour of the integrated \cat
feature as a function of the relevant stellar population parameters in old-aged
SSPs. In \S~\ref{sec:intermediate} we will present a discussion for some
younger ages. We also discuss here some of the main differences seen in the
full spectra predicted by the models, but we refer the reader to C03
and Cenarro (2002), for a detailed description of the behaviour of
other features and the slope of the characteristic continuum around the \cat
feature, which is largely determined by the contributions of the TiO molecular
bands. In \S~\ref{sec:catagemet} we describe the time evolution of the \cat
feature and its dependence on metallicity, whereas in \S~\ref{sec:catimf} we
discuss the effects of the IMF.

\begin{figure}
\centerline{\psfig{file=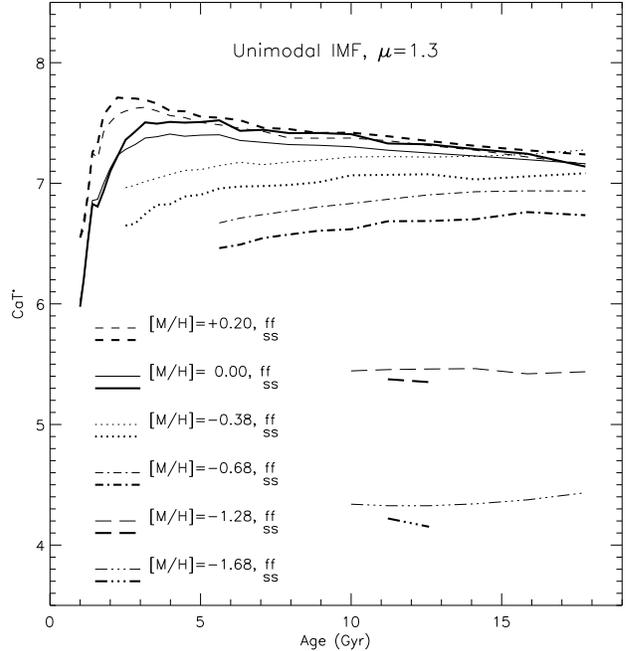,width=3.6in}}
\caption{
The CaT$^{*}$ index in SSPs as a function of age and metallicity. Different
line types refer to the metallicity of the SSP, as quoted in the plot. All the
models are calculated with a unimodal IMF of slope $\mu=1.3$. We plot the
CaT$^{*}$ index as calculated on the basis of the fittings functions of
Paper~III (ff, thin lines), and as measured on the synthesized spectra (ss,
thick lines). The spectral resolution is 1.5\AA~(FWHM) (i.e. $\sigma$ = 22.2
km s$^{-1}$) for the two set of predictions.
}
\label{fig:cat_met_age}
\end{figure}

\begin{figure*}
\centerline{\psfig{file=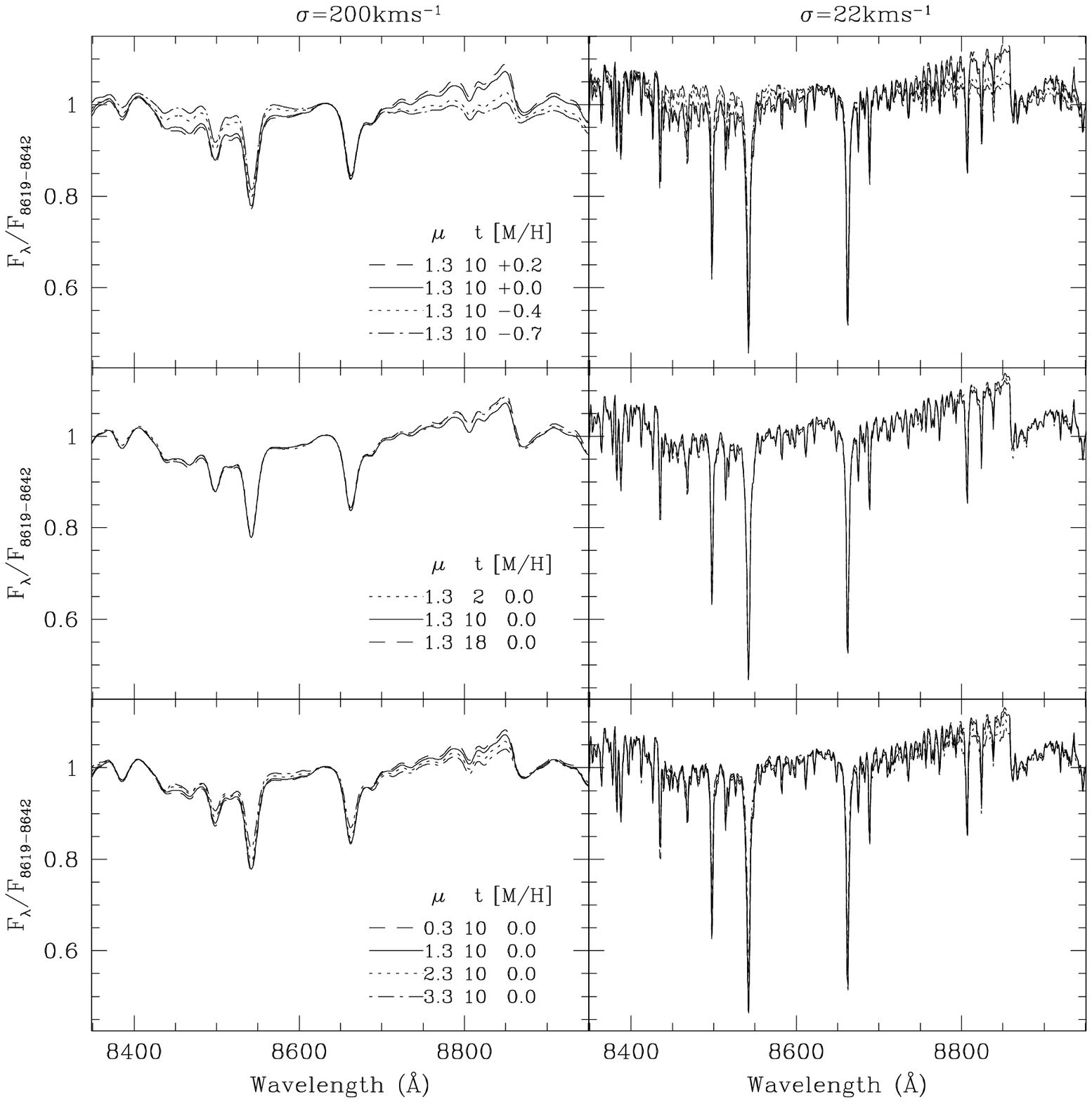,width=7.6in}}
\caption{
Model SSP spectra at $\sigma = 200$ km~s$^{-1}$ and $\sigma = 22.2$ km~s$^{-1}$
(i.e. the nominal resolution) for SSPs of different metallicities (upper plot),
ages (middle) and IMF slopes  (bottom). The models were calculated using a
unimodal IMF. All the spectra were normalized by the flux in the range
$\lambda\lambda$8619-8642\AA. 
}
\label{fig:cat_espectros}
\end{figure*}

\subsection{The effects of the age and metallicity}
\label{sec:catagemet}

In Figure~\ref{fig:cat_met_age} we plot the time evolution of the CaT$^{*}$
index as calculated on the basis of the fittings functions of Paper~III (see
\S~\ref{sec:ff}) and as measured on the synthesized spectra  (\S~\ref{sec:ss}).
The two set of predictions are given at resolution 1.5\AA~(FWHM), i.e. $\sigma$
= 22.2 km s$^{-1}$ (the behaviour of the \cat feature as a function of spectral
resolution will be discussed in \S~\ref{sec:veldisp}).
Fig.~\ref{fig:cat_met_age} reveals three major  features: {\it i)} the strength
of the CaT$^{*}$ index does not vary much for stellar populations older than
$\sim3$~Gyr, for all metallicities, {\it ii)} there is a strong sensitivity of
the CaT$^{*}$ index as a function of metallicity for values below
[M/H] $\sim -0.5$ and {\it iii)} for higher metallicities, the CaT$^{*}$ index does
not show any significant dependence on the metallicity and the strength of this
feature tends to saturate. 

The observed saturation of the CaT$^{*}$ index can be understood in terms of
comparing the contributions of the MS and RGB to the total luminosity of an SSP
in this spectral range (see Fig.~\ref{fig:cat_contr}). If we focus on the
second row of panels of this figure, i.e. for $\mu=1.3$ (we will discuss the
effects of the IMF in \S~\ref{sec:catimf}), we see that the relative
contribution of the MS with respect to the RGB gets larger as metallicity
increases. Dwarf stars provide smaller strengths for this feature than giants
according to the results obtained in Paper~III. Moreover, in metal-rich SSPs
the RGB is populated with an increased fraction of very cool stars. For
temperatures cooler than $\sim$ 3500~K, the CaT$^{*}$ index does not depend on
metallicity and decreases very rapidly with decreasing temperature (see Fig.~7
of Paper~III). The saturation of the \cat for large metallicities is a result
that differs from the predictions of previous authors (see
\S~\ref{sec:comparison}). This result might be supported by the CaT$^{*}$ index
measurements obtained for a large sample of ellipticals (C03)
and bulges of spirals (Falc\'on-Barroso et al. 2002), as well
as for the most metal-rich galactic globular clusters (see \S~\ref{sec:GC}).
Furthermore, Cohen (1979), Bica \& Alloin (1987), Terlevich et al. (1990b) and
Houdashelt (1995) found that the strength of the \cat did not vary much among
early-type galaxies of different absolute magnitudes and colours.

Fig.~\ref{fig:cat_met_age} shows a reasonably good agreement between our
CaT$^{*}$ index predictions based on the empirical fitting functions and the
ones measured on the synthesized spectra. We note, however, a better agreement
for larger metallicities, where the CaT$^{*}$ strengths provided by the
spectral synthesis are marginally larger than the ones predicted on the basis
of the empirical fitting functions. For [M/H] $=-0.38$ and [M/H]$=-0.68$, the
spectral synthesis approach provides smaller CaT$^{*}$ values, and the obtained
differences between the predictions based on the two approaches can be as large
as 0.2~\AA~ (i.e. $\sim3$\% of the CaT$^{*}$ index value) for [M/H] $=-0.68$.
Also, the SSP spectra for [M/H] $=-1.28$ and $-1.68$ provide smaller CaT$^{*}$
strengths than the values obtained on the basis of the fitting functions,
although the observed differences are smaller than those obtained for
[M/H] $=-0.68$. These differences must be attributed to differences in the
methods employed by the two approaches to provide, for a given set of stellar
atmospheric parameters, a representative average spectrum and index value,
respectively. One of the main differences between these two methods is that the
spectral synthesis works at much higher resolution in the stellar parameter
space than the one used for calculating the fitting functions. In the
boxes-method employed in the spectral synthesis those stars whose parameters
resemble more the requested ones are assigned significantly larger weights when
computing a representative spectrum (see Appendix~\ref{ap:boxes}). Therefore,
since the spectral synthesis emphasizes the contribution of the most likely
stars, the obtained differences between the two set of model predictions are
expected to be larger when a given set of stars, with a relevant contribution
within a stellar population, is lacking. In fact the lack of giant stars of
temperatures lower than $\sim$~4200~K (see Fig.~\ref{fig:PARAM_CAT}) causes the
main differences seen for [M/H] $=-0.38$ and $-0.68$. However, for [M/H] $<-1.0$ these
differences are smaller because the coolest stars along the isochrones have
larger temperatures than those for more metal-rich stellar populations and,
therefore, the lack of these stars constitutes a less severe problem. However, we
note that these differences in the predictions of the two model approaches are
of the order of typical observational errors in galaxy spectra (see
Paper~I and C03). Moreover, these differences are considerably
smaller than the ones obtained from comparisons between different author
predictions (see \S~\ref{sec:comparison}).

It is worth noting that the SSP spectra obtained for [M/H] $=-1.28$ and
[M/H] $=-1.68$ are less reliable than the ones for higher metallicities due to
the lack, in our stellar library, of spectra representing important
evolutionary phases. In particular, this is the case for stars corresponding to
the MS turnoff (i.e. dwarfs with temperatures larger than 6000~K), or the Blue
Horizontal Branch, which is more prominent for SSPs of ages above 13~Gyr
(see Fig.~\ref{fig:PARAM_CAT}). This is the reason why we only show in
Fig.~\ref{fig:cat_met_age} the CaT$^{*}$ values measured in the spectra
corresponding to SSPs for ages in the range 10-13~Gyr, where the contribution
of these stars is smaller.

Figure~\ref{fig:cat_espectros} shows the effects of the metallicity (top
panels), age (middle) and IMF (bottom) on the synthesized SSP spectra at
resolution $\sigma = 200$ km~s$^{-1}$ (left panels) and $\sigma =
22.2$ km~s$^{-1}$ (right panels). We overplot in the upper panels various
representative SSP spectra of similar age (i.e. 10~Gyr) and IMF (unimodal,
$\mu=1.3$), and with different metallicities. We have chosen to normalize all
the spectra to the continuum at $\lambda\lambda$8619-8642\AA\ (see
Table~\ref{eq:defCaTPaT}), to be able to see the effect of the metallicity on
the third line of the \cat feature, which can be taken as representative of the
overall \cat feature. We see that the depth of this line is virtually constant
in the metallicity range (i.e. $-0.7<{\rm [M/H]}<0.2$). We obtain similar
results for the other lines if normalizing to the appropriate continua. We note,
however, a strong variation of the slope of the continuum around the \cat
feature. For a given age and IMF, the larger the metallicity the larger the
slope of this continuum. The overall shape of the continuum for SSPs of high
metallicities shows strong similarities to those of early M-type stars. In this
spectral range the contribution of these cool stars to the total luminosity is
significantly larger (see Fig.~\ref{fig:cat_contr}) than in the
visible, where the integrated spectra resemble those of K-giants. The slope
of these spectra and its comparison to galaxy spectra are discussed in C03.

The effect of the age on SSP spectra of the same metallicity (i.e. solar) and
IMF (unimodal with slope 1.3) is shown in the second row of panels of 
Fig.~\ref{fig:cat_espectros}, where we vary the SSP age from 2.0 to
17.8~Gyr. We do not see any significant variation either on the
\cat feature or on the overall shape of the spectrum.

\begin{figure*}
\centerline{\psfig{file=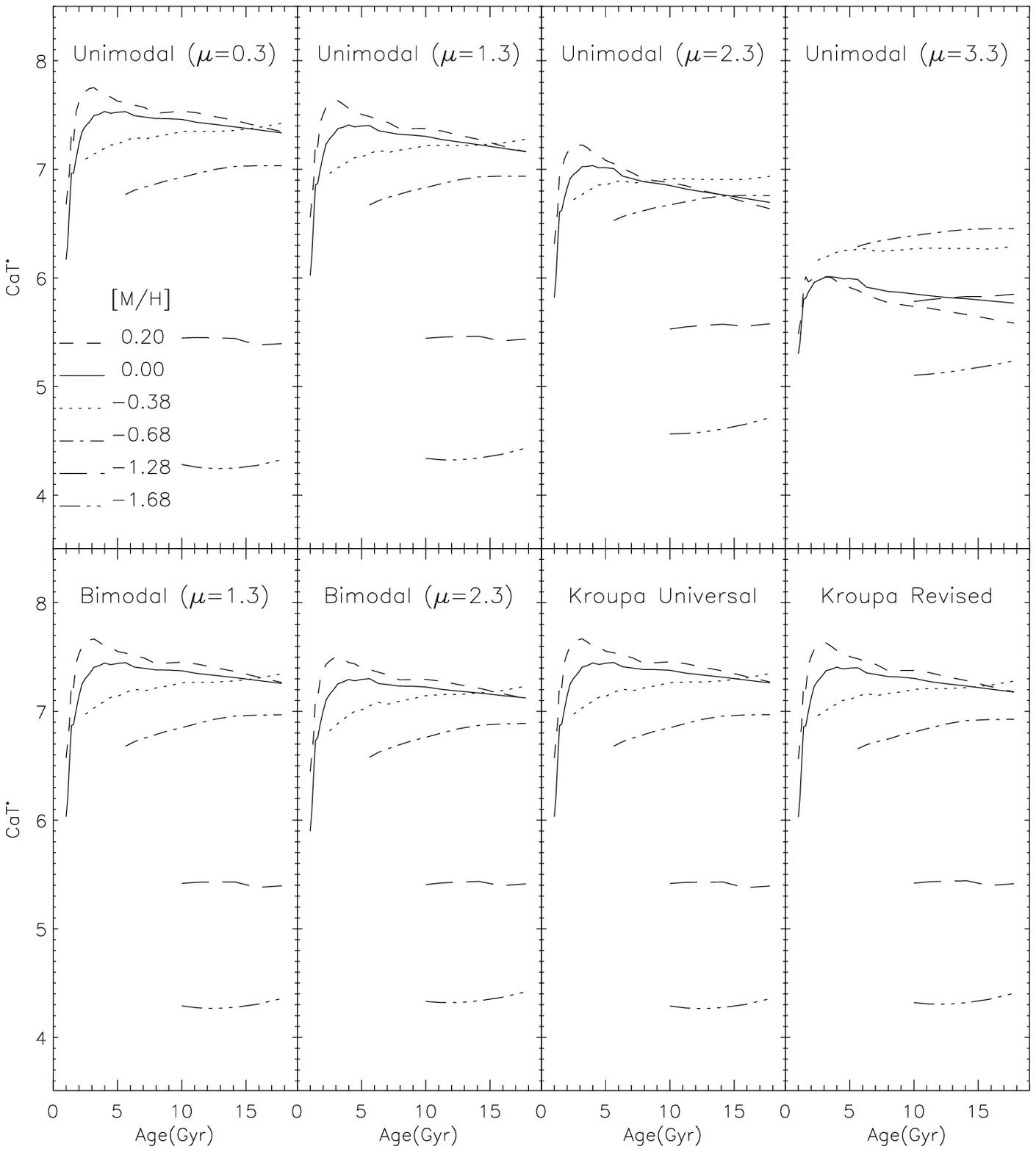,width=7.6in}}
\caption{
The CaT$^{*}$ index as a function of the IMF. All the models
shown here are calculated on the basis of the fitting functions.
}
\label{fig:cat_imf}
\end{figure*}

\subsection{The effects of the IMF}
\label{sec:catimf}

The third row of panels in Fig.~\ref{fig:cat_espectros} illustrates the effect
of the IMF on the SSP spectra. In these plots we keep constant the age
(10~Gyr) and the metallicity (solar) and vary the slope of the IMF from
$\mu=0.3$ to 3.3. We see that the third line of the \cat feature is not
significantly affected when the slope of the IMF varies from $\mu=0.3$ to
$\sim2.0$, but the line rapidly weakens for extremely dwarf-dominated IMFs.
This result is quantified in Figure~\ref{fig:cat_imf}, where we show the
CaT$^{*}$ index for different IMF types and slopes. The top panels show the
values obtained for different slopes and a unimodal IMF. The larger the
metallicity the larger the weakening of the CaT$^{*}$ index as a function
of increasing the IMF slope. This effect is not remarkable for lower
metallicities, although we do see an opposite trend for metallicities below
[M/H] $<-1.0$. Fig.~\ref{fig:cat_contr} shows the reason for this behaviour:
the larger the IMF slope the larger the relative contribution of the MS stars
(with lower \cat strengths, see Paper~III) to the total luminosity and to the
contribution of the RGB phase (see from top to bottom panels). 

In the two bottom panels of Fig.~\ref{fig:cat_imf} starting from the left we
show the results for a bimodal IMF for two different slopes, i.e. $\mu=1.3$
and 2.3. These plots show that the effect of the IMF slope on the CaT$^{*}$
index is much less pronounced for this IMF. This result is expected since for
the bimodal IMF the contribution of the very low MS stars is decreased with
respect to a unimodal IMF of the same slope. Finally, in the last two
panels we plot the results for the universal and revised IMF of K01.
This figure and the index values tabulated in Table~\ref{tab:index-values}
indicate that the CaT$^{*}$ values obtained from the unimodal and bimodal IMFs
of slope $\mu=1.3$ and the two IMFs of K01 are very similar. However,
the CaT$^{*}$ values obtained with the universal IMF match closer the ones of
the bimodal IMF, whilst those of the revised IMF match those of the unimodal
shape. This result is easily understood by looking at Fig.~\ref{fig:IMF}.

\section{Ca{\sc II} Triplet Feature Dependence on Spectral Resolution}
\label{sec:veldisp}

\begin{figure}
\psfig{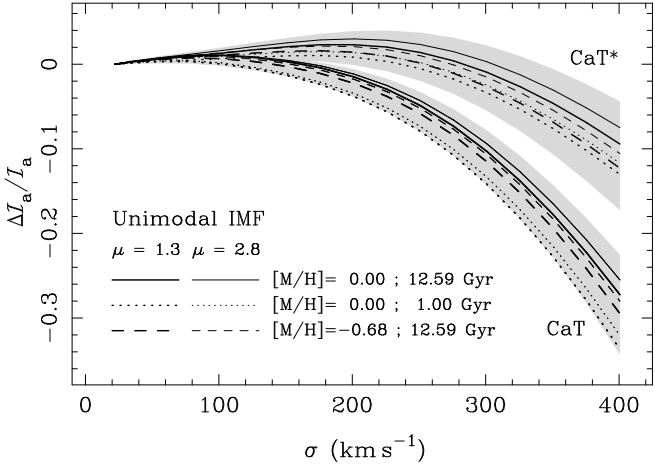}
\caption{
Broadening correction $\Delta$CaT$^{*}$/CaT$^{*}$ and $\Delta$CaT/CaT for the
\cat feature in SSPs. To calculate these corrections we have broadened the
model spectra by convolving with gaussians from $\sigma=25$ km~s$^{-1}$ up to
$\sigma=400$ km~s$^{-1}$ in steps of 25 km~s$^{-1}$. $\Delta I/I$ is zero for
$\sigma$ = 22.2 km s$^{-1}$ (the spectral resolution of the stellar library).
We plot with different line types the broadening corrections for a set of
representative models of different ages (1, 12.59~Gyr), metallicities (0.0,
$-0.68$) and IMFs (unimodal, $\mu=1.3$, $\mu=2.8$). In grey we show the region
covered by the broadening corrections obtained for the whole model SSP spectral
library. The upper envelope for the two indices corresponds to the model
$\mu=3.3$, [M/H] $=+0.2$ and 17.78~Gyr, whereas the lower envelope refers to the
model $\mu=0.3$, [M/H] $=-1.3$ and 14.12~Gyr. 
}
\label{fig:pols_sigma}
\end{figure}

In order to study the sensitivity of the \cat indices to the spectral
resolution, or galaxy velocity dispersion broadening ($\sigma$), we have
broadened the whole SSP model spectral library by convolving with gaussians
from $\sigma=25$ km~s$^{-1}$ up to $\sigma=400$ km~s$^{-1}$ (in
steps of 25 km~s$^{-1}$). The \cat indices were measured for the full set
of broadened spectra and we fit, for each model, a third-order
polynomial to the relative changes of the index values as a function
of velocity dispersion

\begin{equation} 
{\frac{{\cal I}(\sigma) - {\cal I}
(\sigma_{0})}{{\cal I}(\sigma)} 
= a + b\sigma + c\sigma^{2} + d\sigma^{3} \equiv  p(\sigma)},
\label{poly}
\end{equation}

\noindent 
where $\sigma_{0}=22.2$ km~s$^{-1}$ is the nominal resolution of the models
(FWHM = 1.50 \AA). Finally, making use of this formula it is straightforward
to relate the indices measured at two different velocity dispersions,
$\sigma_1$ and $\sigma_2$.

Table~\ref{polcoefdisp} tabulates the derived coefficients for the CaT$^{*}$
and CaT indices measured over a number of representative SSP models. The
velocity dispersion corrections for the PaT index can be inferred from those of
the CaT$^{*}$ and CaT indices. We plot in Figure~\ref{fig:pols_sigma} the
obtained  $\Delta$CaT$^{*}$/CaT$^{*}$ and $\Delta$CaT/CaT values for this set
of representative models. The grey region represents the locus of broadening
corrections for the whole SSP model spectral library. Note that the CaT index
is significantly more sensitive to the velocity dispersion than the CaT$^{*}$
index. This result is in agreement with that obtained in Paper~I for stars. In
that paper, it was shown that PaT is quite sensitive to velocity dispersion
broadening. The improvement in the CaT$^{*}$ sensitivity is explained since the
effects of the broadening on CaT and PaT indices are partially compensated when
the CaT$^{*}$ index is computed. Paper~I also shows that the CaT$^{*}$ index is
less sensitive to resolution than most of the other popular index definitions
for this feature. 

\begin{table*}                                                              
\centering{
\caption{Coefficients of the broadening correction polynomials ($\Delta
{\cal I}/{\cal I} = a + b\sigma + c\sigma^{2} +
d\sigma^{3}$) for the CaT$^{*}$ and CaT indices for a set of 
representative SSP models ($\mu$,[M/H],$t$).}
\label{polcoefdisp}                                                            
\begin{tabular}{@{}ccr@{}lr@{}lr@{}lc@{}}
\hline
 ${\cal I}$ & Model & \multicolumn{2}{c}{$a(\times10^{-3})$} & 
\multicolumn{2}{c}{$b(\times10^{-5})$} & 
\multicolumn{2}{c}{$c(\times10^{-7})$} & $d(\times10^{-9})$\\
\hline     
CaT*&1.3,  0.0,  1.0&\ $-$2.&058826 &   8.&056297 &6.&404771 &$-$4.142633\\
    &1.3,  0.0, 12.6&\ $-$3.&987821 &  16.&893273 &5.&690300 &$-$3.923405\\
    &1.3, $-0.7$, 12.6&\ $-$3.&632453 &  15.&659963 &4.&025051 &$-$3.878037\\
    &2.8,  0.0,  1.0&\ $-$2.&989543 &  12.&448730 &5.&462115 &$-$3.954834\\
    &2.8,  0.0, 12.6&\ $-$4.&815550 &  20.&660642 &5.&485660 &$-$3.790087\\
    &2.8, $-0.7$, 12.6&\ $-$4.&428785 &  19.&288911 &3.&811699 &$-$3.766459\\
&&&&&&&&\\
CaT &1.3,  0.0,  1.0&\ $-$2.&958673 &  15.&256039 &$-$7.&750521 &$-$4.221853\\
    &1.3,  0.0, 12.6&\ $-$6.&121178 &  29.&992397 &$-$0.&101206 &$-$3.505068\\
    &1.3, $-0.7$, 12.6&\ $-$4.&687171 &  23.&040400 &$-$7.&787181 &$-$4.022836\\
    &2.8,  0.0,  1.0&\ $-$3.&215187 &  16.&386110 &$-$7.&673347 &$-$4.054138\\
    &2.8,  0.0, 12.6&\ $-$6.&122020 &  29.&828926 &$-$9.&389365 &$-$3.405152\\
    &2.8, $-0.7$, 12.6&\ $-$5.&524840 &  27.&065372 &$-$8.&985443 &$-$3.732187\\
\hline
\end{tabular}                                                              
}                                                                          
\end{table*}

The fact that the broadening correction depends on the model parameters shows
us to what extent one can match the effects of galaxy velocity broadening by
using a number of stellar templates. In fact, the stellar spectra show even
larger variations than those shown by the models (as it can be seen by
comparing Fig.~\ref{fig:pols_sigma} to Fig.~5a of Paper~I). Note that this method has
been widely used in the literature to match as well the resolution
requirements of the Lick/IDS system of indices in the optical spectral range
in order to compare galaxy line-strength measurements with the index
predictions of models based on that system (Worthey 1994; V96).
The reader is referred to the extensive review of WO97
for a detailed description of the method. It is evident that the method
that we propose here to smooth the new model spectra to galaxy velocity
dispersion (and resolution of the data) does not suffer from these broadening
correction uncertainties. The advantages of this approach over the traditional
method, based on the stellar templates, has also been proven by Falc\'on-Barroso et
al. (2003), who have used the new SSP models presented here as templates for
deriving accurate stellar velocities, velocity dispersions and higher order
Gauss-Hermite moments profiles.

\begin{figure}
\centerline{\psfig{file=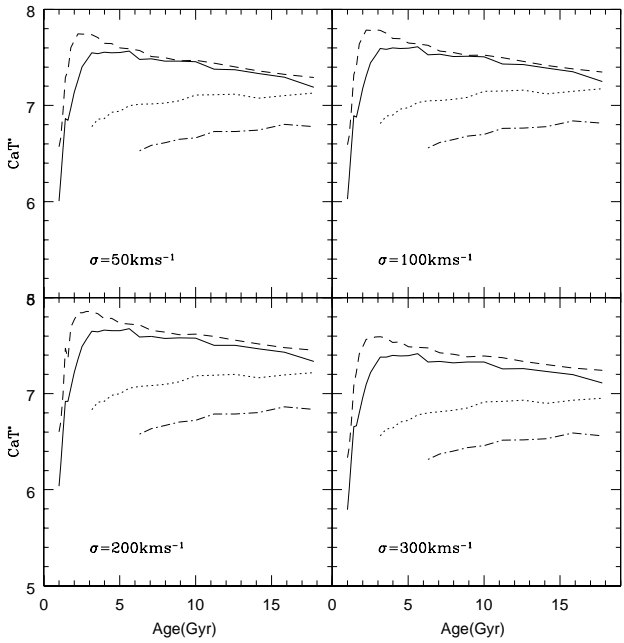,width=3.6in}}
\caption{
The CaT$^{*}$ index measured on the model SSP spectral library smoothed
to $\sigma$ = 50, 100, 200 and 300~km~s$^{-1}$. We use a unimodal IMF of slope
1.3.
}
\label{fig:vdisp}
\end{figure}

It is interesting to see whether the information provided by the CaT$^{*}$
index, in terms of the relevant parameters of the stellar populations, is
varying as a function of galaxy velocity dispersion. For this purpose we plot
in Figure~\ref{fig:vdisp} the CaT$^{*}$ index measured on the model SSP spectral
library smoothed from $\sigma$ = 50  to 300~km~s$^{-1}$. We see some relative
variations between the model lines of different metallicities, e.g. the
separation between the lines corresponding to [M/H] $=-0.68$ and [M/H] $=+0.2$ is
larger at $\sigma$ = 300~km~s$^{-1}$ than at $\sigma$ = 50~km~s$^{-1}$.
However, these differences are small in comparison to typical observational
errors (see Paper~I and C03) and, therefore, the stellar
population parameter estimates obtained on the basis of this index do not
depend significantly on the adopted resolution.

\section{Solar Metallicity Models in the Age Interval 0.1-1~Gyr}
\label{sec:intermediate}

\begin{figure}
\centerline{\psfig{file=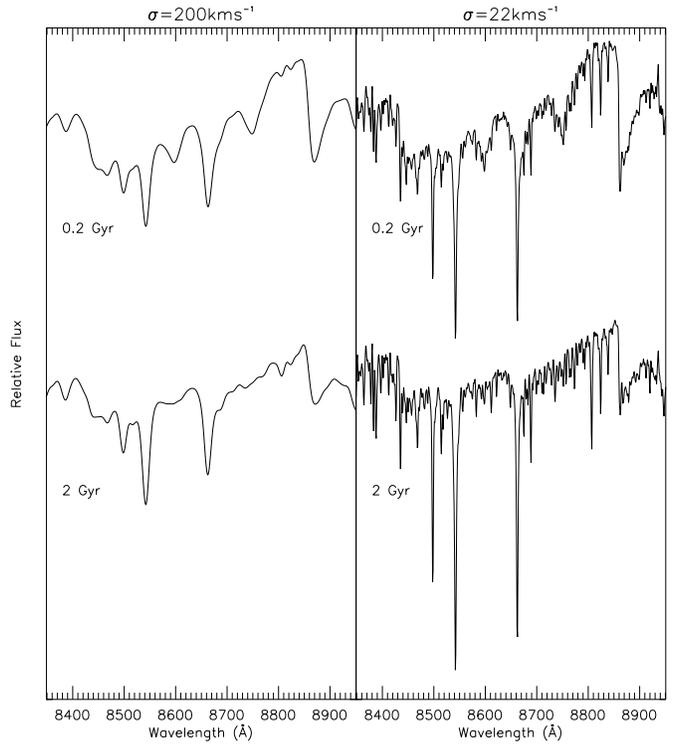,width=3.6in}}
\caption{
From top to bottom: SSP model spectra of 0.2 and 2.0~Gyr for solar metallicity
and unimodal IMF with slope 1.3. The left panel spectra were smoothed to
$\sigma = 200$ km~s$^{-1}$, whilst the right panel spectra are kept at the
nominal resolution of the models, i.e. $\sigma = 22.2$ km~s$^{-1}$.
}
\label{fig:cat_spectra_young}
\end{figure}

\begin{table}
\begin{center}
\caption{
Predicted CaT$^*$ and PaT indices for intermediate-age SSPs for solar
metallicity and different IMF shapes (for the unimodal and bimodal IMFs we
adopt $\mu$=1.3). The tabulated index values are calculated on the basis of
the empirical fitting functions given in Paper~III. The spectral
resolution is 1.5\AA~(FWHM). The CaT index can be obtained from the relation 
CaT = CaT$^*$ + 0.93~PaT
}
\label{tab:intermediate}
\begin{tabular}{@{}cc@{$\;$}c@{$\;$}cc@{$\;$}cc@{$\;$}c@{$\;$}c@{$\;$}cc@{$\;$}c@{}}
\hline
\multicolumn{1}{c}{Age}&\multicolumn{2}{c}{Unimodal}&&
\multicolumn{2}{c}{Bimodal}&&\multicolumn{2}{c}{K.Universal}
&&\multicolumn{2}{c}{K.Revised}\\
\cline{2-3}\cline{5-6}\cline{8-9}\cline{11-12}
(Gyr)& CaT$^*$&   PaT    &&  CaT$^*$&   PaT  && CaT$^*$&   PaT  && CaT$^*$& PaT  \\
\hline
 0.10  &  5.753  &  3.946   &&   5.755  &  3.952 && 5.755 &  3.953 &&  5.755  & 3.942 \\
 0.11  &  5.558  &  4.003   &&   5.560  &  4.010 && 5.560 &  4.010 &&  5.561  & 3.999 \\
 0.13  &  5.055  &  3.468   &&   5.056  &  3.473 && 5.056 &  3.473 &&  5.057  & 3.466 \\
 0.14  &  4.898  &  3.580   &&   4.899  &  3.586 && 4.898 &  3.586 &&  4.901  & 3.578 \\
 0.16  &  4.720  &  3.735   &&   4.720  &  3.741 && 4.720 &  3.742 &&  4.723  & 3.732 \\
 0.18  &  4.597  &  3.840   &&   4.596  &  3.847 && 4.596 &  3.848 &&  4.600  & 3.836 \\
 0.20  &  4.502  &  3.954   &&   4.502  &  3.963 && 4.502 &  3.963 &&  4.507  & 3.950 \\
 0.22  &  4.472  &  4.033   &&   4.471  &  4.042 && 4.471 &  4.043 &&  4.477  & 4.028 \\
 0.25  &  4.483  &  4.001   &&   4.482  &  4.011 && 4.482 &  4.011 &&  4.488  & 3.996 \\
 0.28  &  4.474  &  3.985   &&   4.473  &  3.995 && 4.473 &  3.996 &&  4.479  & 3.979 \\
 0.32  &  4.498  &  3.958   &&   4.497  &  3.969 && 4.497 &  3.970 &&  4.504  & 3.952 \\
 0.35  &  4.579  &  3.871   &&   4.578  &  3.882 && 4.578 &  3.883 &&  4.585  & 3.865 \\
 0.40  &  4.657  &  3.784   &&   4.656  &  3.796 && 4.656 &  3.797 &&  4.662  & 3.778 \\
 0.45  &  4.791  &  3.644   &&   4.791  &  3.657 && 4.791 &  3.658 &&  4.797  & 3.638 \\
 0.50  &  4.992  &  3.435   &&   4.993  &  3.447 && 4.993 &  3.448 &&  4.998  & 3.429 \\
 0.56  &  5.196  &  3.190   &&   5.198  &  3.202 && 5.198 &  3.203 &&  5.202  & 3.184 \\
 0.63  &  5.347  &  2.990   &&   5.350  &  3.002 && 5.350 &  3.003 &&  5.353  & 2.985 \\
 0.71  &  5.516  &  2.762   &&   5.520  &  2.774 && 5.520 &  2.775 &&  5.522  & 2.757 \\
 0.79  &  5.698  &  2.520   &&   5.703  &  2.531 && 5.703 &  2.531 &&  5.704  & 2.515 \\
 0.89  &  5.859  &  2.295   &&   5.865  &  2.305 && 5.865 &  2.306 &&  5.865  & 2.290 \\
 1.00  &  6.024  &  2.068   &&   6.031  &  2.078 && 6.031 &  2.078 &&  6.029  & 2.064 \\
\hline
\end{tabular}
\end{center}
\end{table}

\begin{figure}
\centerline{\psfig{file=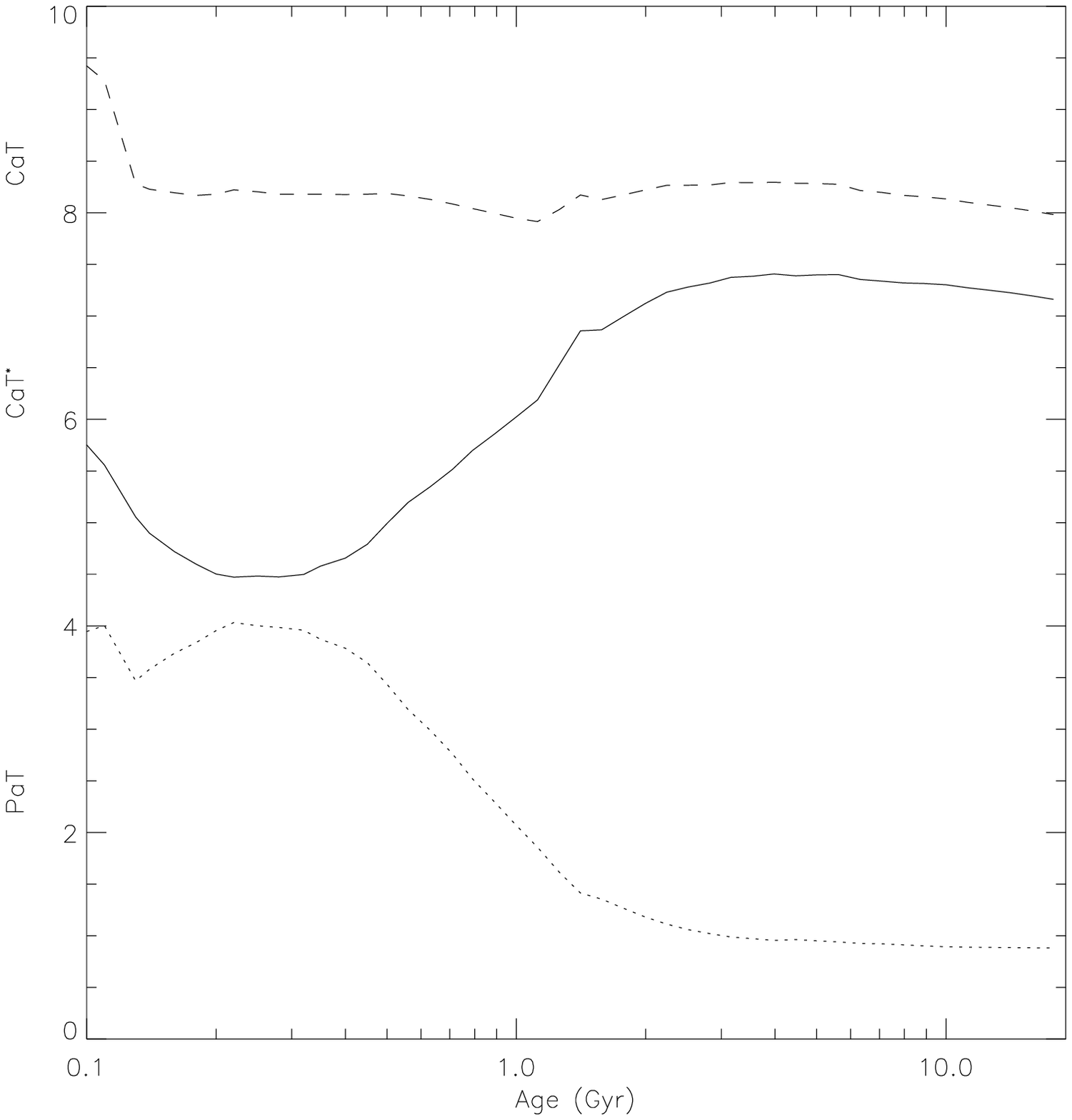,width=3.6in}}
\caption{
Time evolution of the CaT (dashed line), CaT$^{*}$ (solid line) and PaT (dotted
line) indices for intermediate aged SSPs. These predictions are calculated on
the basis of a unimodal IMF of slope 1.3 and solar metallicity. The spectral
resolution is 1.5\AA~(FWHM).
}
\label{fig:cat_young}
\end{figure}

Although the predictions for young stellar populations will be described in
detail in a forthcoming paper, in this section we extend the age of our SSP
models down to 0.1~Gyr for solar metallicity. When compared to old stellar
populations here are several major changes in the relative contributions to the
total luminosity of the different evolutionary stages in this age range. In
fact, the MS, HB and AGB phases increase their contributions, whilst the RGB
mostly disappears. In Figure~\ref{fig:cat_spectra_young} we plot a set of
representative SSP spectra corresponding to 0.2 and 2~Gyr for solar metallicity
and unimodal IMF of slope 1.3 (for two different spectral resolutions). As it
was expected, there is a significant strengthening of the Paschen series
towards younger ages (see for example the change in the strength of the lines
at either side of the third \cat line). This result is a consequence of the
increased contribution of hotter MS stars to the total luminosity in this
spectral range. We also see that the continuum around the \cat feature is
significantly affected by the molecular band absorptions, which constitute a
typical feature of the M-type stars. This continuum shape with its
characteristic slope is attributed to the contribution of the AGB phase (see
Appendix~\ref{ap:uncertain}), whilst similar shapes are obtained for older
stellar populations due to the contribution of the RGB.

In Figure~\ref{fig:cat_young} we show the time evolution of the PaT, CaT$^{*}$
and CaT indices. The plot shows the strengthening of the PaT index and the
weakening of the CaT$^{*}$ index towards younger SSPs. Interestingly, these
changes are compensated in a such a way that the CaT index is virtually
constant in this age range, and have similar values to the ones obtained for
older stellar populations. Finally, it is worth noting that the Paschen series
cannot be detected for SSPs of ages larger than $\sim1.5$~Gyr. In
Table~\ref{tab:intermediate} we list the predicted CaT$^{*}$ and PaT index
values corresponding to this age interval for several IMF types.

\section{Comparison with Previous Predictions}
\label{sec:comparison}
\begin{figure}
\centerline{\psfig{file=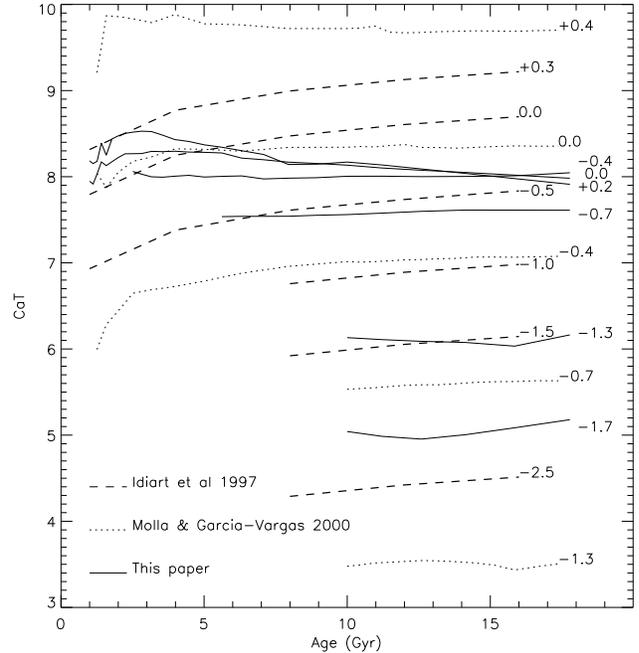,width=3.6in}}
\caption{
Comparison of the CaT index predictions by different authors for SSPs of
different metallicities and ages. All the models are calculated with a Salpeter
IMF. The dotted lines refer to the predictions of MGV, 
the dashed lines refer to ITD and the solid line
represents our predictions. To perform this comparison we have transformed
their predictions to our system on the basis of the formulae given in Paper~I
(see the text).
}
\label{fig:cat_comp}
\end{figure}

In Figure~\ref{fig:cat_comp} we compare our CaT index predictions (which
superseed those of V96) with those of MGV (which superseed those of
Garc\'{\i}a-Vargas, Moll\'a \& Bressan 1998) and ITD.
These authors employ different \cat
index definitions (see Paper~I for an extensive comparison between different
systems). However, we can transform their predictions to our system on the
basis of the relations given in Table~8 of Paper~I. In that Paper we presented
two sets of calibrations to convert between different systems. The first set
accounted for the effect of different \cat index definitions, while the second
set can be used to correct from flux calibration effects and differences in
spectral resolution. The composite calibrations to convert the predictions of
MGV (CaT(MGV), which are on the system by D\'{\i}az et al. (1989), and 
ITD (CaT(ITD), which employ the system by AZ88,
to our CaT index are the following:

\begin{equation}
\label{eq:conversions}
 \begin{array}{r@{}c@{}l@{}}
{\rm CaT}\ & = &\ 0.533+1.103\ {\rm CaT(MGV)}\\
{\rm CaT}\ & = &\ -0.155+1.183\ {\rm CaT(ITD)}.\\
 \end{array}
\end{equation}

In Fig.~\ref{fig:cat_comp} we compare the tabulated predictions of these
authors (converted to our system) with ours. Among the main differences, we
must remark that their CaT index predictions increase as a function of
metallicity for all metallicity regimes, whilst we obtain a saturation of this
index for metallicities above $\sim-0.5$. This saturation can be explained in
part by the fact that the larger the metallicity the lower is the difference
between the contributions of the Main Sequence (lower CaT values) and the Red
Giant Branch (larger CaT values) to the total luminosity of an SSP, as it can
be seen in Fig.~\ref{fig:cat_contr} (second row of panels,
i.e. $\mu=1.3$). Moreover, in metal-rich SSPs the RGB is populated with an
increasing fraction of cold stars, and, for temperatures cooler than
$\sim3750$~K, the CaT index decreases dramatically with decreasing temperature
(see Fig.~7 of Paper~III). The main point is that this behaviour for the M
giants is not predicted by the fitting functions used by these authors since
their stellar libraries do not include these stars.

In the case of ITD, the coolest stars included in their library are of K3
spectral type. MGV extrapolate to lower temperatures the theoretical fitting
functions of J{\o}rgensen et al. (1992), valid only for $T_{\rm
eff}\geq 4000$~K. Note that although MGV try to justify this extrapolation in
their Figure~1, their argument is not completely correct. For instance, for a
giant star of $T_{\rm eff}=3000$~K (which has a gravity of $\log g\sim 0$),
the J{\o}rgensen et al. extrapolated fitting functions predict CaT $=9.84$ 
\AA\ (11.39 \AA\ when corrected to our system), whereas the actual CaT 
measured values in our library stars of similar parameters range from 2.3 
to 5.0~\AA.

These effects may also explain the fact that these authors obtain slightly
larger CaT values for increasing ages, whilst our predictions for the time
evolution of the CaT index is towards slightly lower values for SSPs of large
metallicities.

For lower metallicities, our CaT predictions are in better agreement with the
predictions of ITD, whilst those of MGV show much lower values. This can be
explained on the basis of Figure~12 of Paper~III, where it can be seen that
the fitting functions of J{\o}rgensen et al. predict a larger dependence on
metallicity than what is actually observed.

 \begin{figure}
\centerline{\psfig{file=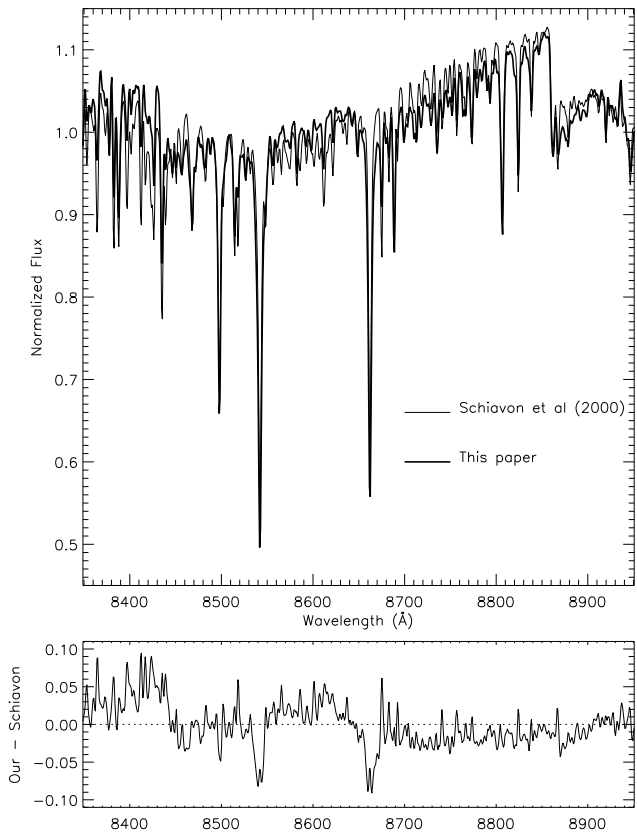,width=3.6in}}
\caption{
Normalized representative SSP model spectra of SBB
(thin line) and this work (thick line). The two models have a
Salpeter IMF, solar metallicity and 13~Gyr. The spectra were smoothed to
FWHM $=2$\AA. The difference between these two spectra is shown in the lower
panel.
}
\label{fig:cat_schiavon} 
\end{figure}

The first paper in synthesizing SSP spectra at high resolution in this spectral
range is that of SBB. These authors follow a
completely different approach than the one used in this paper. They employ a
fully synthetic stellar spectral library calculated by Schiavon \& Barbuy
(1999), which is based on model photospheres and molecular and atomic line
lists. Another important difference is that they adopt the old Padova set of
isochrones (B94), whilst we make use of the new Padova set.
Figure~\ref{fig:cat_schiavon} shows a comparison of a representative SSP model
spectrum, kindly provided to us by these authors, to an equivalent spectrum of
our model library. These spectra correspond to a Salpeter IMF, solar
metallicity and 13~Gyr and have a resolution 2~\AA~(FWHM). They were normalized
according to the flux in the spectral range covered by our models. In the lower
panel of Fig.~\ref{fig:cat_schiavon} we plot the difference between these two
models. Although we see some differences in the pseudocontinua, particularly in
the bluest one, the main difference is in the depth of the \cat feature. We
have measured the CaT index of the SBB
spectra and find that their \cat strengths are 2.5-3~\AA\ lower than the values
we obtain for the SSP spectra of our model library. In particular, for the
spectra plotted in Fig.~\ref{fig:cat_schiavon} we obtain 4.25~\AA\ for the
SBB spectrum and 7.33~\AA\ for our spectrum. Schiavon (private
communication) suggests that the difference in the \cat strength can be
attributed in part to the fact that their synthetic stellar spectral library was
calculated without adopting a NLTE approach, which it is needed to properly
model the depth of the lines. In fact we have compared the empirical
stellar spectra of our library to synthetic stellar spectra of similar
atmospheric parameters of their model grid and found that their \cat values are
systematically lower for the giant stars. These differences may be as high
as 5~\AA. On the other hand, we find a reasonably good agreement for dwarf stars.
This comparison shows that further work on the theoretical libraries, including
appropriate input physics and updated opacities, is required for this spectral
range in order to be able to make use of these libraries for stellar population
synthesis modeling.

\section{Comparison to Globular Clusters and Early-Type Galaxies}
\label{sec:GC-G}

\subsection{Globular clusters}
\label{sec:GC}

\begin{figure}
\centerline{\psfig{file=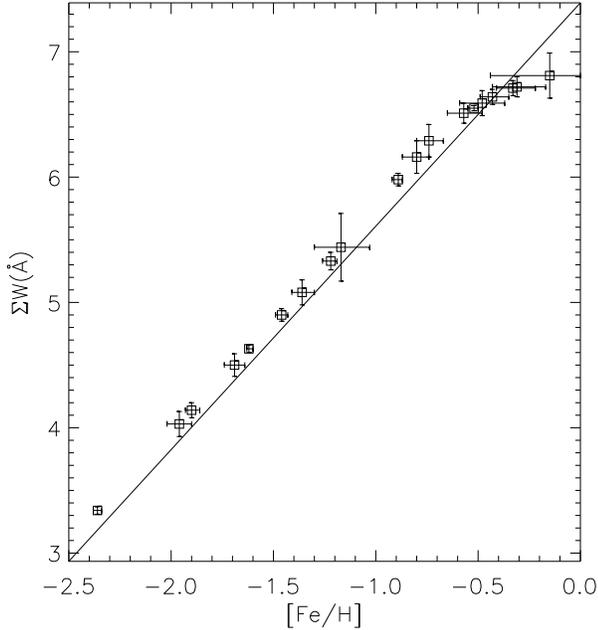,width=3.6in}}
\caption{
Galactic globular cluster \cat index measurements given by AZ88
(i.e. $\sum W$) versus the metallicities predicted by our models, when
transforming from their system to ours, by means of the CaT index. We assumed a
Salpeter IMF and that the clusters are 14~Gyr old (see the text). For clusters
with [Fe/H] $<-1.68$ we extrapolated our model predictions. The solid line
represents the AZ88 calibration.
}
\label{fig:cat_AZ88}
\end{figure}

Globular clusters are the ideal candidates to check our model predictions
since they can be treated as single-age, single-metallicity, stellar
populations. We follow here the test suggested by ITD
to calibrate our \cat predictions versus the
semi-empirical metallicity calibration of AZ88. For this
purpose, we need to transform their galactic globular cluster \cat index
measurements (i.e. $\sum W$, following the notation of these authors) to our
system by means of the CaT index. This conversion is required since these
authors used different spectral resolution and dispersion and did not correct
their data to a relative flux scale. Therefore we first applied the
broadening correction given in Table~7 of Paper~I to translate their
measurements, i.e. at resolution 4.8~\AA, to ours (1.5~\AA).
The second step is to transform their $\sum W$ index (on their instrumental
response curve) to our CaT index (on a relative flux scale) by making use of
Eq.~\ref{eq:conversions}.

The fact that our models show that the \cat feature exhibits a negligible
dependence on the age of the stellar population, for the age range that is
usually assumed for the globular clusters, makes it possible to neglect this
parameter. Figure~\ref{fig:cat_AZ88} shows our metallicity estimates obtained
from a comparison of the corrected CaT measurements to our predictions for
models of 14~Gyr and unimodal IMF of slope 1.3. We note that we extrapolate the
models for [Fe/H] $<-1.68$. The solid line represents the AZ88
calibration (${\rm [Fe/H]}_{\rm Ca\sc{II}}=-4.146 + 0.561\sum W$). 
Fig.~\ref{fig:cat_AZ88} shows the good agreement achieved for metallicities
smaller than $\sim-0.5$. This result probes the validity of using the \cat of
integrated spectra as an alternative method to derive metallicities in globular
clusters of [M/H]$<-0.5$.

Fig.~\ref{fig:cat_AZ88} also shows a small shift between the obtained
metallicities $\Delta{\rm [Fe/H]} < 0.15$~dex. In part, this shift may be
explained by the large sensitivity of the $\sum W$ index of AZ88
to the spectral resolution, as it was shown in Paper~I. In fact, we have
tested that for a resolution mismatch of $\sim 20$ km~s$^{-1}$ the
obtained offset vanishes. This mismatch could either be originated if the
AZ88 resolution differs from 4.8~\AA~ (these authors
obtained their observations in different telescopes and used different
instrumental setups) or by our assumption of taking the broadening correction
representing a M0 star, according to Paper~I. An important factor to be taken
into account for explaining the obtained metallicity shift is the fact that we
adopted the Carretta and Gratton (1997) scale for the cluster stars of our
stellar spectral library (see Paper~II). Moreover, for the stars of M~71 we
adopted here an even lower metallicity value (i.e. [Fe/H] $=-0.84$) than
that predicted by Carretta \& Gratton (1997) (i.e. [Fe/H] $=-0.70$), as discussed
in \S~\ref{sec:library}. It has been largely discussed in the literature that
this scale, based on high-dispersion spectra, shows
significant deviations from the Zinn \& West (1984) scale, which was
adopted by AZ88. This effect is more pronounced for the
largest metallicities. We refer the reader to the discussion presented in
Rutledge et al. (1997).

Finally, we note that for metallicities larger than $\sim-0.5$ the measured CaT
index does not increase as a function of metallicity, departing from the
metallicity scale relation of AZ88. Therefore, these
measurements are in agreement with our model predictions. However, as a result
of this saturation, we are not able to provide accurate metallicity estimates
for the most metal-rich globular clusters on the basis of the \cat. 

\subsection{Early-type galaxies}
\label{sec:G}

In C03 we have applied these models to a large sample of
early-type galaxies of different luminosities. In that paper we analyse the
\cat as well as newly defined features, such as the slope of the continuum
around the \cat lines and the Mg$_{\sc I}$ feature at 8807\AA. We therefore
refer the reader to that paper for a complete description of the galaxy sample
and for an extensive discussion of their stellar populations. In this section,
we aim at showing the potential use of the new models by selecting two
representative galaxies from C03 sample that have already been
analysed in the blue spectral range (V01A) on the basis of
very high quality spectra that allowed us to measure the new age indicator
H${\gamma_{\sigma}}$ of VA99, which does not depend on
metallicity. These galaxies are NGC~4478 and NGC~4365. The galaxy spectra that
we use here correspond to the central 4~arcsec aperture.

The synthesized model spectra both in the blue (V99) and in the spectral
region of the \cat can be used to analyse the observed galaxy spectra in a
very easy and flexible way, allowing us to adapt the theoretical predictions
to the characteristics of the data, instead of proceeding in the opposite
direction as, for example, we must do when we have to compare our data
with models based on the widely used Lick/IDS system (see V99 for an extensive
discussion of the advantages of the new approach). After smoothing the
synthetic SSP spectra, with flux-calibrated spectral response curves, to the
measured resolution (i.e.  $\sigma_{\rm total}^{2}=\sigma_{\rm
galaxy}^{2}+\sigma_{\rm instr}^{2}$), we can analyse the galaxy spectrum in
its own system. The spectral resolution of the near-IR spectra that we use
here is $\sigma_{\rm instr}=42$ km~s$^{-1}$. We then measure our favorite
spectral features, such as the Ca\,{\sc ii} triplet, in both the models and
the data in order to build up index-index diagrams that can help us to obtain
the most relevant stellar population parameters.

\begin{figure*}
\centerline{\psfig{file=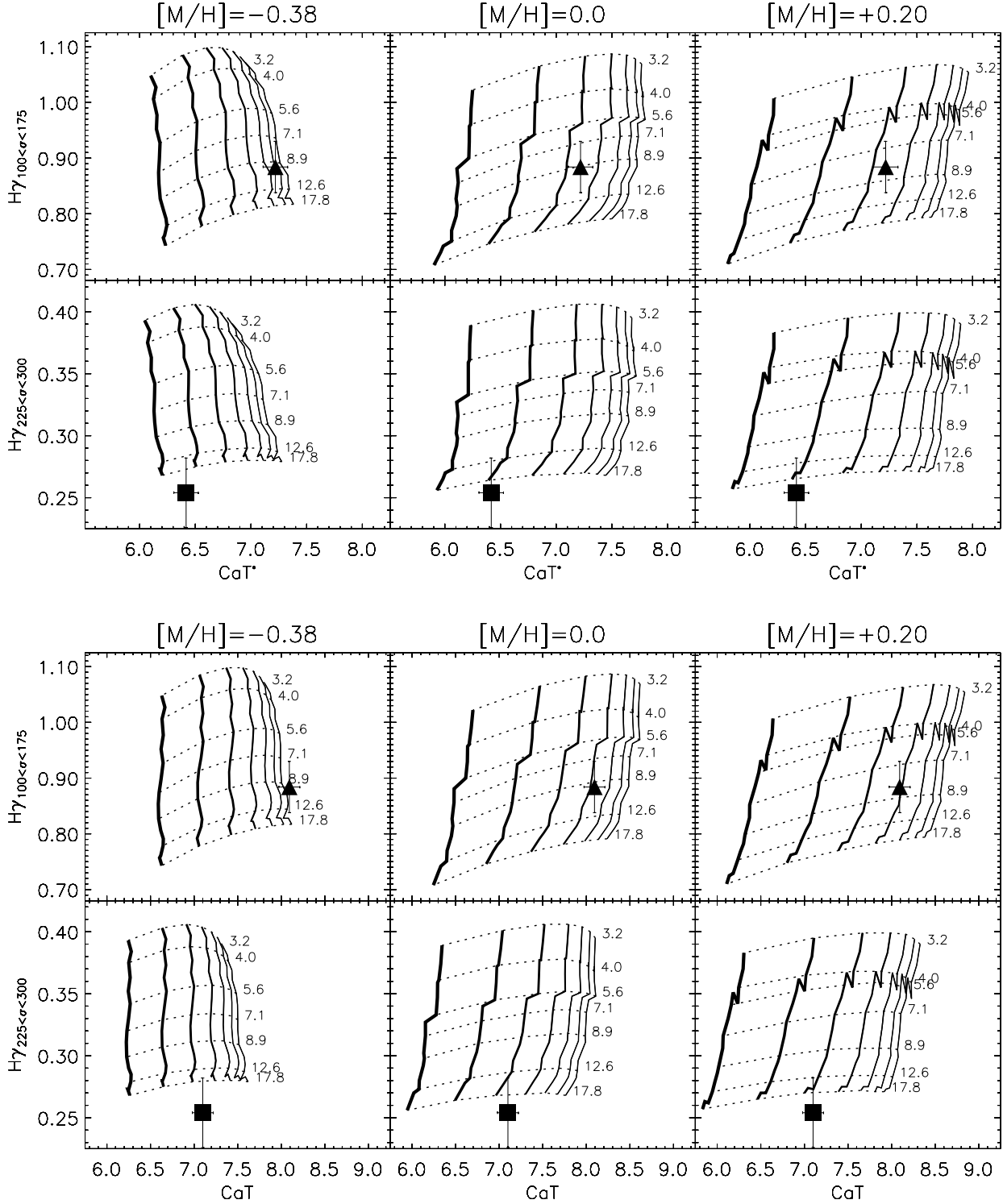,width=6.6in}}
\caption
{
Plots of CaT$^*$ and CaT indices versus the H$\gamma_{\sigma}$ age indicators
of VA99. Filled symbols represent the index measurements
for the central apertures (4~arcsec) of two galaxies in common between the
samples of C03 (for the \cat indices) and V01A
(for the H$\gamma$ indices). The triangle represents NGC~4478, whilst
the square represents NGC~4365. Overplotted are the models by V99
for the H$\gamma$ indices and those from this paper for the \cat feature. All
index measurements were performed after smoothing the model SEDs to the
appropriate galaxy velocity dispersion and instrumental resolution. We increase
the metallicity of the model grids from left to right panels as quoted in the
upper plots. We use a unimodal IMF, varying in each panel its slope from
$\mu=0.3$ (thinnest vertical line) to $\mu=3.3$ (thickest) by steps of
$\Delta\mu=0.5$. The third vertical line starting from the thinnest (i.e. right)
corresponds to the Salpeter (1955) value (i.e. $\mu=1.3$). Finally, thin dotted
horizontal lines represent models of constant ages, which are quoted in
gigayears. 
}
\label{fig:cat_G_HG}
\end{figure*}

\begin{figure*}
\centerline{\psfig{file=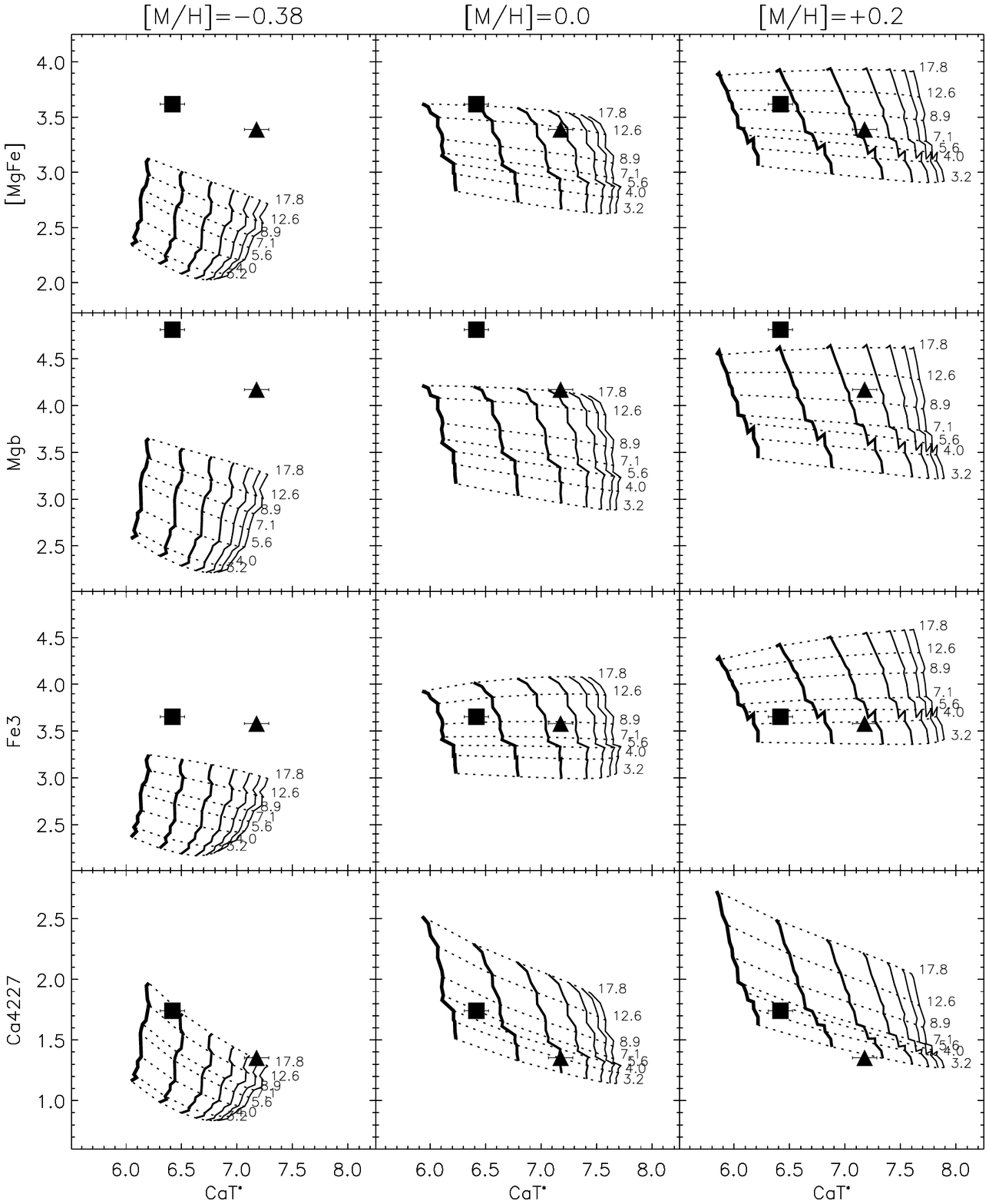,width=7.0in}}
\caption
{
Plots of CaT$^*$ versus different metallicity indicators in the blue spectral
region. Lines and symbols have the same meaning as in Fig.~\ref{fig:cat_G_HG}.
In these plots all index measurements were obtained at resolution
$\sigma_{\rm total}=260$ km~s$^{-1}$. The blue indices are taken from V01A.
}
\label{fig:cat_G_METALS}
\end{figure*}

In Figure~\ref{fig:cat_G_HG} we combine our new \cat index measurements 
(C03) with the blue index measurements of V01A. The
CaT$^*$ and CaT indices are plotted versus the H$\gamma_{100<\sigma<175}$ and
H$\gamma_{225<\sigma<300}$ indices of VA99. Each
H$\gamma_{\sigma}$ index definition provides stable and sensitive age predictions
within the $\sigma$ ranges quoted in the subindices. Therefore, whereas the
first index is the one appropriate to study NGC~4478 (i.e.
$\sigma_{\rm total}=140$ km~s$^{-1}$), the second is adequate to study NGC~4365
(i.e.  $\sigma_{\rm total}=260$ km~s$^{-1}$). The metallicity of the models
increases from the left to the right panels as quoted above the top panels. We
use models of unimodal IMF characterized by its slope $\mu$, which is constant
along the vertical solid lines. We vary this slope, indicated by the strength
of the line, from $\mu=0.3$ (thinnest) to $\mu=3.3$ (thickest) in steps of
$\Delta\mu=0.5$. The third vertical line starting from the thinnest (i.e.
right) corresponds to the Salpeter (1955) (i.e. $\mu=1.3$). Models of constant
age are shown by thin dotted horizontal lines.

All the model grids of Fig.~\ref{fig:cat_G_HG} look approximately orthogonal, which
indicates that for a given metallicity we are able to separate the effects of
the age from the ones of the IMF slope. According to these plots, smaller
H$\gamma_{\sigma}$ values mean larger ages and smaller CaT$^*$ or CaT indices
mean steeper IMF slopes (i.e. dwarf-dominated). We see that the inferred ages
do not depend on the metallicity of the model grid in use due to the
insensitivity of H$\gamma_{\sigma}$ to this parameter. Moreover, the fact that
the \cat indices are insensitive to the metallicity for the range of
metallicities covered in these plots ($-0.38\le{\rm [M/H]}\le+0.2$) makes it
possible to provide almost unique IMF slope solutions. All the diagrams
indicate a mean luminosity weighted age of $\sim$9~Gyr for NGC~4478 and
$\simeq15$~Gyr for the giant elliptical galaxy NGC~4365. It is worth noting
that the signal-to-noise of the NGC~4365 spectrum was not high enough for an
accurate measurement of the H$\gamma_{\sigma}$ index. However, the obtained age
is not different from the one derived on the basis of the H$\beta$ index (see
V01A), and the age estimate given in Davies et al. (2001) from
integral field spectroscopy. From the H$\gamma_{\sigma}$ versus CaT$^*$
diagrams we find $\mu\sim2.3$ for NGC~4478 and $\mu\sim2.8$ for NGC~4365, which
indicate rather steep IMFs in comparison to the Salpeter value. However, it is
worth recalling that variations in the theoretical prescriptions adopted for
building-up the stellar tracks and isochrones might drive to important changes
in the predicted IMF slope. In fact, if we replace the model grids of
Fig.~\ref{fig:cat_G_HG} with another set synthesized on the basis of the old
Padova isochrones (i.e. B94), which yield smaller CaT$^*$
strengths (i.e. $\sim-0.5$~\AA) as a result of a slightly cooler RGB phase (see
Appendix~\ref{ap:isochrones} for details on the differences in the adopted
theoretical prescriptions), we would have obtained $\mu\sim1.3$ for NGC~4478 and
$\mu\sim1.8$ for NGC~4365 (the H$\gamma_{\sigma}$ age indicator does not vary
significantly as a result of this replacement, see V01B).
Therefore these results must be taken into account on a relative basis, whereas
the discussion about galaxy trends is more secure (C03).

In Figure~\ref{fig:cat_G_METALS} we plot the CaT$^*$ index versus different
metallicity indicators in the blue spectral range, i.e. [MgFe] (Gonz\'alez 1993),
Mg$b$ (W94), Fe3 (K00) and Ca4227 (W94).
In order to be able to show on a single plot the two galaxies we have
smoothed the spectrum of NGC~4478 to match the resolution of NGC~4365. Each
panel shows an approximately orthogonal model grid, with the position of the grid
varying from the left to the right panels as a function of increasing
metallicity. We, therefore, are unable to obtain a robust age determination
on the basis of this figure. However, if we adopt for these galaxies the ages
inferred from Fig.~\ref{fig:cat_G_HG} we are in position to obtain the
metallicity. The [MgFe] -- CaT$^*$ diagrams suggest [M/H]$_{\rm [MgFe]}\sim+0.1$
for NGC~4478 and solar metallicity for NGC~4365. From the Mg$b$ -- CaT$^*$
diagrams we obtain [M/H]$_{\rm Mgb}\sim+0.2$ for the two galaxies. The Fe3 --
CaT$^*$ plots suggest [M/H]$_{\rm Fe3}\sim0.0$ for NGC~4478 and
[M/H]$_{\rm Fe3}\sim-0.1$ for NGC~4365. Finally, from the Ca~4227 -- CaT$^*$
diagrams we obtain [M/H]$_{\rm Ca4227}\sim-0.2$ for NGC~4478 and
[M/H]$_{\rm Ca4227}\sim-0.4$ for NGC~4365. Interestingly, despite the fact that NGC~4365 is a
larger galaxy, its metallicity seems to be slightly lower than the one obtained
for NGC~4478.

Fig.~\ref{fig:cat_G_METALS} shows us that the obtained metallicities are
strongly influenced by non-solar abundance ratios in agreement with previous
determinations (e.g. see, for recent references, K00; T00;
V01A, and PS02). Particularly
interesting is the fact that the lowest metallicities are inferred when using
the Ca~4227 -- CaT$^*$ diagrams. Despite the fact that Ca, like Mg, is an
$\alpha$-element, Ca~4227 does not track Mg$b$ (see V97;
Peletier et al. 1999; T00; V01A;
PS02). Although this result is in disagreement with our present
knowledge of the nucleosynthesis theory (Woosley \& Weaver 1995), if these
galaxies were deficient in Ca, causing the Ca~4227 line to be smaller than
predicted by scaled-solar models, we should consider the possibility that the
\cat feature might be reflecting this deficiency as well. This suggests, as an
alternative scenario, that the IMF is Salpeter-like and the low \cat values are
due to Ca deficiencies rather than from the steepening of the IMF. 

A possible approach that can be followed for interpreting these results is to
build-up SSP predictions based on non solar abundance ratios. Such isochrones
were predicted by, e.g. SW98, S00, VandenBerg et al. (2000), Kim et
al. (2002), for several $\alpha$-enhancement ratios. These calculations were
motivated by the finding that elliptical galaxies show an enhancement of Mg
over Fe when compared to scaled-solar stellar population model predictions
(e.g. Peletier 1989; Worthey et al.  1992; K00; V01A). However to predict
stellar population models of different $\alpha$-enhancement ratios we require
stellar tracks built-up on the basis of appropriate opacity tables and energy
generations, and the corresponding $\alpha$-enhanced stellar spectral
libraries. It is not yet clear what ratios should be adopted for the different
$\alpha$-elements. In fact such $\alpha$-enhancement stellar models have been
calculated by various authors assuming a constant enhancement for each
$\alpha$-element (e.g. VandenBerg et al. 2000; Kim et al. 2002), or somewhat
more empirically-based element mixture (e.g. SW98; S00). Furthermore Ca~4227
does not track Mg (e.g. V97).  However, such theoretical stellar spectral
libraries are not yet available and all the empirical libraries such as the
one used here mostly follow the Galactic disk element ratios, particularly for
the metallicity regime characteristic of the galaxies shown in
Fig.~\ref{fig:cat_G_METALS} (i.e. around solar). Amongst the most important
effects of adopting such isochrones is the fact that, for a given total
metallicity, the $\alpha$-enhanced mixtures lead to lower opacities, which
translates into an increase of the temperature of the stars in both the Main
Sequence (MS) and the RGB phases. However, for a given [Fe/H] metallicity, we
obtain the opposite trend since the total metallicity of the $\alpha$-enhanced
models is larger. Another important result from these studies is that, for
metal-rich stellar populations, $\alpha$-enhanced isochrones cannot be
mimicked with a scaled-solar isochrone of different metallicity.

Although it would be interesting to have models with varying [Ca/Fe] ratios as
well, stellar libraries and isochrones calculated with such abundance ratios
are not yet available. Moreover, it is not clear how we expect [Ca/Fe] to
behave. It is worth noting that we did not obtained any significant correlation
of the [Ca/Fe] abundance ratios with the residuals of the predictions of the
fitting functions and the observed values for the \cat in our stellar library
(Paper~III). Despite the fact that we are not in position to build-up such
fully self-consistent non-solar element ratios SSP models, we have tested in
Appendix~\ref{ap:isochrones} the effects of adopting the $\alpha$-enhanced
isochrones of S00. These isochrones are calculated on the
basis of the same input physics as that used in G00 for old
stellar populations. For total metallicity [M/H] $\ge0$ we obtain larger
CaT$^*$ strengths ($\Delta{\rm CaT}^*\sim0.5$~\AA). If we plot these model
predictions on Fig.~\ref{fig:cat_G_HG} we obtain even steeper IMF slopes.
Alternatively, if a Salpeter-like IMF is used, the $\alpha$-enhanced isochrones
lead to a larger calcium underabundance problem.

\begin{figure}
\centerline{\psfig{file=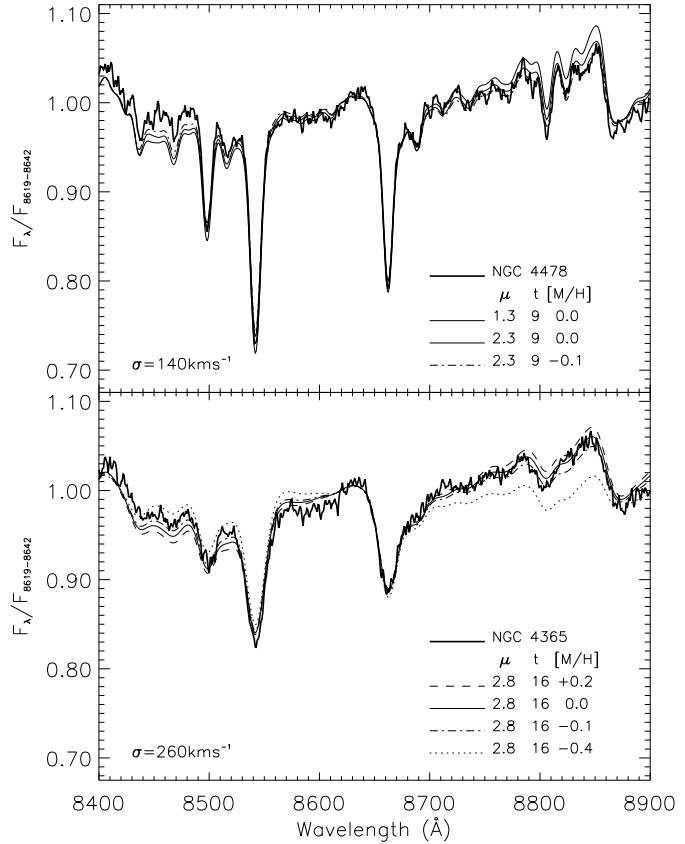,width=3.6in}}
\caption{
Spectra of NGC~4478 (upper panel) and NGC~4365 (lower panel). Overplotted are
several model spectra selected on the basis of Fig.~\ref{fig:cat_G_HG} and
Fig.~\ref{fig:cat_G_METALS} (see text for details). All the 
spectra were normalized according to the flux measured in the spectral 
region $\lambda\lambda$8619-8642\AA\
}
\label{fig:cat_spectra_g}
\end{figure}

It is however interesting to see how the model spectra of the adopted IMF
slopes, ages and metallicities match the full spectral region around the \cat
feature for these galaxies. In Figure~\ref{fig:cat_spectra_g} we show the spectra
of NGC~4478 (top panel) and NGC~4365 (bottom panel). For NGC~4478 we overplot
several models of age 9~Gyr as suggested by Fig.~\ref{fig:cat_G_HG}. In order
to illustrate the effect of the IMF slope we have chosen two model spectra
corresponding to $\mu=1.3$ (as a reference) and $\mu=2.3$ (the most suitable
fit, according to Fig.~\ref{fig:cat_G_HG} and Fig.~\ref{fig:cat_G_METALS}). The
selected metallicity is around solar according to the [MgFe] and Fe3 indices.
All the spectra were normalized according to the flux measured in the spectral
region $\lambda\lambda$8619-8642~\AA. It is apparent that the reference model
with $\mu=1.3$ provides slightly deeper \cat lines than observed in the galaxy
spectrum. We also see that the model with $\mu=2.3$ provides a better fit to
the whole spectral region, including the slope of the continuum around the \cat
feature. Finally we overplot another model spectrum of $\mu=2.3$ and
[M/H] $=-0.1$, which seems to provide a somewhat better fit to the overall galaxy
spectrum. 

In the lower panel of Fig.~\ref{fig:cat_spectra_g} we plot the spectrum of
NGC~4365 and overplot several model spectra of similar age (15.85~Gyr) and IMF
slope ($\mu=2.8$). The four overplotted models were selected to have the
metallicities inferred on the basis of the four metallicity indicators studied
in Fig.~\ref{fig:cat_G_METALS}. Overall, we see that the best fits are achieved
for the metallicities in the range $-0.2\le{\rm [M/H]}\le+0.1$, in good agreement
with the predictions obtained on the basis of the [MgFe] (i.e. $\sim0.0$) and
Fe3 (i.e. $\sim-0.1$) indices. On the other hand, the models of metallicities
obtained from the Mg$b$ (i.e. $\sim+0.2$) and Ca4227 (i.e. $\sim-0.4$)
provides slightly steep and rather flat continua around the \cat feature
respectively. This result suggests that the metallicity inferred on the basis
of the Fe3 and [MgFe] indices provide reasonably good fits to the overall
spectrum in this spectral range.

\section{Conclusions}
\label{sec:conclusions}

We have presented a new evolutionary stellar population synthesis model in the
near-IR spectral region covering the range $\lambda\lambda$8350-8950~\AA\ at a
spectral resolution of 1.5~\AA~(FWHM). The ultimate aim of the models is to
derive reliable predictions of the \cat strength for stellar populations in a
wide range of ages, metallicities and IMF shapes. Apart from spectral energy
distributions, the model predicts measurements in a set of new line indices,
namely CaT$^{*}$, CaT and PaT, which we have defined in Paper~I to overcome
some of the limitations of previous \cat index definitions in this spectral
range. In particular, the CaT$^{*}$ index is specially suited to remove the
contamination from Paschen lines in the integrated spectra of galaxies. 

The stellar population model presented here is a revised version of the model
of V96 and V99. Several aspects of the model
have been updated in order to produce a state-of-the-art output. The isochrones
of B94 have been replaced by the newest Padova isochrones
(G00), based on solar abundance ratios. The theoretical
parameters of the predicted stars have been transformed to fluxes and colours
on the basis of almost fully empirical relations based on extensive photometric
stellar libraries. Apart of the IMF shapes used in V96 we
use the two multi-part power-law IMFs recently proposed by K01. The
main ingredient for the predictions presented here, is the new extensive
empirical stellar spectral library presented in Paper~I and II, from which a
subsample composed of more than 600 stars has been carefully selected. The
sample shows an unprecedented coverage of the stellar atmospheric parameters
(T$_{\rm eff}$, log $g$ and [M/H]) for stellar population synthesis modeling. 

Two main products of interest for stellar population analysis are presented:
{\it i)} a spectral library for SSPs with metallicities $-1.7<{\rm [Fe/H]}<+0.2$, a
large range of ages (0.1-18~Gyr) and IMF types, and  {\it ii)} line-strengths
calculated on the basis of the empirical fitting functions presented in
Paper~III for the CaT$^*$, CaT and PaT indices, and for which conversion
formulae to previous definitions can be found in Paper~I. Tables and model
spectra are available electronically at the web pages given in
\S~\ref{sec:modelsIR}, together with the \cat stellar library. 

The newly synthesized model spectra can be used to analyze observed galaxy
spectrum in a safe and flexible way, allowing us to adapt the theoretical
predictions to the nature of the data instead of proceeding in the opposite
direction. The synthetic SSP library, with flux-calibrated spectral response,
can be smoothed to the same resolution of the observations or to the measured
internal galaxy velocity dispersion, making it possible to fully use all the
information in the data. This opens the way (as do the models of V99
for the blue spectral region) for new applications. For example, if part of the
spectrum is corrupted or affected by strong sky-lines, one can easily define
new absorption lines on both models and data. Also, these models can be used as
templates for determining stellar kinematics in galaxies (see Falc\'on-Barroso
et al. 2003). Moreover, the models offer us a great opportunity to accurately
study galaxies at larger redshifts. 

We have analyzed in detail the behaviour of the \cat feature in old-aged SSPs,
finding the following results: {\it i)} the strength of the CaT$^*$ index does
not vary much for ages larger than $\sim3$~Gyr, for all metallicities, {\it
ii)} this index shows a strong dependence as function of metallicity for values
below [M/H] $\sim-0.5$ and {\it iii)} for higher metallicities, this index does
not show any significant dependence either on age or on the metallicity, being
more sensitive to changes in the slope of power-like IMF shapes. These models
open up the analysis of a suitable dwarf-giant discriminator in the analysis of
stellar populations in galaxies. It is worth noting that the saturation of the
\cat feature for metallicities above $-0.5$ has not been predicted by previous
models. This prediction might be supported by the most metal-rich Galactic
globular clusters and galaxy data analyzed here. Furthermore, it might also be
supported by the measurements of the CaT$^*$ and CaT indices in large samples
of elliptical galaxies (Saglia et al. 2002; C03) and bulges
(Falc\'on-Barroso et al. 2002). 

For stellar populations in the age interval 0.1-1~Gyr we find that the Paschen
lines, as defined by the PaT index, become more prominent and the CaT$^*$ index
decreases significantly. However the two effects are compensated in the CaT in
such a way that that this index is virtually constant along this age range.
Another interesting result is that the overall shape of the continuum around
the \cat feature for large metallicities resembles more that of M-type stars
rather than that of K-giants (as it is the case for the blue spectral range).

By comparing globular cluster data (\S~\ref{sec:GC}) we show that the derived
metallicities are in excellent agreement with the ones from the  literature,
probing the validity of using the integrated \cat feature for determining
metallicities for these stellar systems. It is worth noting that this is
possible since the models predict that the \cat indices are virtually constant
for stellar populations of ages larger than $\sim$3~Gyr and strongly dependent
on metallicity for [M/H] $<-0.5$.

In \S~\ref{sec:G} we have applied the models to two early-type galaxies of
different luminosities, NGC~4478 and NGC~4365 (from C03), for which a detailed
stellar population analysis based on the optical spectral range was published
in V01A. We propose several index-index diagrams, i.e. CaT$^*$ versus several
age indicators, and CaT$^*$ versus several metallicity indicators, which
provide virtually orthogonal model grids from which to disentangle,
unambiguously, the relevant parameters of their stellar populations. The
Ca-abundance as inferred from the \cat does not follow Mg, as nucleosynthesis
calculations predict, in agreement with measurements of the Ca~4227 line in
the blue (V97). We should, however, note that \cat measurements cannot be
fitted unless a very dwarf-dominated IMF is imposed, or unless the
Ca-abundance, as measured from the Ca\,{\sc ii} triplet, is even lower than
the Fe-abundance. However, it is worth recalling that variations in the
theoretical prescriptions adopted for building-up the stellar tracks and
isochrones might lead to significantly lower IMF slope estimates and,
therefore, these results must be taken into account on a relative basis, being
the discussion about galaxy trends more secure (see e.g. Saglia et
al. 2002; C03). Moreover if we adopted $\alpha$-enhanced isochrones we would obtain
steeper IMF slopes compared to scaled-solar. Alternatively, if a Salpeter-like
IMF is used, the $\alpha$-enhanced isochrones lead to a larger calcium
underabundance problem. It is also important to note that the overall shape of
the spectrum around the \cat feature and in particular the characteristic
slope of the continuum is significantly better represented by choosing SSPs of
metallicities as inferred from Fe or [MgFe] indices, rather than the ones
based on the Mg$b$ or Ca~4227 lines. More details can be found in C03, 
where a large sample of ellipticals are discussed in detail revealing
very interesting trends.

\section*{Acknowledgments}

We are indebted to the Padova group for making available their isochrone
calculations. We are grateful to R. Schiavon for providing us with a set of
synthetic spectra of SSPs and stars, as well as for very interesting
discussions. This work was supported in part by a British Council grant within
the British/Spanish Joint Research Programme (Acciones Integradas) and by the
Spanish Programa Nacional de Astronom\'{\i}a y Astrof\'{\i}sica under grant
No. AYA2000-974. This work is based on observations at the JKT, INT and WHT on
the island of La Palma operated by the Isaac Newton Group at the Observatorio
del Roque de los Muchachos of the Instituto de Astrof\'{\i}sica de Canarias.


\label{lastpage}

\appendix

\section{The IMF Types}
\label{ap:IMF} 

In this appendix we summarize all the IMFs used in this paper, which include
the two IMF shapes given by V96, i.e. unimodal and bimodal,
and the new IMFs proposed by K01, i.e. universal and revised. These
four IMF shapes are plotted in Fig.~\ref{fig:IMF}. 

The unimodal IMF has a power-law form characterized by its slope $\mu$ as a
free parameter

\begin{equation}
\Phi(m)=\beta m^{-(\mu+1)}.
\end{equation}

\noindent
Therefore, the Salpeter (1955) solar neighbourhood IMF is obtained when
$\mu=1.3$.

The bimodal IMF is similar to the unimodal case for stars with masses above
$0.6{\rm M}_{\odot}$, but decreasing the weight of the stars with lower masses
by means of a transition to a shallower slope, which becomes flat in the $\log
(\Phi (\log (m)))$ -- $\log (m)$ diagram for masses lower than $0.2{\rm
M}_{\odot}$.  The main motivation for this IMF is to achieve a reasonable good
fit to the observational data of Scalo (1986) and Kroupa, Tout \& Gilmore
(1993) on the basis of a single free parameter, i.e. $\mu$, rather than using
several segments as it is the case for the IMFs of K01. The main advantage of
the bimodal IMF is that it allows us to vary the parameter $\mu$ in the same
way as for the unimodal case. The form of this IMF\footnote{Note the erratum
present in Eq.~4 (case $m\leq0.2 {\rm M}_{\odot}$) of V96, which should be divided
by $m$.} is given by

\begin{equation}
\Phi(m) = \beta \left\{
\begin{array}{@{\;}l@{\;}c@{\;}l}
\frac{m_{1}^{-\mu}}{m}&,& m \leq m_0\\
p(m)&,& m_0 < m \leq m_2\\
m^{-(\mu+1)} &,& m > m_2,\\
\end{array}\right.
 \label{eq:bimodal}
\end{equation}

\noindent
where $m_0$, $m_1$ and $m_2$ are 0.2, 0.4 and 0.6~M$_{\odot}$ respectively.
$p(m)$ is a spline for which we calculate the corresponding coefficients
solving for the following system obtained by the boundary conditions

\begin{equation}
\begin{array}{r@{\;}c@{\;}l}
p(m_0) &=&m_{1}^{-\mu},\\
p'(m_0)&=&0,\\
p(m_2) &=&m_{2}^{-\mu},\\
p'(m_2)&=&-\mu m_{2}^{-(\mu+1)}.\\
\end{array}
\end{equation}

We have introduced in this paper the two IMF shapes recently proposed by K01.
The universal IMF is a multi-part power-law IMF, which has the
following form  

\begin{equation}
\Phi(m) = \beta \left\{
\begin{array}{@{\;}l@{\;}c@{\;}l}
\left(\frac{m}{m_0}\right)^{-0.3}&,& m \leq m_0\\
\left(\frac{m}{m_0}\right)^{-1.3}&,& m_0 < m \leq m_1\\
\left(\frac{m_1}{m_0}\right)^{-1.3}\left(\frac{m}{m_1}\right)^{-2.3}&,& m > m_1,\\
\end{array}\right.
 \label{eq:kuniversal}
\end{equation}

\noindent
where $m_0$ and $m_1$ are 0.08 and 0.5~M$_{\odot}$, respectively.

We have also implemented the revised multi-part power-law IMF of K01, which
tries to correct for the systematic effects due to unresolved binaries on the
single-star IMF. The main effect is that the slope of the IMF is steeper than
the universal IMF by $\Delta\mu\sim0.5$ for the mass range $0.08 < m <
1~{\rm M}_{\odot}$. Its shape is summarized in the following equation:

\begin{equation}
\Phi(m) = \beta \left\{
\begin{array}{@{\;}l@{\;}c@{\;}l}
\left(\frac{m}{m_0}\right)^{-0.3}&,& m \leq m_0\\
\left(\frac{m}{m_0}\right)^{-1.8}&,& m_0 < m \leq m_1\\
\left(\frac{m_1}{m_0}\right)^{-1.8}\left(\frac{m}{m_1}\right)^{-2.7}&,& m_1 < m \leq m_2\\
\left(\frac{m_1}{m_0}\right)^{-1.8}\left(\frac{m_2}{m_1}\right)^{-2.7}
\left(\frac{m}{m_2}\right)^{-2.3}&,& m > m_2,\\
\end{array}\right.
 \label{eq:krevised}
\end{equation}

\noindent
where $m_0$, $m_1$ and $m_2$ are 0.08, 0.5 and 1~M$_{\odot}$, respectively.

Finally the constant $\beta$ is calculated via the normalization

\begin{equation}
\beta \int_{m_l}^{m_u} \Phi(m)dm=1,  
\end{equation}

\noindent
where $m_l$ and $m_u$ are the lower and upper mass cutoffs. We adopt
0.01~M$_{\odot}$ and 120~M$_{\odot}$ for these stellar masses, respectively.

\section{Computation of a Representative Stellar Spectrum}
\label{ap:boxes}

We provide here a detailed description of how we interpolate spectra in the
database and compute the spectrum corresponding to a given set of requested
parameters ($\theta_{0}$, $\log g_{0}$ and [M/H]$_0$). The most
straightforward approach to follow is to let the code find all the stars
enclosed within a given cube centered on these atmospheric parameters. However,
since in many cases there is a lack of symmetry in the distribution of stars
around a given point in the 3-parameter space, we have divided the original box
in 8 cubes all with one corner at $\theta_{0}$, $\log g_{0}$, [M/H]$_{0}$. In
this way we make sure that there are stars in all the directions with respect
to the required point. The code finds all the stars $N_j$ enclosed within a
cube $j$ (where $j=1,\ldots,8$) and combines their spectra according to

\begin{equation}
{S_{\lambda}}^{j} = \frac{\sum_{i=1}^{N_j} {S_{\lambda}}_{i}^{j} W^{j}_{i}}
{\sum_{i=1}^{N_j} W^{j}_{i}}, 
\label{eq:Sj}
\end{equation}

\noindent
where $W_i^j$ is the weight assigned to the star $i$ within the cube $j$:

\begin{eqnarray}
W_i^j&=&\left[e^{-\left(\frac{\theta_i-\theta_0}{\sigma_{\theta_0}}\right)^2}
e^{-\left(\frac{\log g_i-\log g_0}{\sigma_{\log g_0}}\right)^2} 
e^{-\left(\frac{{\rm [M/H]}_i-{\rm [M/H]}_0}
{\sigma_{{\rm [M/H]}_0}}\right)^2}\right]\nonumber\\
&&SN_{n_i},
\label{eq:Wij}
\end{eqnarray}

\noindent 
where $SN_{n_i}$ is a measure of the quality of the empirical spectrum 
${S_{\lambda}}_{i}^{j}$ estimated by its signal-to-noise normalized
in the following way

\begin{equation}
\label{eq_weight_sn_star}
 \begin{array}{r@{}@{\;\;}c@{}@{\;\;}l@{}@{\;\;}l@{}}
 SN_{n_i} & = & 1,                         & SN_i \ge SN_{99.7}\\
 SN_{n_i} & = & \frac{SN_i^2}{SN_{99.7}^2},& SN_i   < SN_{99.7},\\
 \end{array}
\end{equation}

\noindent
where $SN_{99.7}$ is the limiting $SN$ value where the cumulative SN
distribution fraction of our stellar sample reaches the upper 99.7 percentile
(${\sum SN_i(SN_{i}<SN_{99.7})}{/\sum SN_i}$). We note, however, that a
spectroscopic binary or a star with anomalous signature of high variability
that is present in the stellar library (see \S~\ref{sec:library}) would be
able to contribute significantly within a given cube if its SN$_{n_i}$ is
high. To prevent this possibility we have chosen to get down the SN$_{n_i}$ of
these stars to the value where the cumulative SN distribution fraction of our
stellar sample reaches the lower 0.3 percentile. 

The term within the square brackets in Eq.~\ref{eq:Wij} represents the weight
resulting from the position of the star $i$ in the parameter space ($\theta$,
$\log g$, [M/H]). We have assumed a  gaussian-like function, which assigns
larger weights to the stars closer to the requested point ($\theta_{0}$, $\log
g_{0}$, [M/H]$_{0}$). This term, as well as $SN_{n_i}$, varies from 0 to 1.
The ${\sigma_{\theta_0}}$, ${\sigma_{\log g_0}}$ and ${\sigma_{{\rm [M/H]}_0}}$
values (generically ${\sigma_p}_0$) are estimated on the basis of the density
of stars at the requested point in the parameter space $\rho_{0}$. We assumed
an inversely proportional gaussian-like function  of the form:

\begin{equation}
\label{eq_density-sigma}
{\sigma_p}_0 \propto e^{{\frac{1}{2}}\left (\frac{\rho_0 - \rho_{\rm M}}
{{\sigma_{\rho}}_0}\right )^2},
\end{equation}

\noindent
where ${\rho_{\rm M}}$ is the maximum density. Assuming that the smallest
${\sigma_p}_{\rm m}$ is obtained at the point of maximum density ($\rho_{\rm M}$)
and that the largest ${\sigma_p}_{\rm M}$ is reached when ${\rho}$ $\rightarrow$
$0$ we derive ${\sigma_{\rho}}_0$ and then Eq.~\ref{eq_density-sigma} can be
written as:

\begin{equation}
\label{eq:density-sigma_ok}
 \begin{array}{r@{}@{\;}c@{}@{\;}l@{}}
{\sigma_{\theta}}_0 &=& {{\sigma_{\theta}}_{\rm m}} e^{\left 
( \frac{\rho_0 - \rho_{\rm M}}{\rho_{\rm M}}\right )^2 \ln
\frac{{\sigma_{\theta}}_{\rm M}}
{{\sigma_{\theta}}_{\rm m}}}\\
{\sigma_{\log g}}_0 &=& {{\sigma_{\log g}}_{\rm m}} e^{\left 
( \frac{\rho_0 - \rho_{\rm M}}{\rho_{\rm M}}\right )^2 \ln \frac{{\sigma_{\log
g}}_{\rm M}}
{{\sigma_{\log g}}_{\rm m}}}\\
{\sigma_{\rm [M/H]}}_0  &=& {{\sigma_{\rm [M/H]}}_{\rm m}} e^{\left ( \frac{\rho_0 -
\rho_{\rm M}}{\rho_{\rm M}}\right )^2 \ln \frac{{\sigma_{\rm [M/H]}}_{\rm M}}
{{\sigma_{\rm [M/H]}}_{\rm m}}}.\\
 \end{array}
\end{equation}

We have assumed ${\sigma_{\theta}}_{\rm m}$, ${\sigma_{\log g}}_{\rm m}$ and
${\sigma_{\rm [M/H]}}_{\rm m}$ to be the minimum uncertainty in the
determination of $\theta$, $\log g$ and [M/H] respectively. We adopted
${\sigma_{\theta}}_{\rm m} = 0.009$, ${\sigma_{\log g}}_{\rm m} = 0.18$ and
${\sigma_{\rm [M/H]}}_{\rm m} = 0.09$ according to the values given in
Paper~II. It is worth noting that the points where the uncertainties are the
smallest coincide with the most populated regions of the parameter space,
e.g. dwarfs of $\sim5800$~K and giants of $\sim4800$~K of solar
metallicity. On the other hand, for ${\sigma_{\theta}}_{\rm M}$,
${\sigma_{\log g}}_{\rm M}$ and ${\sigma_{\rm [M/H]}}_{\rm M}$ we assumed a
range of 1/10 of the total covered by these parameters in our stellar
library. We adopted ${\sigma_\theta}_{\rm M} = 0.17$, ${\sigma_{\log g}}_{\rm
M} = 0.51$ and ${\sigma_{\rm [M/H]}}_{\rm M} = 0.41$. For ${\sigma_{\theta}}$
we also assumed the condition that $60\le T_{\rm eff} \le 3350$~K
(corresponding to the adopted ${\sigma_{\theta}}$ limiting values when
transforming to the T$_{\rm eff}$ scale).

\begin{figure}
\centerline{\psfig{file=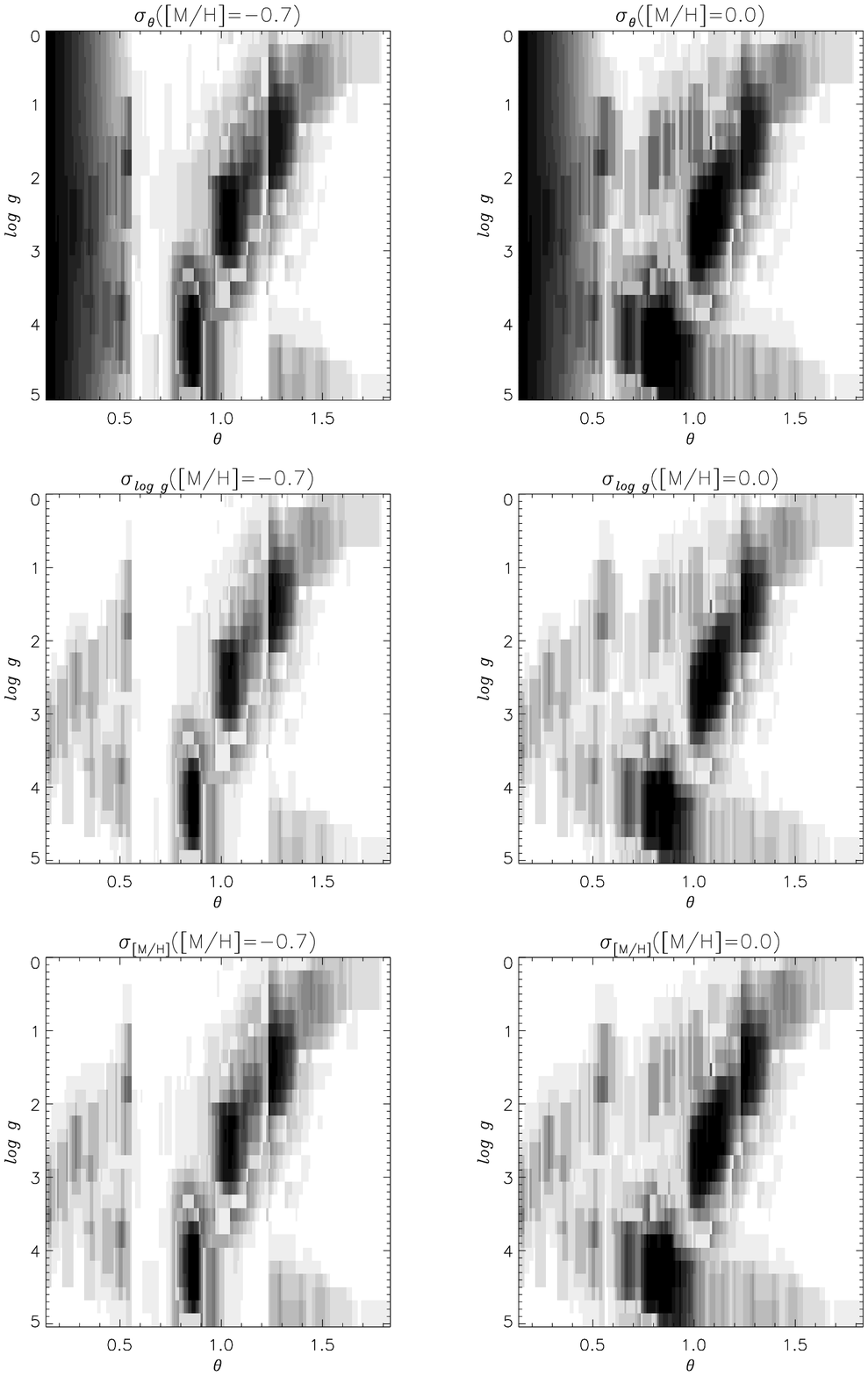,width=3.6in}}
\caption{
Grey levels representing the values of ${\sigma_{\theta}}$, ${\sigma_{\log g}}$
and ${\sigma_{\rm [M/H]}}$ (from top to bottom plots), calculated for [M/H] $=-0.7$
(left) and [M/H] $=0.0$ (right) in the $\theta-\log g$ plane. Black means
${\sigma_{\theta}}=0.009$, ${\sigma_{\log g}}=0.18$ and
${\sigma_{\rm [M/H]}}=0.09$, whereas white means ${\sigma_{\theta}}=0.17$,
${\sigma_{\log g}}=0.51$ and ${\sigma_{\rm [M/H]}}=0.41$ in the corresponding
plots.  
}
\label{fig:sigmas}
\end{figure}

At a given point $\rho_{0}$ is calculated by counting the number of stars
present in a box of size 3$\times{\sigma_p}_{\rm m}$. If no stars are found, the box
is simultaneously enlarged in its three dimensions, in steps of
1$\times{\sigma_p}_{\rm m}$, until at least one single star is reached. For stars
with temperatures larger than 9000~K and lower than 4000~K we do not take into
account the metallicity parameter, since its uncertainty is very large and
since we adopted [Fe/H] $=0$ for stars with unknown metallicity in these
temperature ranges. In practise, we adopt ${\rho_{\rm M}}$ as the density value where
the cumulative density fraction reaches the upper percentile 99.7. For larger
densities we adopt this value as well. An example of the resulting
${\sigma_{\theta}}$, ${\sigma_{\log g}}$ and ${\sigma_{\rm [M/H]}}$ is shown in
Figure~\ref{fig:sigmas} for two values of the metallicity. Black means
${\sigma_{\theta}}=0.009$, ${\sigma_{\log g}}=0.18$ and
${\sigma_{\rm [M/H]}}=0.09$, whereas white means ${\sigma_{\theta}}=0.17$,
${\sigma_{\log g}}=0.51$ and ${\sigma_{\rm [M/H]}}=0.41$ in the corresponding
plots. Overall, the grey levels for the plots corresponding to [M/H] $=0.0$ are
significantly darker than the ones for [M/H] $=-0.7$ due to the fact that our
library is more populated with solar metallicity stars. Note that the MS and
RGB phases are clearly visible. An interesting feature is that for the
${\sigma_{\theta}}$ plots the grey levels become gradually darker towards
hotter stars as a result of the fact that we assumed that the largest
${\sigma_{\theta}}$ value cannot be larger than 1/10 of total range in the
$T_{\rm eff}$ scale. Note the jump in the grey levels when $\theta$ reaches values
around 0.55 and 1.25 due to the fact that for very hot or very cool stars we
did not consider their metallicities. Obviously this jump is more pronounced
for the plots corresponding to [M/H] $=-0.7$.

Once the density is known and ${\sigma_{\theta}}_0$, ${\sigma_{\log g}}_0$ and
${\sigma_{\rm [M/H]}}_0$ have been estimated according to
Eq.~\ref{eq:density-sigma_ok}, we find all the stars enclosed in each of the
eight cubes of size $1.5{\sigma_p}_0$. For the most populated regions of the
parameter space we find stars in all these eight cubes. However, if no stars
are found in a given cube, we simultaneously enlarge its three sizes in steps
of 0.5$\times{\sigma_p}_0$. Since one of the corners of the cube is always
located on the required parameter point itself, the box is enlarged to one
side only. In the worst case, this iterative procedure ensures finding stars
in, at least, three cubes. Finally, a similar criterion to that applied for
calculating $\rho_{0}$ is taken into account for stars cooler than 4000~K and
hotter than 9000~K, i.e. neglecting their metallicity.

Having obtained a representative spectrum for each cube,
${S_{\lambda}}^{j}$, we are now in a position to calculate
$S_{\lambda}(m,t,{\rm [M/H]})$. The weight to be assigned to ${S_{\lambda}}^{j}$
is estimated as follows

\begin{equation}
\label{eq:Wj}
W^j=e^{-\left(\frac{\theta^j-\theta_0}{\sigma_{\theta_0}}\right)^2}
e^{-\left(\frac{\log g^j-\log g_0}{\sigma_{\log g_0}}\right)^2}
e^{-\left(\frac{{\rm [M/H]}^j-{\rm [M/H]}_0}{\sigma_{{\rm [M/H]}_0}}\right)^2}, 
\end{equation}
 
\noindent
where $\theta^j$, $\log g^j$ and [M/H]$^j$ are given by

\begin{equation}
 \begin{array}{r@{}@{\;}c@{}@{\;}l@{}}
\theta^j          &=& \frac{\sum_{i=1}^{N_j} \theta^{j}_{i}
W^{j}_{i}}{\sum_{i=1}^{N_j} 
W^{j}_{i}}\\
\log g^j          &=& \frac{\sum_{i=1}^{N_j} \log g^{j}_{i}
W^{j}_{i}}{\sum_{i=1}^{N_j} 
W^{j}_{i}}\\
{\rm [M/H]}^j&=& \frac{\sum_{i=1}^{N_j}{\rm [M/H]}^{j}_{i}
W^{j}_{i}}
{\sum_{i=1}^{N_j} W^{j}_{i}}.\\
 \end{array}
\end{equation}

We then solve for the following undetermined linear system of equations

\begin{equation}
 \begin{array}{r@{}@{\;}c@{}@{\;}l@{}}
\sum_{j=1}^{k} \alpha^j W^j (\theta^j - \theta_0) &=& 0\\
\sum_{j=1}^{k} \alpha^j W^j (\log g^j - \log g_0) &=& 0\\
\sum_{j=1}^{k} \alpha^j W^j ({\rm [M/H]}^j  - {\rm [M/H]}_0)  &=& 0,\\
 \end{array}
 \label{eq:linear-eq}
\end{equation}

\noindent 
where $k$ represents the number of available cubes that contain stars after
the iterative procedure explained above ($3\le k \le8$), and $\alpha^j$ are
coefficients to be applied for correcting the weights $W^j$ to be assigned to
the spectra ${S_{\lambda}}^{j}$ when combining them to obtain
$S_{\lambda}(m,t,{\rm [M/H]})$. Only values in the range $0\le\alpha^j<1$ are
acceptable for these coefficients. Among the solutions we choose the one that
minimizes the modification of the intrinsic weights $W^j$'s. This is performed
by calculating the projection of the solution hyperplane onto the point
($\alpha^j=1$, $j=1$, $k$) of the $k$-dimensional space. When no  non-negative
$\alpha^j$ solutions   can be found taking into account the $k$ cubes, we
start a procedure where we neglect the contribution of a given cube in
Eq.~\ref{eq:linear-eq}. This is performed following the criterion of removing
first the cubes with lower intrinsic weights according to ${SN_{n}}^j W^j$
where  ${SN_{n}}^j$\footnote{Note that we do not adopt the true SN
corresponding to ${S_{\lambda}}^{j}$ since the most populated cubes would be
assigned considerably higher weights, an effect which tends to emphasize the
asymmetries present in several regions of the parametrical space.} is given by

\begin{equation}
{SN_{n}}^{j} = \frac{\sum_{i=1}^{N_j} {SN_{n}}^{j}_{i} W^{j}_{i}}
{\sum_{i=1}^{N_j} W^{j}_{i}}. 
\end{equation}

If no satisfactory solutions are found, we select the one which
minimizes the difference between the required atmospheric parameters
and those resulting from the solution star having taken into account
$\alpha^j$'s as in Eq.~\ref{eq:linear-eq}. The final spectrum,
$S_{\lambda}(m,t,{\rm [M/H]})$, is obtained as follows

\begin{equation}
{S_{\lambda}}(m,t,{\rm [M/H]}) = \frac{\sum_{j=1}^{k} {S_{\lambda}}^j \alpha^j W^j}
{\sum_{j=1}^{k} \alpha^j W^j}, 
\end{equation}

\noindent   
where ${S_{\lambda}}^j$ are normalized to the selected wavelength reference
interval $\Delta\lambda_{\rm ref}$ and scaled according to the derived
$\alpha^jW^j$.

\section{Model Uncertainties}
\label{ap:uncertain}

In this Appendix we perform a number of tests to evaluate the major
uncertainties affecting the obtained SSP model predictions presented in this
paper. We first focus on the uncertainties derived from varying the model
calculation details, particularly in the way that we compute a representative
stellar spectrum. We then explore the effects of varying the adopted
theoretical prescriptions for calculating the stellar evolutionary tracks and
isochrones. To illustrate how significant a given effect is, we have chosen to
measure the variation of the most important feature in this spectral range,
i.e. the Ca\,{\sc ii} triplet, by means of the CaT$^{*}$ index.

\subsection{Uncertainties derived from the gridding of the stellar library}
\label{ap:griding}

\begin{table*}                                                              
\centering{
\caption{Uncertainties of model calculation details on the CaT$^*$ index for 
SSPs of age 12.59~Gyr, unimodal IMF of slope 1.3 and different metallicities}
\label{tab:uncertain}                                                            
\begin{tabular}{lccr@{$\;$}r@{$\;$}r@{$\;$}r}
\hline     
Test & Reference &  Adopted &\multicolumn{4}{c}{CaT$^*$(adopted)$-$CaT$^*$(reference) (\AA)}\\
\cline{4-7}
\multicolumn{3}{c}{}&[M/H]=$-$0.68 & [M/H]=$-$0.38 & [M/H]=0.0 & [M/H]=+0.20\\
\hline
T$_{\rm eff}$ where [M/H] is neglected                     &   4000~K& 3750~K &$-$0.080& 0.059& 0.046& 0.052\\
T$_{\rm eff}$ where [M/H] is neglected                     &   4000~K& 4250~K & 0.020&$-$0.005&$-$0.012& 0.003\\
\\
${\sum \rho_i(\rho_{i}<\rho_{\rm M})}{/\sum \rho_i}$&99.7$\%$&100$\%$&$-$0.037&$-$0.020& 0.000& 0.006\\
${\sum \rho_i(\rho_{i}<\rho_{\rm M})}{/\sum \rho_i}$&99.7$\%$& 90$\%$&$-$0.012& 0.015& 0.028& 0.008\\
\\
Starting finding box size ($\times{\sigma_p}_0$)&1.5      &0.5     &$-$0.090&$-$0.118& 0.000& 0.016\\
Starting finding box size ($\times{\sigma_p}_0$)&1.5      &1.0     &$-$0.071&$-$0.020& 0.000& 0.001\\
Starting finding box size ($\times{\sigma_p}_0$)&1.5      &3.0     &$-$0.037&$-$0.047& 0.009& 0.033\\
\\
Finding box size enlargement per iteration ($\times{\sigma_p}_0$)&0.5&1.5&$-$0.005& 0.002& 0.015&$-$0.021\\
\\
$p$ range fraction adopted for ${\sigma_p}_{\rm M}$ & 1/10&1/6      &$-$0.166& 0.002& 0.016& 0.059\\
$p$ range fraction adopted for ${\sigma_p}_{\rm M}$ & 1/10&1/8      &$-$0.083&$-$0.010&$-$0.008& 0.030\\
$p$ range fraction adopted for ${\sigma_p}_M$ & 1/10&1/12     & 0.000& 0.029&$-$0.003&$-$0.001\\
\hline
\end{tabular}                                                              
}                                                                          
\end{table*}

Here we study the effects of varying the way in which we compute a stellar
spectrum according to the description given in \S~\ref{sec:ss} and, particularly,
in Appendix~\ref{ap:boxes}. We select for this purpose a representative set of
SSP models of age 12.59~Gyr (old stellar populations are more sensitive to the
lack of cool stars) and metallicities [M/H] $=-0.68, -0.38, 0.0 and +0.2$.

We first test to vary the lower temperature limit, below which the
metallicities of the stars are not taken into account when calculating a
representative spectrum corresponding to a given set of atmospheric parameters.
Table~\ref{tab:uncertain} lists the obtained residuals for the CaT$^{*}$ index
when increasing this temperature from the adopted value, i.e. 4000~K, to
4250~K, and when decreasing it to 3750~K. We note that above 4000~K the
obtained differences change very softly. Varying this limit
towards temperatures lower than 3750~K has virtually no effect when compared to
the values obtained for this temperature limit, due to the fact that most of
the stars have either solar metallicity or have been assumed to be solar when
no metallicity values were available in the literature. We conclude that this
temperature limit is mostly affecting the integrated spectrum of SSPs of
[M/H] $=-0.68$, but the largest variation in the CaT$^{*}$ index is $\sim1$\% of
the CaT$^{*}$ index strength.

We also explored the effect of varying the density value ${\rho_{\rm M}}$
where the cumulative density fraction of stars reaches the upper percentile
from 99.7 to 100 and 90 (see Table~\ref{tab:uncertain}). In practice for an upper 
percentile value of 100, this
assumption means that there is a single point in the stellar parametric space
where ${\sigma_{\theta}}_{\rm m}$, ${\sigma_{\log g}}_{\rm m}$ and ${\sigma_{\rm
[M/H]}}_{\rm m}$ adopt the values corresponding to the minimum uncertainties in the
determination of the stellar parameters.

Another interesting test is to explore the effect of varying the size of the
parametrical cubes in which we start to find stars. We adopted in this paper
1.5$\times{\sigma_p}_0$ but we list in the same table the values obtained for
0.5 1.0 and 3$\times{\sigma_p}_0$. A different, but
related test, is to vary the applied enlargement to the size of these cubes
when no stars are found after each iteration. The obtained residuals when
adopting 1.5$\times{\sigma_p}_0$, rather than 0.5$\times{\sigma_p}_0$, are
tabulated in Table~\ref{tab:uncertain}. It is clear that the starting finding
box size plays a more relevant role than the way in which the code expands this
box when no stars are found. The largest residual, i.e. $-0.12$\AA, is found for
[M/H] $=-0.38$ when adopting $0.5\times{\sigma_p}_0$. However, for larger starting
size values the obtained differences decrease.

Finally, we show the effect of adopting for ${\sigma_{\theta}}_{\rm M}$,
${\sigma_{\log g}}_{\rm M}$ and ${\sigma_{\rm [M/H]}}_{\rm M}$ several
fractions of the total range covered by these parameters. Our reference value
is 1/10, but Table~\ref{tab:uncertain} lists the residuals when we set this
limit to 1/6, 1/8 and 1/12. By increasing this fraction we increase the
weights assigned to the stars in the less populated regions of the
parametrical space. This allow us to include stars whose parameters are more
distant from the requested point.  The largest fractions (e.g. 1/6) provides
the largest effect on the \cat feature for the smallest and largest
metallicities (i.e. [M/H] $=-0.68$ and [M/H] $=+0.2$), where the CaT$^{*}$ weakens
by 0.17\AA\ for [M/H] $=-0.68$ and strengthen by 0.06\AA\ for [M/H] $=+0.2$ with
respect to our reference value.  However, these differences become smaller as
we approach to our reference value, whilst for smaller fractions (e.g. 1/12)
the obtained differences are almost negligible for all metallicities. Overall
we conclude that the SSP model predictions for the largest and smallest
metallicities are subject to the largest uncertainties.

\subsection{Effects of varying the theoretical prescriptions of the stellar
tracks}
\label{ap:isochrones}

Here we focus on the effects caused by varying the theoretical
prescriptions adopted for calculating the stellar evolutionary tracks and
isochrones. We first discuss the effects derived from the most uncertain
stellar evolutionary phase, i.e. the AGB. Then we discuss the effects of
specific theoretical prescriptions affecting intermediate-aged stellar
populations, such as the adopted convective overshooting scheme. Next we focus
on those prescriptions affecting old-aged stellar populations. Finally we
explore the effects of adopting stellar isochrones build-up on the basis of
non-solar element ratios.

We first show the effects of the AGB stellar evolutionary phase. We illustrate
this study by means of performing a very crude approach, which is to remove
this phase from the isochrones. Table~\ref{tab:uncertainAGB} lists the obtained
residuals for the CaT$^*$ index for SSPs of different ages and metallicities.
For old stellar populations the effect of this evolutionary phase is very
small, and the obtained residuals are of the order of the uncertainties derived
in Appendix~\ref{ap:griding}. As expected, the largest residuals, which can
reach as high as $\Delta{\rm CaT}^*\sim0.5$~\AA, are obtained for ages below
$\sim1$~Gyr. These residuals are significantly larger than the ones listed in
Table~\ref{tab:uncertain}. Obviously, this is a very crude test, which
completely neglects our current knowledge of this late stellar evolutionary
phase. In practise one expects to obtain smaller residuals due to uncertainties
in the AGB. This test shows us that the results presented in this paper for old
stellar populations are stable against variations in the input physics of
the most uncertain stellar evolutionary phase.

\begin{figure}
\centerline{\psfig{file=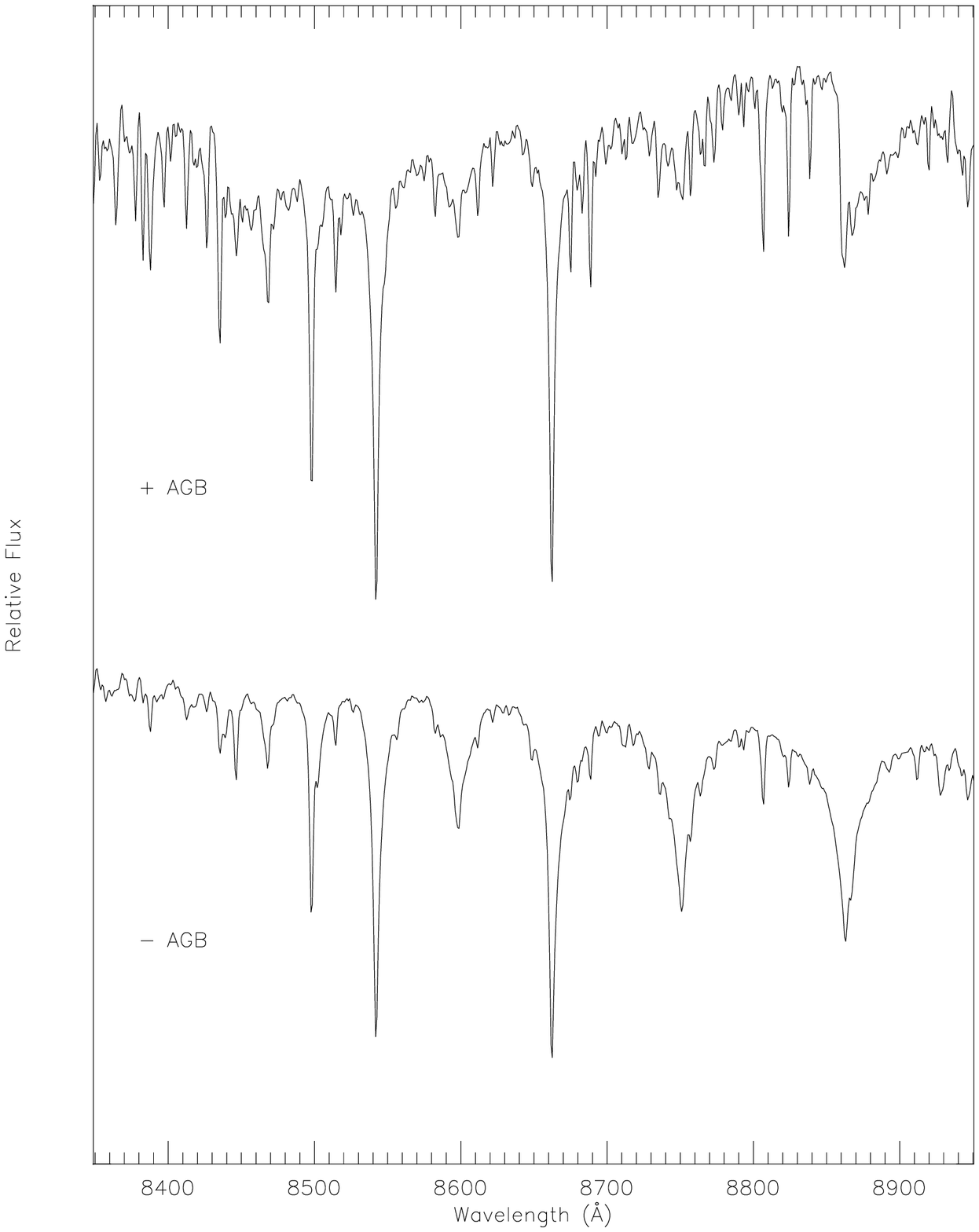,width=3.6in}}
\caption{
Synthetic SSP spectra for solar metallicity, 0.2~Gyr and unimodal IMF
($\mu=1.3$). For the upper spectrum we have included the AGB phase, whilst for
the bottom spectrum we have excluded this stellar evolutionary phase. The
spectral resolution is 1.5~\AA~(FWHM).
}
\label{fig:cat_spectra_agb}
\end{figure}

We plot in Figure~\ref{fig:cat_spectra_agb} the obtained spectra for an SSP of
solar metallicity and 0.2~Gyr synthesized on the basis of including and
excluding the whole AGB stellar evolutionary phase. We see that the overall
shapes of these spectra are remarkably different. For these intermediate-aged
SSPs the AGB stars are able to shape the continuum around the \cat feature in a
way which is characteristic of M-type stars. Furthermore, without the AGB phase
the Paschen lines clearly dominate the synthesized SSP spectra. This exercise
also shows us to what extent is important to work with an appropriate stellar
spectral library, with a great coverage of the stellar atmospheric parameters
including cool stars.

\begin{table}                                                              
\centering{ 
\caption{The effect of the AGB on the CaT$^*$ for SSPs of different
ages and metallicities. We use models of unimodal IMF of slope 1.3.
}
\label{tab:uncertainAGB}
\begin{tabular}{cc@{$\;$}c@{$\;$}c@{$\;$}c} 
\hline 
Age(Gyr) &\multicolumn{4}{c}{$\Delta$CaT$^*$(AGB[no--yes]) (\AA)}\\ 
\cline{2-5}
 &[M/H]=-0.68 & [M/H]=-0.38 & [M/H]=0.0 & [M/H]=+0.20\\ 
\hline 
0.20 &       &             &$-$0.499     &            \\ 
1.00 &       &             &$-$0.087     & 0.157      \\ 
3.16 &       &        0.000& 0.185     & 0.120      \\ 
10.00& 0.000 &       $-$0.058& 0.035     & 0.029      \\ 
\hline
\end{tabular}                                                              
}                                                                          
\end{table}

\begin{figure}
\centerline{\psfig{file=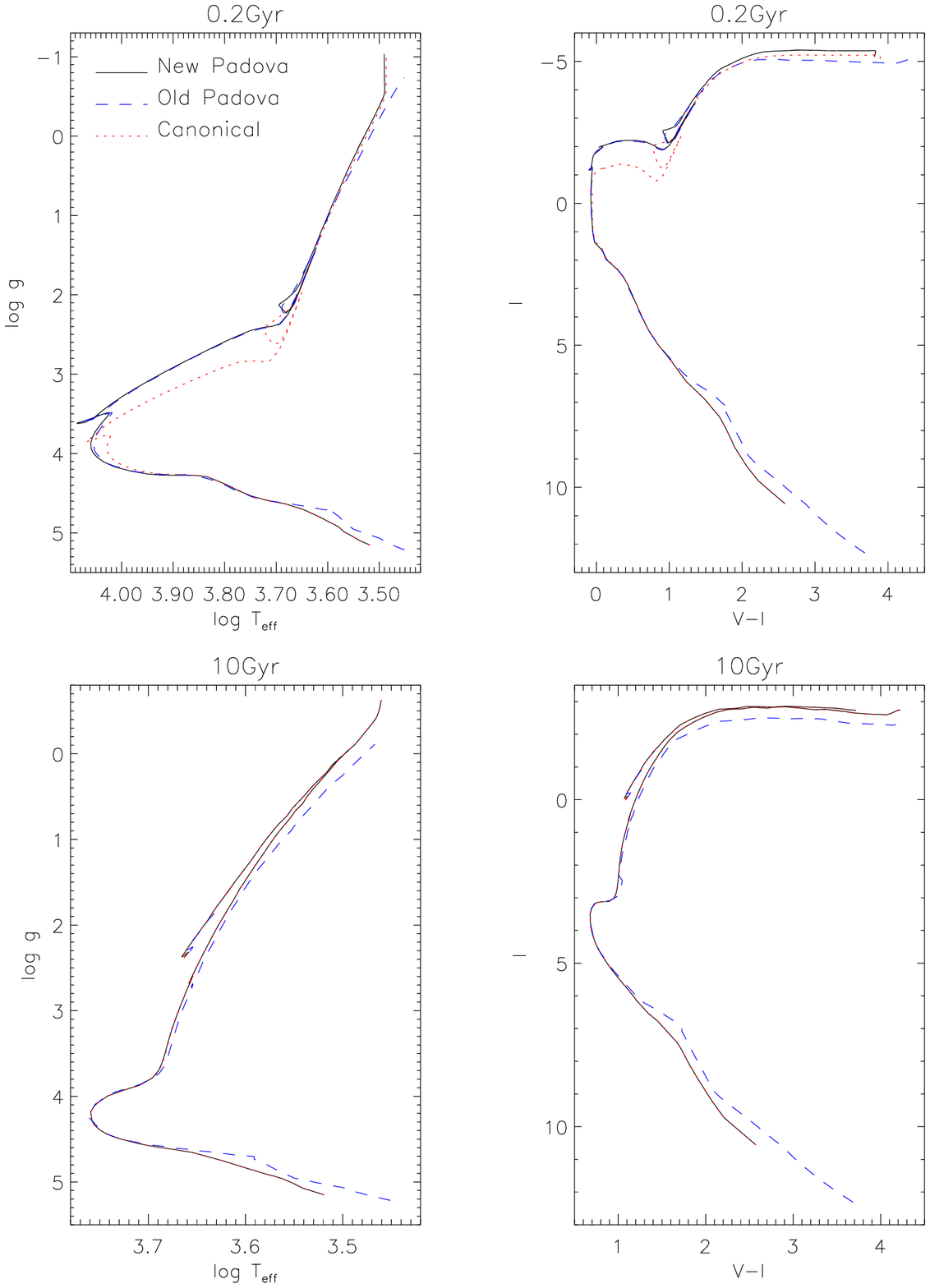,width=3.6in}}
\caption{
Plot of the new, old and canonical (i.e. without overshooting scheme) Padova
isochrones for solar metallicity and 0.2 and 10~Gyr. See text for details.
}
\label{fig:isochrones}
\end{figure}

We have also explored the effects of varying other important theoretical
prescriptions of the stellar evolutionary tracks, such as the adoption of a
convective overshooting scheme. The new Padova isochrones (G00)
have been updated on the basis of a significantly milder overshooting
than that adopted for the old Padova isochrones (B94).
Furthermore, G00 also provide a canonical set of solar metallicity
isochrones, following the classical Schwarzschild criterion for the convective
boundaries (i.e. without overshooting). The overshooting is only important for
stellar masses larger than $\sim1.2$~M$_\odot$, therefore old-aged stellar
populations are insensitive to the adopted scheme. We have transformed all
these isochrones to the observational plane following the same prescriptions
described in \S~\ref{sec:transformations}. We then computed SSPs for different
ages (i.e. from 0.2 to 10~Gyr) for solar metallicity and unimodal IMF
($\mu=1.3$). In Table~\ref{tab:uncertain_isoc} we list the obtained CaT$^*$
residuals. For SSPs of 0.2 and 1.0~Gyr the CaT$^*$ values provided by the
canonical isochrones are larger than those based on the new Padova set, which
are larger than those obtained from the old Padova isochrones. In fact we
obtain a total residual $\Delta{\rm CaT}^*\sim1$~\AA. As expected, no significant
differences are found for old stellar populations due to the adopted
overshooting scheme (see the first column of residuals in
Table~\ref{tab:uncertain_isoc}), since stars with masses smaller than
$\sim1.1{\rm M}_\odot$ present a radiative core during the MS.
Fig.~\ref{fig:isochrones} shows the main differences between these three sets
of isochrones for different ages. 

\begin{table}                                                              
\centering{ 
\caption{
CaT$^*$ index residuals due to different theoretical prescriptions in the
stellar evolutionary tracks. The obtained residuals are calculated for SSPs of
different ages (indicated in gigayears), solar metallicity and unimodal IMF
($\mu=1.3$).
}
\label{tab:uncertain_isoc}                                                            
\begin{tabular}{@{}cc@{$\;$}c@{}} 
\hline 
Age&$\Delta$CaT$^*$(Overshoot[no--yes])&$\Delta$CaT$^*$(Padova[old--new])\\ 
\hline 
0.20 &  0.481 &$-$0.600\\
1.00 &  0.826 &$-$0.146\\
3.16 & $-$0.016 &$-$0.631\\
10.00& $-$0.003 &$-$0.533\\
\hline
\end{tabular}                                                              
}                                                                          
\end{table}

For old-aged SSPs we find CaT$^*$ residuals larger than 0.5~\AA~ when
comparing the old and new Padova isochrones. The CaT$^*$ values provided by
the old Padova set are smaller because they predict slightly cooler RGBs as
shown in Fig.~\ref{fig:isochrones}. For giant stars with temperatures smaller
than $\sim3500$~K the \cat decreases with decreasing temperature (see
Paper~III). An exhaustive comparison of the theoretical parameters used to
build-up these two sets of isochrones is out of the scope of this paper and we
refer the interested reader to the G00 paper. However, it is
worth recalling that if we replaced the model grids shown in
Fig.~\ref{fig:cat_G_HG} with those synthesized on the basis of the old Padova
set of isochrones, which yield smaller CaT$^*$ strengths, 
we would derive smaller IMF slopes, i.e.
$\Delta\mu\sim1$ for the two galaxies plotted in this figure (see
\S~\ref{sec:G}). Therefore, users of these SSP model predictions should be
warned that, for this reason, IMF slope estimates must be taken into
account on a relative basis. Despite the caveats, we believe that our
predictions are still useful for obtaining robust conclusions, particularly
when discussing possible trends among galaxies (see C03).

\begin{table}                                                              
\centering{ 
\caption{
CaT$^*$ residuals obtained for SSPs of different ages and metallicities
(unimodal IMF of slope 1.3) calculated on the basis of the $\alpha$-enhanced
and scaled-solar isochrones of S00. We use a unimodal IMF
($\mu=1.3$).
}
\label{tab:uncertain_alpha}                                                            
\begin{tabular}{cc@{$\;$}c@{$\;$}c} 
\hline 
\hline 
Age(Gyr)&\multicolumn{3}{c}{$\Delta$CaT$^*$($\alpha$-enhanced -- scaled-solar) (\AA)}\\ 
\cline{2-4}
 & [M/H] $=-0.38$ & [M/H] $=0.0$ & [M/H] $=+0.20$\\ 
\hline 
 0.20&     &0.954&     \\
 1.00&     &0.692&0.555\\
 3.16&0.299&0.475&0.435\\
10.00&0.269&0.541&0.517\\
\hline
\end{tabular}                                                              
}                                                                          
\end{table}

Elliptical galaxies show an enhancement of Mg over Fe when compared to
scaled-solar SSPs model predictions (e.g. Peletier 1989; Worthey et al.  1992;
V97; K00; V01A).  Therefore, it is worth exploring the \cat strengths in SSPs
synthesized on the basis of enhanced $\alpha$-elements mixtures. Such models
require both stellar tracks calculated with appropriate opacity tables and
stellar spectral libraries of similar $\alpha$-enhanced ratios. However, no
consensus has yet been reached for the degree of enhancement of different
$\alpha$-elements. For example, a constant enhancement for each
$\alpha$-element was adopted by VandenBerg et al. (2000) and Kim et al.
(2002), whilst a somewhat more empirically-guided mixtures have been used by
SW98 and S00. These authors find that, for a given total metallicity, the
adoption of $\alpha$-enhanced mixtures leads to higher effective temperatures
with respect to scaled-solar for all evolutionary phases, and that, at
relatively high metallicities and old ages, an $\alpha$-enhanced isochrone
cannot be mimicked by using a scaled-solar isochrone of different
metallicity. It is worth noting that all these stellar models adopt calcium
enhanced ratios as it is an $\alpha$-element.  However the Ca~4227 line does
not seem to track Mg but Fe in elliptical galaxies (V97; Trager et al. 2000a;
T00; V01A; PS02). On the other hand we do not expect any significant variation
in the isochrone as a result of varying the degree of enhancement of this
element, as its contribution to the total metallicity is lower than 0.5\%. The
other important problem in synthesizing $\alpha$-enhanced SSPs is the lack of
stellar spectral libraries of appropriate $\alpha$-enhanced mixtures. In
particular, the empirical library used by our model follows the Galactic disk
element ratios.

In spite of the fact that we are unable to build up such fully
self-consistent $\alpha$-enhanced SSP models for this spectral range, we are
in position to evaluate the isochrone effects on the synthesized spectra. For
this purpose we have made use of both $\alpha$-enhanced and scaled-solar
isochrones of S00, which are constructed on the basis of
almost the same input physics as that used in G00.
Table~\ref{tab:uncertain_alpha} lists the residuals obtained for the CaT$^*$
index for SSPs of different ages and total metallicities. We find that the
$\alpha$-enhanced isochrones provide larger CaT$^*$ strengths for all the
tabulated ages and metallicities. This is due to the fact that the RGB of
these isochrones is hotter and therefore there is a smaller number of stars
falling into the cool temperature regime, where the strength of the CaT$^*$
decreases as a function of decreasing temperature (see \S~\ref{sec:catagemet}
and Paper~III for more details). We also see that the obtained residuals
decrease as function of increasing age. In particular for old stellar
populations of total metallicities [M/H] $\ge0$ we obtain
$\Delta{\rm CaT}^*\sim0.5$~\AA, whilst for [M/H] $=-0.38$ we obtain
$\Delta{\rm CaT}^*\sim0.3$~\AA. Therefore, if we plotted these results on
Fig.~\ref{fig:cat_G_HG} we would obtain steeper IMF slopes for these two
elliptical galaxies. Put in other way, if a Salpeter IMF is used, the
$\alpha$-enhanced isochrones lead to a larger calcium underabundance problem.

\end{document}